\documentclass[11pt,oneside,notitlepage]{article}

\usepackage[letterpaper, total={6.5in, 8.5in}]{geometry}

\usepackage{amsmath}
\usepackage{amssymb}
\usepackage{amsthm}
\usepackage{array}
\usepackage{bm}
\usepackage{mathrsfs}
\usepackage{mathtools}

\usepackage[font=small]{caption} 
\usepackage[font=small]{subcaption}

\usepackage{cite}
\usepackage{comment}
\usepackage{enumitem}
\usepackage{float}
\usepackage[T1]{fontenc}
\usepackage{graphicx}
\usepackage[
colorlinks=true,
citecolor=MidnightBlue,
linkcolor=BrickRed,
urlcolor=MidnightBlue,
breaklinks=true,
linktoc=all
]{hyperref}
\usepackage{listings}
\usepackage{multicol}
\usepackage{stmaryrd}
\usepackage{textgreek}
\usepackage{titlesec}
\usepackage{tocloft}
\usepackage{wrapfig}
\usepackage[usenames,dvipsnames]{xcolor}
\usepackage{xy}
\usepackage{amsmath}
\usepackage{relsize}
\usepackage{empheq}
\usepackage{comment}
\usepackage{booktabs}
\usepackage{layouts}
\usepackage{environ}
\usepackage{pdflscape}
\usepackage{afterpage}
\usepackage{placeins} % floatbarrier
\usepackage[normalem]{ulem} % for strikeout

\usepackage{pgfplots}
\pgfplotsset{compat=1.15}
\usepackage{pgfmath}
\usetikzlibrary{patterns, calc, shapes.geometric, positioning, arrows.meta,backgrounds}
\usepackage{tikz-imagelabels}
\usepackage{tikz}

\newboolean{showcaption}
\setboolean{showcaption}{true}
\newcommand{\ifcap}[2]{\ifthenelse{\boolean{showcaption}}{#1}{#2}}

\def\imagebox#1#2{\vtop to #1{\vfill\null\hbox{#2}\vfill}}

\makeatletter
\newsavebox{\measure@tikzpicture}
\NewEnviron{scaletikzpicturetowidth}[1]{%
  \def\tikz@width{#1}%
  \begin{lrbox}{\measure@tikzpicture}%
  \BODY
  \end{lrbox}%
  \pgfmathparse{#1/\wd\measure@tikzpicture}%
  \BODY
}
\makeatother

\let\originalleft\left
\let\originalright\right
\newcommand{\leftR}{\mathopen{}\mathclose\bgroup\originalleft}
\newcommand{\rightR}{\aftergroup\egroup\originalright}

\newcommand{\diff}[1]{~\mathrm{d}#1}
\newcommand{\ddiff}[1]{\mathrm{d}#1}
\newcommand{\norm}[2]{\lvert \lvert #1 \rvert \rvert_{#2}}
\newcommand{\ChebT}[2]{P_{#1}\leftR(#2\rightR)}
\newcommand{\ChebTR}[1]{P_{#1}}
\newcommand{\curl}[1]{\mathrm{curl}\leftR(#1\rightR)}
\newcommand{\divv}[1]{\mathrm{div}\leftR(#1\rightR)}
\newcommand{\grad}[1]{\mathrm{grad}\leftR(#1\rightR)}
\newcommand{\laplaces}[1]{\Delta_\mathrm{s} #1}

\newcommand{\phik}[1]{\phi_{#1} \leftR(\zeta^\alpha \rightR)}
\newcommand{\xk}[1]{\bm{x}_{#1} \leftR(\zeta^\alpha \rightR)}

\newcommand{\textspace}{\smallskip}

\tikzset{
    position/.style args={#1:#2 from #3}{
        at=(#3), xshift=#1, yshift=#2
    }
}

\titleformat{\paragraph}[hang]{\normalfont\normalsize\bfseries}{\theparagraph}{1em}{}
\titlespacing*{\paragraph}{0pt}{3.25ex plus 1ex minus .2ex}{0.5em}

\definecolor{dkgreen}{rgb}{0,0.6,0}
\definecolor{gray}{rgb}{0.5,0.5,0.5}
\definecolor{mauve}{rgb}{0.58,0,0.82}

\newcolumntype{A}{ >{$} r <{$} @{} >{${}} l <{$} } % A for "align"
\allowdisplaybreaks

\renewcommand{\thefootnote}{\fnsymbol{footnote}}

\usepackage{sepfootnotes}

\makeatletter
\def\@fnsymbol#1{%
	\ensuremath{\ifcase#1\or
	\ddagger\or			% 1
	\mathsection\or		% 2
	*\or		% 3
	\dagger\or			% 4
	\mathflat\or				% 5
	\|\or				% 6
	\ddagger\ddagger\or	% 7
	\dagger\dagger\or	% 8
	**					% 9
	\else\@ctrerr\fi}}
\makeatother

\pgfdeclarelayer{bg0}    
\pgfdeclarelayer{bg1}
\pgfsetlayers{bg0,bg1,main}  % set the order of the layers (main is the standard layer)

\begin{document}

	\begin{center}
    
		{\textbf{
                {\large The $(2+\delta)$-dimensional theory of the  electromechanics of lipid membranes: \\[4pt] I. Electrostatics}
            }} \\

		\vspace{0.21in}

            % email addresses
            \sepfootnotecontent{YO}{\href{mailto:yannick.omar@berkeley.edu}{yannick.omar\textit{@}berkeley.edu}} 
            %%%%
            \sepfootnotecontent{ZL}{\href{mailto:zlipel@berkeley.edu}{zlipel\textit{@}berkeley.edu}}
            \sepfootnotecontent{KM}{\href{mailto:kranthi@berkeley.edu}{kranthi\textit{@}berkeley.edu}}

		{\small
			Yannick A. D. Omar\textsuperscript{1,}\sepfootnote{YO}\,, Zachary G. Lipel\textsuperscript{1,}\sepfootnote{ZL}\,,
			and Kranthi K. Mandadapu\textsuperscript{1,2,}\sepfootnote{KM}\\
		}
		\vspace{0.25in}

		{\footnotesize
			{
				$^1$
				Department of Chemical \& Biomolecular Engineering,
				University of California, Berkeley, CA 94720, USA
				\\[3pt]
				$^2$
				Chemical Sciences Division, Lawrence Berkeley National Laboratory, CA 94720, USA
				\\
			}
		}
	\end{center}

	\vspace{13pt}
	%
	% *** Abstract
	%

	\begin{abstract}
             The coupling of electric fields to the mechanics of lipid membranes gives rise to intriguing electromechanical behavior, as, for example, evidenced by the deformation of lipid vesicles in external electric fields. Electromechanical effects are relevant for many biological processes, such as the propagation of action potentials in axons and the activation of mechanically-gated ion channels. Currently, a theoretical framework describing the electromechanical behavior of arbitrarily curved and deforming lipid membranes does not exist. Purely mechanical models commonly treat lipid membranes as two-dimensional surfaces, ignoring their finite thickness. While holding analytical and numerical merit, this approach cannot describe the coupling of lipid membranes to electric fields and is thus unsuitable for electromechanical models. In a sequence of articles, we derive an \textit{effective} surface theory of the electromechanics of lipid membranes, named a $(2+\delta)$-dimensional theory, which has the advantages of surface descriptions while accounting for finite thickness effects. The present article proposes a new, generic dimension-reduction procedure relying on low-order spectral expansions. This procedure is applied to the electrostatics of lipid membranes to obtain a $(2+\delta)$-dimensional theory that captures potential differences across and electric fields within lipid membranes. This new model is tested on different geometries relevant for lipid membranes, showing good agreement with the corresponding three-dimensional electrostatics theory. 
	\end{abstract}   

	\vspace{15pt}
	%
	% *** TABLE OF CONTENTS
	%

	{ \hypersetup{linkcolor=black} \tableofcontents }
	\vspace{20pt}

	%
	% *** CONTENT 
	%

        % swtich back to numeric footnotemarkers
        \renewcommand{\thefootnote}{\arabic{footnote}}
        \setcounter{footnote}{0} 
        
        \section{Introduction}

Lipid membranes separate the interior and exterior of a biological cell and its organelles, serving as barriers that regulate the transport of proteins, ions, and other molecules. They exist in various, dynamically-changing shapes, with radii of curvature ranging from tens to hundreds of nanometers. In contrast, they are comprised of only two layers of lipid molecules, forming a bilayer structure with a thickness of just $3-5~\mathrm{nm}$. This makes lipid membranes exceptionally thin materials. \textspace 

The thin, bilayer structure of lipid membranes gives rise to peculiar mechanical behavior.
In-plane stretch and out-of-plane bending indicate elastic behavior, while the in-plane flow of lipids shows signatures of viscous behavior.
In addition, lipid membranes exhibit an intricate coupling between out-of-plane elastic deformations and in-plane viscous flows. Consequently, lipid membranes are considered viscous-elastic materials \cite{arroyo2009relaxation,rangamani2013interaction, sahu2017irreversible}. \textspace

Lipid membranes also exhibit coupled electrical and mechanical behavior. 
For instance, under the action of an electric field, membrane vesicles deform into prolates, oblates, and other shapes \cite{winterhalter1988deformation, kummrow1991deformation,dimova2007giant, dimova2009vesicles,portet2012destabilizing,perrier2017lipid, vlahovska2019electrohydrodynamics} and even form pores \cite{stampfli1957membrane,coster1975mechanism,mehrle1985evidence,needham1989electro,hibino1991membrane,riske2005electro,pavlin2008chapter, dimova2007giant,portet2012destabilizing,perrier2017lipid}.
In addition, the bulk fluid surrounding lipid membranes is often an electrolyte with varying ionic concentrations across the boundaries and within the interior of cells and organelles. Such concentration differences can give rise to electro-osmotic flows and expose lipid membranes to local electric fields, thereby inducing Maxwell stresses and deformations.\textspace% 

Understanding the electromechanics of lipid membranes is relevant across various disciplines. For example, electroporation, the creation of pores by an external electric field, is employed in novel procedures for non-thermal food processing, non-thermal tumor ablation, and the delivery of cancer treatment drugs into cells \cite{yarmush2014electroporation,arshad2020electrical,arshad2021pulsed}. Furthermore, electroporation of nearby lipid membranes leads to their subsequent fusion. This so-called electrofusion is used to facilitate cell-hybridization \cite{kanduvser2014cell} and the creation of microreactors \cite{perrier2017lipid}. \textspace

The electromechanics of lipid membranes is also essential for understanding many biological phenomena. One fascinating example is the propagation of an action potential through an axon. Action potentials constitute localized and transient depolarizations of the axon caused by ionic currents through the lipid membrane. They travel along the axon to propagate signals---for example, between sensory neurons and the brain \cite{purves2019neuroscience}. Despite evidence of thermal and mechanical effects \cite{bernstein1906untersuchungen, abbott1958positive,iwasa1980mechanical,tasaki1980mechanical,nguyen2012piezoelectric,yang2018imaging}, the perspective of action potentials as purely electric phenomena prevails. However, recent attempts challenge the purely electrical description by accounting for coupled thermodynamic, electrical and mechanical aspects\cite{el2015mechanical,mussel2019sounds, jerusalem2019electrophysiological}.\textspace

Theoretical models of the electromechanics of lipid membranes are indispensable to understanding the above phenomena. They often involve long time scales and large length scales, necessitating the development of continuum models. Due to their small thickness, continuum theories commonly model lipid membranes as two-dimensional surfaces\cite{helfrich1973elastic,steigmann1999fluid,sahu2017irreversible}, an approach well-established for the mechanics of lipid membranes. However, a surface description may not be suitable for the electromechanics of lipid membranes. \textspace

An electromechanical theory that treats lipid membranes as surfaces suffers from multiple shortcomings. 
First, surface descriptions do not resolve potential drops across lipid membranes but instead yield continuous potentials---contradictory to what is observed in action potentials, for instance.
Second, arbitrary surface charge densities on the two interfaces between lipid membranes and their surroundings cannot be accounted for correctly. Lastly, the electric field in the interior of lipid membranes is not well-defined when treated as surfaces. Yet, the aforementioned aspects are all required to capture the Maxwell stresses acting on lipid membranes. Hence, a suitable electromechanical theory cannot treat lipid membranes as surfaces but needs to resolve effects arising from their finite thickness.  \textspace

Three-dimensional models naturally account for the finite thickness of lipid membranes. However, three-dimensional models are complex and quickly become intractable for deforming geometries ---finding analytical solutions is often unwieldy, and even their numerical treatment is challenging. In this work, we propose an \textit{effective} two-dimensional theory to describe the electromechanics of lipid membranes. Starting from a three-dimensional continuum picture, we introduce a new dimension reduction procedure using low-order spectral expansions. This approach leads to an effective two-dimensional theory, which explicitly retains the thickness information required to capture {potential differences and} Maxwell stresses.  
At the same time, the resulting equations are analytically and numerically less challenging than those of three-dimensional models and can be analyzed using the tools developed for two-dimensional surface theories. 
Thus, the proposed theory combines the advantages of three-dimensional and surface descriptions of lipid membranes and is referred to as the $(2+\delta)$-dimensional theory. \textspace

This article is the first in a series of three that derives the $(2+\delta)$-dimensional theory of the electromechanics of lipid membranes. The series of articles is structured as follows. 
\begin{description}
\item [Part 1: Electrostatics ] We introduce a new dimension reduction technique for partial differential equations based on low-order spectral expansions of the solution. We then apply the dimension reduction procedure to the electrostatics of thin films and show the effectiveness of the new, dimensionally-reduced theory.
\item [Part 2: Balance laws] We apply the dimension reduction procedure to the three-dimensional mechanical balance laws of thin films, while accounting for Maxwell stresses arising from electric fields. This yields dimensionally-reduced, constitutive model-independent mass, angular momentum, and linear momentum balance equations. 
\item [Part 3: Constitutive models] We propose three-dimensional elastic and viscous constitutive models for lipid membranes and derive the governing equations of the $(2+\delta)$-dimensional theory for the electromechanics of lipid membranes. 
\end{description}

The remainder of this article is structured as follows. In Sec.~\ref{sec:electrostatics_thin_films}, we revisit the well-known equations describing an electric field under quasi-static conditions and introduce the problem of a thin film embedded in a bulk domain. 
In Sec.~\ref{sec:dim_reduction_prod}, we take an abstract perspective and introduce the new dimension reduction procedure for a general differential equation.
A more physically-inclined reader may skip Sec.~\ref{sec:dim_reduction_prod} and immediately proceed to Sec.~\ref{sect:appl_electrostatics}, wherein we apply the proposed dimension reduction method to the electrostatics of thin films. Section~\ref{sect: comp_to_3DGauss} concludes the article with analytical and numerical comparisons of the three-dimensional and dimensionally-reduced electrostatic theories for different geometries relevant for lipid membranes.

\section{Electrostatics of a Thin Film} \label{sec:electrostatics_thin_films}

We begin this section by recalling the theory of continuum electrostatics with discontinuities \cite{kovetz2000electromagnetic}. Subsequently, we describe the electrostatics equations governing a thin film embedded in a three-dimensional bulk domain. \textspace

Under the conditions of electrostatics, Maxwell's equations for a linear dielectric material with constant permittivity reduce to\footnote{\textit{Check} symbols are used to distinguish corresponding quantities in the dimensionally-reduced theory, which do not carry a dedicated symbol.}\cite{kovetz2000electromagnetic} (see SM, Sec.~1 for details) 
\begin{align}
    \varepsilon\divv{  \check{\bm{e}}} &= q~, \quad \forall \check{\bm{x}} \in \mathcal{B}~, \label{eq:ME_divv}\\
    \curl{\check{\bm{e}}} &= \bm{0}~, \quad \forall \check{\bm{x}} \in \mathcal{B}~, \label{eq:ME_curl} \\[5pt]
    \bar{\bm{n}} \cdot \llbracket \varepsilon \check{\bm{e}} \rrbracket &= \sigma~, \quad \forall \check{\bm{x}} \in \mathcal{S}~, \label{eq:ME_jump_dot}\\
    \bar{\bm{n}} \times \llbracket \check{\bm{e}} \rrbracket &= \bm{0}~, \quad \forall \check{\bm{x}} \in \mathcal{S}~, \label{eq:ME_jump_times}
\end{align}
where $\mathcal{B}$ denotes the bulk domain, $\varepsilon$ is the permittivity, $\check{\bm{e}}$ is the electric field, and $q$ is the free charge density in $\mathcal{B}$. Additionally, $\mathcal{S}$ denotes an oriented surface, with normal $\bar{\bm{n}}$ and surface charge density $\sigma$, where the electric field is discontinuous. 
The notation $\llbracket \bullet \rrbracket$ denotes the jump across a surface of discontinuity, i.e. $\llbracket \bullet \rrbracket = \bullet^+ - \bullet^-$, where $\bullet^\pm$ denotes the value above and below $\mathcal{S}$, respectively. \textspace

By Helmholtz' theorem \cite{arfken1999mathematical}, Eq.~\eqref{eq:ME_curl} is satisfied by construction if we define the electric potential $\check{\phi}$ such that
\begin{align}
    \check{\bm{e}} = - \grad{\check{\phi}}~, 
\end{align}
which further simplifies Maxwell's equations to
\begin{alignat}{4}
    \Delta \check{\phi} &= -q/\varepsilon~&, \quad &\forall \check{\bm{x}} \in \mathcal{B}~&, \label{eq:ME_Gaussphi} \\
    \llbracket \check{\phi} \rrbracket &= 0~&, \quad &\forall \check{\bm{x}} \in \mathcal{S}~&, \label{eq:ME_cont_phi} \\
    \bar{\bm{n}} \cdot \llbracket \varepsilon \grad{\check{\phi}} \rrbracket &= -\sigma~&, \quad &\forall \check{\bm{x}} \in \mathcal{S}~&, \label{eq:ME_jump_dot_phi}\\
    \bar{\bm{n}} \times \llbracket \grad{\check{\phi}} \rrbracket &= \bm{0}~&, \quad &\forall \check{\bm{x}} \in \mathcal{S}~&, \label{eq:ME_jump_times_phi} 
\end{alignat}
where $\Delta$ denotes the Laplacian.
Equation~\eqref{eq:ME_Gaussphi} is Gauss' law written in terms of the potential and Eq.~\eqref{eq:ME_cont_phi} describes continuity of the potential across the surface of discontinuity. The latter follows from Coulomb's law for continuous charges \cite{jackson1999classical}. According to Eq.~\eqref{eq:ME_jump_dot_phi}, the normal component of the gradient of the electric potential is discontinuous at a surface of discontinuity while, according to Eq.~\eqref{eq:ME_jump_times_phi}, components parallel to the surface of discontinuity are continuous. Note that, given Eq.~\eqref{eq:ME_cont_phi}, Eq.~\eqref{eq:ME_jump_times_phi} is trivially satisfied. \textspace

\begin{wrapfigure}{L}{0.58\textwidth}
    \centering
    \vspace{-10pt}
    \tikzset{>=latex}
%\begin{scaletikzpicturetowidth}{0.5\textwidth}
\begin{tikzpicture}[font=\normalfont, x=0.47\textwidth, y=0.47\textwidth]

\pgfdeclarelayer{background}
\pgfdeclarelayer{foreground}
\pgfsetlayers{background,foreground}

\def \h{0.07}

\begin{pgfonlayer}{background}
\coordinate (A) at (0,0.1); 
\coordinate (B) at (0.6,0.1); 
\coordinate (C) at (1,0.3); 
\coordinate (D) at (0.4,0.3); 
\coordinate (topcenter) at (0.55, 0.28);
\coordinate (bottomcenter) at (0.55, 0.109);
\coordinate (nustart) at (0.96, 0.3);

\draw[semithick] 
      (A) .. controls +(0.2, -0.05) and +(-0.25,0.05)  .. (B)
      (B) .. controls +(0.15, 0.15) and +(-0.15, -0.15) .. (C)
      %(C) .. controls +(-0.2, -0.05) and +(0.2, 0.05) .. (D) % remove this line?
      %(D) -- (A)
      (A) -- ++(0,\h) coordinate (E)
      (B) -- ++(0,\h) coordinate (F)
      (C) -- ++(0,\h) coordinate (G)
      (D) -- ++(0,\h) coordinate (H);
      
\end{pgfonlayer}

\begin{pgfonlayer}{foreground}
\filldraw[fill=white, semithick] 
      (E) .. controls +(0.2, -0.05) and +(-0.25,0.05)  .. (F) .. controls +(0.15, 0.15) and +(-0.15, -0.15) .. (G) .. controls +(-0.2, -0.05) and +(0.2, 0.05) .. (H) .. controls +(-0.15, -0.1) and +(0.15, 0.1) .. (E);

\node [position=-7pt:7pt from A] (hnode) {$\delta$};
\node [position=-45pt:-5pt from topcenter] (ltop) {};
\node [position=20pt:16.5pt from B] (rbot) {};
\node [position=32pt:16pt from rbot] (rmid) {};
\node [position=42pt:2pt from A] (fmid) {};

% labels
\draw[] (ltop) .. controls +(-0.1,0) and +(0.1,-0.) .. ++(-0.2,0.08) node[anchor=east] (nodeSp) {$\mathcal{S}^+$};
\draw (rbot) .. controls +(0.07,0) and +(-0.07,-0.) .. ++(0.15,-0.06) node[anchor=west] (nodeSm) {$\mathcal{S}^-$}; 
\draw (rmid) .. controls +(0.07,0) and +(-0.07,-0.) .. ++(0.15,-0.06) node[anchor=west] (nodeS0) {$\mathcal{S}_0$}; 
\draw (fmid) .. controls +(-0.04,0) and +(0.04,0) .. +(-0.1, -0.1) node[anchor=east] (nodeSpar) {$\mathcal{S}_{||}$}; 
\node[position=90pt:32pt from A] (nodeM) {$\mathcal{M}$};
\node[position=-70pt:35pt from C] (nodeBp) {$\mathcal{B}^+$};
\node[position=-50pt:-20pt from B] (nodeBm) {$\mathcal{B}^-$};

% normal vectors
\draw[-{Stealth}, thick] (topcenter) -- ++(-0.01, 0.15);
\node [position=10pt:10pt from topcenter] (nplabel) {$\bar{\bm{n}}^+$};  
\draw[-{Stealth}, thick] (bottomcenter) -- ++(0.003, -0.08);
\node [position=15pt:-10pt from bottomcenter] (nmlabel) {$\bar{\bm{n}}^-$};  

% binormal
\draw[-{Stealth}, thick] (nustart) -- ++(0.15, 0.);
\node [position=30pt:10pt from nustart] (nulabel) {$\bm{\nu}$};  

\draw[dashed] 
    ($(A)+(0.,\h/2)$) .. controls +(0.2, -0.05) and +(-0.25,0.05)  .. ($(B)+(0,\h/2)$)
    ($(B)+(0.,\h/2)$) .. controls +(0.15, 0.15) and +(-0.15, -0.15) .. ($(C)+(0.,\h/2)$);
\end{pgfonlayer}

\end{tikzpicture}
%\end{scaletikzpicturetowidth}
    \caption{Schematic of a thin film $\mathcal{M}$ with thickness $\delta$ that separates the two bulk domains $\mathcal{B}^+$ and $\mathcal{B}^-$. }
    \label{fig:schematic_setup}
\end{wrapfigure}
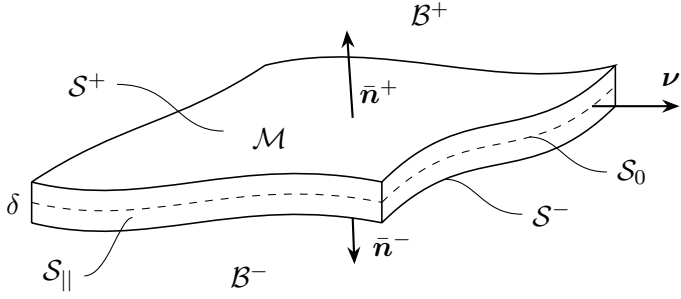
Next, we consider a thin film without any free charge in its interior, as is the case for lipid membranes. The thin film $\mathcal{M}$ has thickness $\delta$ and is embedded in two bulk domains $\mathcal{B}^\pm$ above and below $\mathcal{M}$, as shown in Fig.~\ref{fig:schematic_setup}. The top and bottom bounding surfaces of $\mathcal{M}$ are denoted by $\mathcal{S}^\pm$ and are equipped with surface charge densities $\sigma^\pm$, making $\mathcal{S}^\pm$ surfaces of discontinuity. The outward-pointing normal vectors on $\mathcal{S}^\pm$ are denoted by $\bar{\bm{n}}^\pm$. The three-dimensional electrostatics equations are given by 
\begin{alignat}{4}
     \Delta \check{\phi}_{\mathrm{B}^-} &= -q_{\mathrm{B}^-}/\varepsilon_{\mathrm{B}^-}~&, \quad &\forall \check{\bm{x}} \in \mathcal{B}^-~&, \label{eq:gauss_bulkm} \\
     \llbracket \check{\phi} \rrbracket &= 0~&, \quad & \forall \check{\bm{x}} \in \mathcal{S}^-~&, \label{eq:cont_Sm} \\
     \bar{\bm{n}}^+ \cdot \llbracket \varepsilon \check{\bm{e}} \rrbracket &= \sigma^-~&, \quad &\forall \check{\bm{x}} \in \mathcal{S}^-~&, \label{eq:jumpSm}\\
     \varepsilon_\mathrm{M}\Delta \check{\phi}_\mathrm{M} &= 0~&, \quad &\forall \check{\bm{x}} \in \mathcal{M}~&, \label{eq:gauss_M}  \\ % keeping eps_M here so that we can already define it here! needed later!
     \bar{\bm{n}}^- \cdot \llbracket \varepsilon \check{\bm{e}} \rrbracket &= \sigma^+~&, \quad &\forall \check{\bm{x}} \in \mathcal{S}^+~&, \label{eq:jumpSp}\\
     \llbracket \check{\phi} \rrbracket &= 0~&, \quad & \forall \check{\bm{x}} \in \mathcal{S}^+~&, \label{eq:cont_Sp}  \\
     \Delta \check{\phi}_{\mathrm{B}^+} &= -q_{\mathrm{B}^+}/\varepsilon_{\mathrm{B}^+}~&, \quad &\forall \check{\bm{x}} \in \mathcal{B}^+~&, \label{eq:gauss_bulkp} 
\end{alignat}
where $\varepsilon_{\mathrm{B}^\pm}$ and $\varepsilon_{\mathrm{M}}$ are the permittivity of the bulk regions and thin film, respectively. The jump conditions, Eqs.~\eqref{eq:jumpSm} and~\eqref{eq:jumpSp}, are written in terms of electric fields for later notational convenience. We close the problem with boundary conditions on the lateral surface $\mathcal{S}_{||}$ with outward-pointing normal $\bm{\nu}$, as shown in Fig.~\ref{fig:schematic_setup}:
\begin{alignat}{2}
    \check{\phi}_{\mathrm{M}} &= \bar{\phi}_\mathrm{M}~&&,\quad \forall \check{\bm{x}} \in \mathcal{S}_{||\mathrm{D}}~, \label{eq:DBC_Spar}\\
    -\bm{\nu} \cdot \grad{\check{\phi}_\mathrm{M}} &= \bar{e}~&&,\quad \forall \check{\bm{x}} \in \mathcal{S}_{||\mathrm{N}}~, \label{eq:NBC_Spar}
\end{alignat}
where $\mathcal{S}_{||} = \mathcal{S}_{||\mathrm{D}} \cup \mathcal{S}_{||\mathrm{N}}$, $\mathcal{S}_{||\mathrm{D}}\cap \mathcal{S}_{||\mathrm{N}} = \emptyset$, and $\bar{\phi}_\mathrm{M}$ and $\bar{e}$ are the prescribed potential and electric field component, respectively. The remaining boundary conditions for $\check{\phi}_{\mathrm{B}^\pm}$ are of no consequence for the dimension-reduction procedure and are thus omitted here. 
\textspace

In the following, we refer to Eqs.~\eqref{eq:gauss_bulkm}--\eqref{eq:gauss_bulkp} as the three-dimensional theory. In comparison, an effective, dimensionally-reduced theory replaces Gauss' law on the three-dimensional thin film $\mathcal{M}$, Eq.~\eqref{eq:gauss_M}, by an approximately equivalent equation defined on the two-dimensional mid-surface of $\mathcal{M}$, denoted $\mathcal{S}_0$. To that end, the following section introduces a new dimension reduction procedure that follows ideas used in spectral methods by expanding all unknowns and parameters in terms of orthogonal polynomials.

\section{Spectral Dimension Reduction for Thin Films} \label{sec:dim_reduction_prod}
In this section, we present a new, general approach to deriving dimensionally-reduced differential equations defined on the mid-surface of a thin film. We begin by revisiting spectral expansions in Sec.~\ref{sect:math_prelim} and show how they can be used to derive dimensionally-reduced theories in Sec.~\ref{sect:abstract_setup}. Note that the remaining sections are self-contained and that the reader may immediately proceed to Sec.~\ref{sect:appl_electrostatics} to find the dimensionally-reduced electrostatics equations.

\subsection{Mathematical Preliminaries of Spectral Expansions} \label{sect:math_prelim}
Let $P_k\leftR(\theta\rightR) : (a,b) \rightarrow \mathbb{R}$,\, $k\in \mathbb{N}_0,\, a,b \in \mathbb{R}$ denote a real-valued polynomial and let $\mathbb{P} = \{ P_k\leftR(\theta\rightR) : k\in \mathbb{N}_0\}$ denote the corresponding set of polynomials. For two sufficiently well-behaved functions $f\leftR(\theta\rightR)$, $g\leftR(\theta\rightR)$, $\theta \in (a,b)$, we define the weighted inner product 
\begin{align}
    \langle f\leftR(\theta\rightR), g\leftR(\theta\rightR) \rangle_w = \int_{(a,b)} f\leftR(\theta\rightR)  g\leftR(\theta\rightR) w\leftR(\theta\rightR) \diff \theta~, \label{eq:inner_product}
\end{align}
where $w\leftR(\theta\rightR)$ denotes a weight function. If the polynomials in $\mathbb{P}$ satisfy the relation 
\begin{align}
    \langle P_k\leftR( \theta \rightR), P_l\leftR( \theta \rightR) \rangle_w = c_{k} \delta_{kl}~, \quad \forall P_k,P_l \in \mathbb{P}~, \label{eq:orthogonality}
\end{align}
we call $\mathbb{P}$ an orthogonal set of polynomials. In Eq.~\eqref{eq:orthogonality}, $c_{k}$ is some positive constant and $\delta_{kl}$ denotes the Kronecker delta. Let $\norm{\cdot}{w} = \sqrt{\langle \cdot, \cdot \rangle_w }$ denote the norm induced by the inner product defined in Eq.~\eqref{eq:inner_product} and let $L^2_w$ denote the space of functions bounded in $\norm{\cdot}{w}$. \textspace 

The $N^\mathrm{th}$-order projection of any function $f\leftR(\theta\rightR) \in L^2_w$ onto $\mathbb{P}$, denoted by $f_{\mathbb{P},N}$, is defined as
\begin{equation}
    f_{\mathbb{P},N}\leftR(\theta\rightR) = \sum_{k=0}^N \hat{f}_k P_k\leftR(\theta\rightR)~, \label{eq:fprojection}
\end{equation}
where $\hat{f}_k$ is the $k^\mathrm{th}$ coefficient of the expansion and is given by
\begin{equation}
    \hat{f}_k = \langle f\leftR(\theta\rightR), P_k \leftR(\theta\rightR) \rangle_w~. \label{eq:coeff_projection} 
\end{equation}
The set of polynomials $\mathbb{P}$ is complete with respect to the norm $\norm{\cdot}{w}$ if, for any $f\leftR(\theta\rightR) \in L^2_w$, \cite{canuto2007spectral}
\begin{equation}
    \lim_{N\rightarrow\infty} \norm{f\leftR(\theta\rightR) - f_{\mathbb{P},N}\leftR(\theta\rightR)}{w} = 0~. \label{eq:completeness}
\end{equation}
Equation~\eqref{eq:fprojection} in conjunction with Eq.~\eqref{eq:completeness} allows us to express functions in $L^2_w$ as
\begin{equation}
    f\leftR(\theta\rightR) = \sum_{k=0}^\infty \hat{f}_k P_k\leftR(\theta\rightR)~. \label{eq:f_infexpansion}
\end{equation}%\textspace

Complete and orthogonal polynomials are commonly used in spectral methods to numerically solve differential equations. In the following, we briefly revisit the basics of spectral methods required for our method for dimension reduction; see Refs.~\cite{gottlieb1977numerical, boyd2001chebyshev,canuto2007spectral} for more details.
Let $\mathcal{L}\leftR({v}\leftR(\theta\rightR); \bm{p}\rightR) : \mathcal{U} \rightarrow L^2_w$ denote a scalar-valued differential operator, where $\mathcal{U}$ is some space of sufficiently smooth functions defined on $\left(a,b\right)$ and $\bm{p} = \{p_j\}_{j=1,..,N_p}$ is a set of parameters. We write a generic differential equation as 
\begin{equation}
    \mathcal{L}\leftR(u\leftR(\theta\rightR); \bm{p}\rightR) = 0~, \quad\theta \in (a,b),~ u \in \mathcal{U}~, \label{eq:abstract_diffeq}
\end{equation}
postponing any discussion on the application of boundary conditions to Sec.~\ref{sect:abstract_setup}. Due to the completeness property of $\mathbb{P}$ in Eq.~\eqref{eq:completeness} and Eq.~\eqref{eq:f_infexpansion}, we can expand the solution $u\leftR(\theta\rightR)$ as 
\begin{equation}
    u\leftR(\theta\rightR) = \sum_{k=0}^\infty \hat{u}_k P_k\leftR(\theta\rightR)~. \label{eq:1D_expansion}
\end{equation}
Thus, finding the solution $u\leftR(\theta\rightR)$ is equivalent to finding the constant coefficients $\hat{u}_k$. However, any numerical approach requires truncating the expansion at some finite order $N \in \mathbb{N}$,
\begin{align}
    u_{\mathbb{P},N}\leftR(\theta\rightR) = \sum_{k=0}^N \hat{u}_{N,k} P_k\leftR(\theta\rightR)~, \label{eq:u_truncated_expansion}
\end{align}
where the subscript $N$ on $\hat{u}_{N,k}$ indicates dependence on the truncation order $N$. 
The unknown coefficients $\hat{u}_{N,k}$ in Eq.~\eqref{eq:u_truncated_expansion} are found by replacing $u\leftR(\theta\rightR)$ by $u_{\mathbb{P},N}\leftR(\theta\rightR)$ in Eq.~\eqref{eq:abstract_diffeq} and taking the inner product with the $l^\mathrm{th}$ polynomial, resulting in
\begin{equation}
     \Biggl \langle \mathcal{L}\leftR(\sum_{k=0}^N \hat{u}_{N,k} P_k\leftR(\theta\rightR); \bm{p}\rightR), P_l\leftR(\theta\rightR) \Biggr\rangle_w = 0~, \quad \forall l \in [0,...,N]~. \label{eq:1D_spectral_eqs}
\end{equation}
This yields $N+1$ equations for the $N+1$ unknown coefficients $\{\hat{u}_{N,k}\}_{k=0,...,N}$. Since the spatial dependence is entirely contained in the polynomials $P_k\leftR(\theta\rightR)$, any derivative can be carried out explicitly and the $N+1$ equations in Eq.~\eqref{eq:1D_spectral_eqs} are no longer differential but algebraic equations, independent of $\theta$. Solving the system of algebraic equations resulting from Eq.~\eqref{eq:1D_spectral_eqs} yields the coefficients $\{\hat{u}_{N,k}\}_{k=0,...,N}$ and hence, the approximate solution $u_{\mathbb{P},N}\leftR(\theta\rightR)$. 

\subsection{Dimension Reduction Procedure for Thin Films Using Spectral Expansions} \label{sect:abstract_setup}

Before introducing the new dimension reduction procedure, we revisit the thin film setup described in Sec.~\ref{sec:electrostatics_thin_films}. The arbitrarily curved, thin film $\mathcal{M}$ has constant thickness $\delta$ and the mid-surface $\mathcal{S}_0$ divides $\mathcal{M}$ into two parts of equal thickness. The mid-surface $\mathcal{S}_0$ is equipped with a normal vector $\bm{n}$ and the superscripts ``$+$'' and ``$-$'' indicate quantities associated with the regions above and below $\mathcal{S}_0$, as defined by the orientation of $\bm{n}$. The top, bottom, and lateral bounding surfaces of $\mathcal{M}$ are denoted by $\mathcal{S}^+$, $\mathcal{S}^-$, and $\mathcal{S}_{||}$, respectively (see Fig.~\ref{fig:schematic_setup}). The lateral bounding surface can be expressed as $\mathcal{S}_{||} = \partial \mathcal{S}_0 \times \left( -\delta/2,\delta/2\right)$, where $\partial \mathcal{S}_0$ is the boundary of $\mathcal{S}_0$. The outward-pointing normal on $\mathcal{S}_{||}$ is denoted by $\bm{\nu}$, as shown in Fig.~\ref{fig:schematic_setup}.
Finally, the body $\mathcal{M}$ is embedded into three-dimensional bulk domains, referred to as $\mathcal{B}^+$ above $\mathcal{M}$ and $\mathcal{B}^-$ below $\mathcal{M}$. \textspace
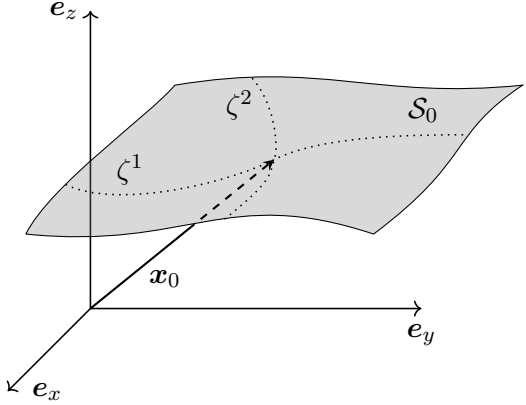
\begin{wrapfigure}{L}{0.45\textwidth}
    \centering
    \begin{tikzpicture}[font=\normalfont, x=0.4\textwidth, y=0.4\textwidth]
% coordinate system
\node(CScenter) at (0.33,0)  {};
\draw[semithick,->] (CScenter.center) -- +(0,0.6) coordinate (ez);
\draw[semithick,->] (CScenter.center) -- +(2/3,0) coordinate (ey);
\draw[semithick,->] (CScenter.center) -- +(-.5/3,-.5/3) coordinate (ex);

\coordinate (A) at (0.1,0.2); 
\coordinate (B) at (0.7,0.2); 
\coordinate (C) at (1.1,0.4); 
\coordinate (D) at (0.5,0.4); 

% patch
\begin{scope}
\draw[smooth, fill=gray,fill opacity=.3] (0.2,0.15) coordinate (BL) .. controls +(0.05,0.1) and +(-0.05,-0.06) ..  +(0.3,0.3) .. controls +(0.3,0.05) and +(-0.3,-0.03) .. +(1,0.3)  .. controls +(-0.15,-0.1) and +(0.2,0.15) .. +(0.7,0)  .. controls +(-0.3,0.1) and +(0.33,-0.033) .. (BL) ;
\end{scope}

% coordinate lines
\draw[dotted, semithick] (0.28,0.25) .. controls +(0.1,-0.05) and +(-0.1,-0.05) .. (0.7,0.3) coordinate (midcross) .. controls +(0.1,0.05) and +(-0.1,0) .. (1.09,0.35);
\draw[dotted, semithick] (0.6,0.18) .. controls +(0.1,0.08) and +(-0.01,-0.03) .. (0.7,0.3) .. controls +(0.01,0.03) and +(0.05,-0.05) .. (0.65,0.47);  

% % a_1
% \draw[->,>=stealth] (1.7,1) -- +(0.25,0.25);
% % a_2
% \draw[->,>=stealth] (1.7,1) -- +(-0.2,0.4);

% % x-vector
 \draw[->,>=stealth, dashed, thick,shorten <=3] (0.5396,0.17) -- (0.7,0.3);
 \draw[thick] (CScenter.center) -- (0.5396,0.17);

% % labels
\node[] at (1,0.4) {$\mathcal{S}_0$};
\node[] at (0.41,0.28) {$\zeta^1$};
\node[] at (0.63,0.41) {$\zeta^2$};
% \node[] at (1.7,1.4) {$\bm{a}_1$};
% \node[] at (2.15,1.2) {$\bm{a}_2$};
\node[position=15pt:0pt from ex.east] (nodeex) {$\bm{e}_x$};
\node[position=0pt:-10pt from ey] (nodeey) {$\bm{e}_y$};
\node[position=-10pt:0pt from ez] (nodeez) {$\bm{e}_z$};
\node[] at (0.48,0.06) {$\bm{x}_0$};

\end{tikzpicture}
    \caption{The mid-surface $\mathcal{S}_0$ is parametrized using the curvilinear coordinates $\{\zeta^1, \zeta^2\} \in \Omega$.}
    \label{fig:S0_parametrization}
\end{wrapfigure}

The thin film $\mathcal{M}$ is arbitrarily curved, making it convenient to formulate the proposed dimension reduction method in a differential geometry framework. To this end, we introduce a parametrization of $\mathcal{M}$. The mid-surface $\mathcal{S}_0$ is endowed with a two-dimensional, curvilinear parametrization $\bigl\{\zeta^1, \zeta^2 \bigr\} \in \Omega$ as shown in Fig.~\ref{fig:S0_parametrization}, where $\Omega$ denotes the parametric domain, such that we can express any position $\bm{x}_0 \in \mathcal{S}_0$ using the mapping $\bm{\chi}_0: \Omega \rightarrow \mathcal{S}_0$, $\left(\zeta^1, \zeta^2\right) \mapsto \bm{x}_0$. Parametrizing the full body $\mathcal{M}$ requires a third parametric direction, $\zeta^3 \in \Xi$, where $\Xi$ denotes the corresponding parametric domain. We can then express any position $\bm{x} \in \mathcal{M}$ using the mapping $\bm{\chi}: \Omega \times \Xi \rightarrow \mathcal{M}$, $\left(\zeta^1,\zeta^2, \zeta^3\right) \mapsto \bm{x}$ with $\bm{\chi}\rvert_{\zeta^3 = 0} = \bm{\chi}_0$. \textspace

We now discuss how spectral expansions can be used to reduce a differential equation defined on $\mathcal{M}$ to a differential equation defined on the mid-surface $\mathcal{S}_0$. For simplicity, we will continue considering only scalar-valued differential operators such as the Laplacian, relevant for electrostatics. However, the ideas presented here can be extended to vector-valued differential differential operators, as will be discussed in part 2 of this sequence of publications where the mechanical balance laws are addressed.\textspace

Let $u\leftR(\zeta^i\rightR) \in \mathcal{U}$ denote a scalar-valued function from a space of sufficiently smooth functions $\mathcal{U}$ defined on $\Omega \times \Xi$, where we used the short-hand notation $\zeta^i \equiv \{\zeta^1,\zeta^2,\zeta^3\}$ for the parametrization. Assume that $u\leftR(\zeta^i\rightR)$ satisfies the differential equation 
\begin{align}
    \mathcal{L}\leftR(u\leftR(\zeta^i\rightR); \bm{p}\rightR) = 0~, \quad \forall \zeta^i \in \Omega \times \Xi~,  \label{eq:3D_diffeq}
\end{align}
where $\mathcal{L}\leftR(u\leftR(\zeta^i\rightR), \bm{p}\rightR)$ is now a \textit{partial} differential operator. A set of appropriate boundary conditions closes Eq.~\eqref{eq:3D_diffeq}: 
\begin{alignat}{2}
    g_m \leftR( u\leftR(\zeta^\alpha\rightR), \zeta^\alpha; \bm{p} \rightR) &= 0~, \quad \forall \zeta^i \in \Omega \times \partial^\pm\Xi~, \quad & m \in [1,...,N_{\partial \Xi}]~,  \label{eq:3D_BC_Spm}\\
    h_n\leftR( u\leftR(\zeta^i\rightR), \zeta^i; \bm{p} \rightR) &= 0~, \quad \forall \zeta^i \in \partial^k\Omega \times \Xi~, \quad & n \in [1,...,N_{\partial \Omega}]~, \label{eq:3D_BC_Slat}
\end{alignat}
where we used the short-hand notation $\zeta^\alpha \equiv \{\zeta^1, \zeta^2\}$\footnote{Greek and Latin letters are used to denote indices taking values $\{1,2\}$ and $\{1,2,3\}$, respectively}. Here, $g_m$ denotes the $m$\textsuperscript{th} boundary condition on either $\mathcal{S}^+$ or $\mathcal{S}^-$ while $h_n$ denotes the $n$\textsuperscript{th} boundary condition on $\mathcal{S}_{||}^n \subseteq \mathcal{S}_{||}$. Note that the number of boundary conditions, $N_{\partial \Xi}$ and $N_{\partial \Omega}$, is determined by the order of the differential operator. \textspace

With Eqs.~\eqref{eq:3D_diffeq}--\eqref{eq:3D_BC_Slat} formulated in terms of the parametrization $\zeta^i$, dimension reduction requires eliminating dependence on the parametric direction $\zeta^3$, implying that the dimensionally-reduced equations only depend on the mid-plane parametrization $\zeta^\alpha$. To this end, the key idea of our proposed method is to express the solution $u\leftR(\zeta^i\rightR)$ as
\begin{equation}
    u\leftR(\zeta^i\rightR) = \sum_{k=0}^\infty \hat{u}_k\leftR(\zeta^\alpha\rightR) \ChebT{k}{\theta\leftR(\zeta^3\rightR)}~, \label{eq:u_zeta3_expansion}
\end{equation} 
where $\hat{u}_k\leftR(\zeta^\alpha\rightR)$ denotes the unknown coefficients of the spectral expansion, and $\theta$ is the mapping $\theta:\Xi \rightarrow (a,b)$. It should be emphasized that the coefficients $\hat{u}_k\leftR(\zeta^\alpha\rightR)$ only depend on the parametrization of the mid-plane $\mathcal{S}_0$ and not on the parametric direction $\zeta^3$ associated with the thickness direction. Instead, the dependence on $\zeta^3$ is entirely contained in the polynomials $\ChebT{k}{\theta}$. Similarly, the parameters $p_j \in \bm{p}$ may also depend on the parametrization $\zeta^i$, and are thus similarly expanded as, 
\begin{equation}
    p_j\leftR(\zeta^i\rightR) = \sum_{k=0}^\infty \hat{p}_{jk}\leftR(\zeta^\alpha\rightR) \ChebT{k}{\theta\leftR(\zeta^3\rightR)}~, \label{eq:p_zeta3_expansion}
\end{equation}
where the coefficients $\hat{p}_{jk}\leftR(\zeta^\alpha\rightR)$ are found by applying Eq.~\eqref{eq:coeff_projection} along the thickness direction. \textspace

To obtain a finite order approximation, the solution expansion in Eq.~\eqref{eq:u_zeta3_expansion} is truncated  at order $N_u$, which reduces Eq.~\eqref{eq:3D_diffeq} to
\begin{equation}
    \mathcal{L}\leftR(\sum_{k=0}^{N_u} \hat{u}_k\leftR(\zeta^\alpha\rightR) \ChebT{k}{\theta\leftR(\zeta^3\rightR)}; \Biggr\{\sum_{k=0}^{\infty} \hat{p}_{jk}\leftR(\zeta^\alpha\rightR) \ChebT{k}{\theta\leftR(\zeta^3\rightR)} \Biggr\}_{j=1,...,N_p} \rightR) = 0~, \quad \forall \zeta^i \in \Omega \times \Xi~.  \label{eq:L_up_expanded}
\end{equation}
As in Eq.~\eqref{eq:1D_spectral_eqs}, we obtain the equations for the unknown coefficients $\hat{u}_k\leftR(\zeta^\alpha\rightR)$ by taking the inner product with the $l^\mathrm{th}$ order polynomial $P_l$ and assuming weighted square integrability:
\begin{align}
    \left\langle \mathcal{L}\leftR(\sum_{k=0}^{N_u} \hat{u}_k\leftR(\zeta^\alpha\rightR) \ChebT{k}{\theta\leftR(\zeta^3\rightR)}; \Biggr\{\sum_{k=0}^{\infty} \hat{p}_{jk}\leftR(\zeta^\alpha\rightR) \ChebT{k}{\theta\leftR(\zeta^3\rightR)} \Biggr\}_{j=1,...,N_p} \rightR), \ChebT{l}{\theta\leftR(\zeta^3\rightR)} \right\rangle_w & = 0~,  \nonumber\\[5pt]
    &\hspace{-7em} \forall \zeta^\alpha \in \Omega~, \quad \forall l \in [0,...,N_u]~. \label{eq:ukcoeff_3D}
\end{align}
In Eq.~\eqref{eq:ukcoeff_3D}, any differentiation with respect to $\zeta^3$ can be carried out explicitly, allowing the evaluation of the inner product. Using the orthogonality condition in Eq.~\eqref{eq:orthogonality}, Eq.~\eqref{eq:ukcoeff_3D} yields $N_u + 1$ partial differential equations that only depend on the mid-surface parametrization $\zeta^\alpha$. Thus, we obtain a set of dimensionally-reduced differential equations for the $N_u+1$ unknown coefficients $\hat{u}_k\leftR(\zeta^\alpha\rightR)$ defined on the mid-surface of the thin film $\mathcal{M}$. \textspace

Due to the potential coupling between higher and lower order coefficients, the coefficients $\hat{u}_k\leftR(\zeta^\alpha\rightR)$ in Eqs.~\eqref{eq:L_up_expanded} and~\eqref{eq:ukcoeff_3D} do not necessarily coincide with the coefficients of the series expansion in Eq.~\eqref{eq:u_zeta3_expansion}. However, for notational simplicity, we use the same symbol $\hat{u}_k\leftR(\zeta^\alpha\rightR)$ throughout. We further note that the series expansions of the parameters $p_j$ are retained in Eq.~\eqref{eq:ukcoeff_3D}. However, truncation of these series can often be physically motivated and might be necessary to obtain an analytically tractable theory, as will be seen in Sec.~\ref{sect:appl_electrostatics} when applying the proposed method to the electrostatics of thin films. \textspace

The original problem, Eq.~\eqref{eq:3D_diffeq}, requires application of the boundary conditions in Eqs.~\eqref{eq:3D_BC_Spm} and~\eqref{eq:3D_BC_Slat}. However, taking the inner product in Eq.~\eqref{eq:ukcoeff_3D} eliminates the derivatives along the $\zeta^3$-direction such that the boundary conditions in Eq.~\eqref{eq:3D_BC_Spm} need to be enforced by discarding $N_{\partial\Xi}$ equations from Eq.~\eqref{eq:ukcoeff_3D} and replacing them by the $N_{\partial\Xi}$ boundary conditions in Eq.~\eqref{eq:3D_BC_Spm}. This, in fact, sets a limit on the minimum order of expansion of the solution, i.e. $N_u \geq N_{\partial \Xi}-1$. 
The $N_{\partial \Omega}$ boundary conditions on the lateral boundary $\mathcal{S}_{||}$ in Eq.~\eqref{eq:3D_BC_Slat} are dimensionally-reduced analogously to Eq.~\eqref{eq:ukcoeff_3D}. Substituting the truncated expansion of the solution into  Eq.~\eqref{eq:3D_BC_Slat} and taking the inner product with the $l^\mathrm{th}$ order polynomial $P_l$, assuming weighted square integrability, yields new boundary conditions defined on the boundary of the mid-surface, $\partial \mathcal{S}_0$:
\begin{align}
    \left\langle h_i\leftR(\sum_{k=0}^{N_u} \hat{u}_k\leftR(\zeta^\alpha\rightR) \ChebT{k}{\theta\leftR(\zeta^3\rightR)}, \zeta^i; \Biggr\{\sum_{k=0}^{\infty} \hat{p}_{jk}\leftR(\zeta^\alpha\rightR) \ChebT{k}{\theta\leftR(\zeta^3\rightR)} \Biggr\}_{j=1,...,N_p} \rightR), \ChebT{l}{\theta\leftR(\zeta^3\rightR)} \right\rangle_w & = 0~,  \nonumber\\[5pt]
    & \hspace{-1cm} \forall l \in [1,...,N_{\partial \Omega}]~. \label{eq:BC_Slat_reduced}
\end{align}
This fully eliminates the parametric direction $\zeta^3$ from Eqs.~\eqref{eq:3D_diffeq}--\eqref{eq:3D_BC_Slat} such that the dimensionally-reduced problem is given by the first $N_u + 1 - N_{\partial \Xi}$ differential equations in Eq.~\eqref{eq:ukcoeff_3D}, the boundary conditions on $\mathcal{S}^\pm$ in Eq.~\eqref{eq:3D_BC_Spm}, and the boundary conditions on $\partial\mathcal{S}_{0}$ in Eq.~\eqref{eq:BC_Slat_reduced}. In the following, we refer to this dimensionally-reduced theory as a $\left(2+\delta\right)$-dimensional theory.\textspace

Note that when deriving a $\left(2+\delta\right)$-dimensional theory, $N_u$ can, in principle, be chosen arbitrarily large. However, the algebraic complexity significantly increases with the order of the expansion. This can be seen in the detailed derivation of the $(2+\delta)$-dimensional theory of the electrostatics of thin films in Sec.~2.3 of the SM. Hence, a $(2+\delta)$-dimensional theory generally only remains analytically tractable for low-order expansions of the solution. Thus, for the proposed method to yield meaningful results, we require the exact solution to be well-approximated by low-order polynomials along the thickness. This, however, is a common and often reasonable approximation for thin bodies, as considered here. 
Thus, given the validity of the low-order expansion of the solution, our proposed method is exact and does not require additional approximations. \textspace

\begin{wrapfigure}{r}{0.45\textwidth}
    \centering%
    \vspace{-10pt}
    \input{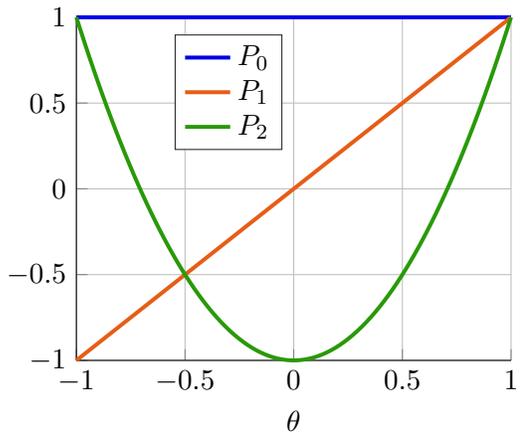}%
    \caption{Plot of the first three Chebyeshev polynomials.}
    \label{fig:T012}
    \vspace{-15pt}
\end{wrapfigure}
For the remainder of this manuscript, we specialize our derivations to Chebyeshev polynomials. Chebyeshev polynomials are defined on the interval $(a,b) = (-1,1)$ and are orthogonal with respect to the inner product in Eq.~\eqref{eq:inner_product} when the weight function is
\begin{equation}
    w\leftR(\theta\rightR) = \frac{1}{\pi} \frac{1}{\sqrt{1-\theta^2}}~,\quad \theta \in \left(-1,1\right)~. 
\end{equation}
The first three Chebyeshev polynomials are 
\begin{align}
    \ChebT{0}{\theta} &= 1~, \\
    \ChebT{1}{\theta} &= \theta~, \\
    \ChebT{2}{\theta} &= 2\theta^2 - 1~,
\end{align}
and are plotted in Fig.~\ref{fig:T012}. Chebyshev polynomials are commonly used in spectral methods and are amenable to analytical derivations. However, the procedure presented in this section is sufficiently general and can be similarly followed using any other set of complete and orthogonal polynomials defined on a bounded domain. 

\section{A Dimensionally-Reduced Theory for the Electrostatics of Thin Films} \label{sect:appl_electrostatics}
In this section, we present the $\left(2+\delta\right)$-dimensional theory of the electrostatics of thin films, obtained by applying the dimension reduction procedure proposed in Sec.~\ref{sec:dim_reduction_prod} to the problem setup in Sec.~\ref{sec:electrostatics_thin_films}, Eqs.~\eqref{eq:gauss_bulkm}--\eqref{eq:gauss_bulkp}. While the detailed derivations are shown in Sec.~2 of the SM, the key assumptions of the $(2+\delta)$-dimensional theory are discussed here. \textspace 

To obtain a dimensionally-reduced equation in place of Eq.~\eqref{eq:gauss_M}, we introduce low-order expansions of the position vector ${\bm{x}} \in \mathcal{M}$ and the electric potential in the membrane ${\phi}_\mathrm{M}$ in terms of Chebyeshev polynomials $P_k$:
\begin{align}
    \bm{x} &= \mathlarger{\sum}_{k = 0}^{1} ~ \xk{k} \ChebT{k}{\theta\leftR(\zeta^3\rightR)}~, \label{eq:chebx} \\
    \phi_{\mathrm{M}} &= \mathlarger{\sum}_{k = 0}^{2} ~ \phik{k} \ChebT{k}{\theta\leftR(\zeta^3\rightR)}~, \label{eq:chebphi}
\end{align}
where $\zeta^\alpha \equiv \{\zeta^1, \zeta^2\} \in \Omega$ indicates the parametrization of the mid-surface $\mathcal{S}_0$ with $\Omega$ being the parametric domain, $\zeta^3 \in (-\delta/2,\delta/2)$ is the parametrization along the thickness direction, and $\theta$ is the mapping $\theta:\, (-\delta/2,\delta/2)\rightarrow (-1,1)$. In Eqs.~\eqref{eq:chebx} and~\eqref{eq:chebphi}, $\bm{x}$ and $\phi_\mathrm{M}$ no longer carry a \textit{check} symbol to distinguish them from their respective exact quantities, $\check{\bm{x}}$ and $\check{\phi}_{\mathrm{M}}$. 
The order of expansion of the position vector $\bm{x}$ is motivated by the common choice of Kirchhoff-Love kinematics, which is suitable for thin materials such as lipid membranes. Kirchhoff-Love kinematics assumes that any point along the normal to the mid-surface remains on the normal to the mid-surface and maintains the same distance to the mid-surface upon deformation \cite{ciarlet2000theory}. Using this kinematic assumption, the expansion of the position vector becomes
\begin{align}
    \bm{x} = \bm{x}_0 \ChebT{0}{\theta\leftR(\zeta^3\rightR)} + \frac{\delta}{2} \bm{n} \ChebT{1}{\theta\leftR(\zeta^3\rightR)}~, \label{eq:x_KL_expansion}
\end{align}
where $\bm{x}_0 \in \mathcal{S}_0$ denotes a point on the mid-surface and $\bm{n}$ is the normal vector of the mid-surface. Equation~\eqref{eq:x_KL_expansion} further implies $\bar{\bm{n}}^+ = -\bar{\bm{n}}^- = \bm{n}$. According to the discussion in Sec.~\ref{sect:abstract_setup}, the electric potential must be expanded to at least first order to enforce two boundary conditions on the top and bottom surfaces, $\mathcal{S}^\pm$, consistent with the differential order of Eq.~\eqref{eq:gauss_M}. However, to preserve the differential nature of Eq.~\eqref{eq:gauss_M}, we expand the potential to second order.
Furthermore, to make the dimensionally-reduced theory tractable, we introduce two further crucial assumptions: 
\begin{align}
    \left(\delta\kappa\right)^2 &\ll 1~, \label{eq:dR_ll_1} \\
    \left(\delta/\ell\right)^2 &\ll 1~, \label{eq:dL_ll_1}
\end{align}
\noindent where $\kappa$ is the principal curvature with the largest magnitude and $\ell$ is a characteristic in-plane length scale for the potential and curvature (see SM, Sec.~2.3 for details). Equation~\eqref{eq:dR_ll_1} implies that the radius of curvature must be much larger than the thickness of the membrane, and Eq.~\eqref{eq:dL_ll_1} implies that the potential and geometry of the thin film change over length scales much larger than the thickness. Thus, Eqs.~\eqref{eq:dR_ll_1} and~\eqref{eq:dL_ll_1} are also conditions for the applicability of the theory proposed here. \textspace

Using Eqs.~\eqref{eq:chebphi}--\eqref{eq:dL_ll_1}, applying the dimension reduction method proposed in Sec.~\ref{sec:dim_reduction_prod} to Eq.~\eqref{eq:gauss_M} yields the dimensionally-reduced equation
\begin{align}
    \varepsilon_\mathrm{M} \laplaces{\phik{0}} - 4 C_\mathrm{M} \phik{1} H +  \frac{16}{\delta} C_\mathrm{M} \phik{2} = 0~, \quad \forall \zeta^\alpha \in \Omega~, \label{eq:surface_gauss} 
\end{align}
where $C_\mathrm{M} = \varepsilon_\mathrm{M}/\delta$ is the membrane capacitance per unit area \cite{tranquillo2008quantitative} and $H$ is the mean curvature of the mid-surface. The surface Laplacian is defined as $\laplaces{\left(\bullet\right)} = \left( \left( \bullet \right)_{,\alpha}\right)_{:\beta} a^{\alpha\beta}$, where the colon indicates the surface covariant derivative of a vector, $a^{\alpha\beta}$ denotes the contravariant components of the identity tensor on the mid-surface, and Einstein's summation convention is used (see SM, Sec.~2 for details). 

We consider Eq.~\eqref{eq:surface_gauss} as an equation for the coefficient $\phik{0}$ and choose the coefficients $\phik{1}$ and $\phik{2}$ such that some of the interface conditions on $\mathcal{S}^\pm$ are satisfied. 
To that end, we can select one of the interface conditions on $\mathcal{S}^-$, Eq.~\eqref{eq:cont_Sm} or~\eqref{eq:jumpSm}, and one of the interface conditions on $\mathcal{S}^+$, Eq.~\eqref{eq:jumpSp} or~\eqref{eq:cont_Sp}. The remaining two interface conditions need to be enforced as boundary conditions for Eqs.~\eqref{eq:gauss_bulkm} and~\eqref{eq:gauss_bulkp} such that all interface conditions are satisfied. In this article, we choose the jump conditions, Eqs.~\eqref{eq:jumpSm} and~\eqref{eq:jumpSp}, to find expressions for $\phik{1}$ and $\phik{2}$, and impose continuity of the potential, Eqs.~\eqref{eq:cont_Sm} and~\eqref{eq:cont_Sp}, as boundary conditions in the two bulk domains, Eqs.~\eqref{eq:gauss_bulkm} and~\eqref{eq:gauss_bulkp}. As detailed in Sec.~2.3 of the SM, this choice yields 
\begin{alignat}{3}
    \phik{1} &= -\frac{1}{2 C_\mathrm{M}} \left(\bm{n} \cdot \langle \varepsilon_\mathrm{B} \bm{e}_{\mathrm{B}} \rangle^\mathrm{M} - \frac{1}{2}\left( \sigma^+ - \sigma^- \right)\right)~&&, \quad &&\forall \zeta^\alpha \in \Omega~, \label{eq:phi1} \\[6pt]
    \phik{2} &=  -\frac{1}{16 C_\mathrm{M}} \left(\bm{n} \cdot \llbracket \varepsilon_\mathrm{B}\bm{e}_{\mathrm{B}} \rrbracket^\mathrm{M} - \left(\sigma^+ + \sigma^-\right)\right)~&&, \quad &&\forall \zeta^\alpha \in \Omega~, \label{eq:phi2}
\end{alignat}
where $\bm{e}_{\mathrm{B}^\pm}$ is the electric field in $\mathcal{B}^\pm$, and $\langle \varepsilon_{\mathrm{B}} \bm{e}_{\mathrm{B}} \rangle^\mathrm{M} = \frac{1}{2}\left(\varepsilon_{\mathrm{B}^+}\bm{e}_{\mathrm{B}^+}\rvert_{\mathcal{S}^+} + \varepsilon_{\mathrm{B}^-}\bm{e}_{\mathrm{B}^-}\rvert_{\mathcal{S}^-} \right)$ and $\llbracket \varepsilon_{\mathrm{B}}\bm{e}_{\mathrm{B}} \rrbracket^\mathrm{M} = \varepsilon_{\mathrm{B}^+}\bm{e}_{\mathrm{B}^+}\rvert_{\mathcal{S}^+} - \varepsilon_{\mathrm{B}^-}\bm{e}_{\mathrm{B}^-}\rvert_{\mathcal{S}^-}$ denote averages and jumps across the thin film $\mathcal{M}$, respectively. \textspace

Deformations of the body $\mathcal{M}$ stretch and compress the bounding surfaces $\mathcal{S}^\pm$, leading to changes in surface charge densities $\sigma^\pm$. 
For a deforming body, it is thus convenient to express the surface charge densities with respect to a flat reference configuration. Using expressions for the change of surface area of $\mathcal{S}^\pm$ under deformations, the surface charge densities are expressed as
\begin{align}
    \sigma^\pm \approx \frac{1}{J_0}\, \sigma_0^\pm \left( 1 \pm H \delta\right)~,
\end{align}
where $J_0$ denotes the relative area change of the mid-surface with respect to a flat reference configuration with charge densities $\sigma_0^\pm$. Substituting these expressions into Eqs.~\eqref{eq:phi1} and~\eqref{eq:phi2}, we find
\begin{alignat}{3}
    \phik{1} &= -\frac{1}{2 C_\mathrm{M}} \left(\bm{n} \cdot \langle \varepsilon_\mathrm{B} \bm{e}_\mathrm{B} \rangle^\mathrm{M} - \frac{1}{2J_0}\left( \sigma_0^+ - \sigma_0^- \right) - \frac{H\delta}{2J_0}\left( \sigma_0^+ + \sigma_0^- \right) \right)~&&, \quad &&\forall \zeta^\alpha \in \Omega~, \label{eq:phi1_sig0} \\[6pt]
    \phik{2} &=  -\frac{1}{16 C_\mathrm{M}} \left(\bm{n} \cdot \llbracket \varepsilon_\mathrm{B}\bm{e}_\mathrm{B} \rrbracket^\mathrm{M} - \frac{1}{J_0}\left(\sigma_0^+ + \sigma_0^-\right) - \frac{H\delta}{J_0} \left(\sigma_0^+ - \sigma_0^-\right)\right)~&&, \quad &&\forall \zeta^\alpha \in \Omega~. \label{eq:phi2_sig0}
\end{alignat}
While Eqs.~\eqref{eq:phi1_sig0} and~\eqref{eq:phi2_sig0} are more useful in practice, Eqs.~\eqref{eq:phi1} and~\eqref{eq:phi2} are used for simplicity in the remainder of this article. \textspace

We now apply the dimension reduction procedure to the boundary conditions on $\mathcal{S_{||}}$ in Eqs.~\eqref{eq:DBC_Spar} and~\eqref{eq:NBC_Spar}. Upon expanding the prescribed potential, $\bar{\phi}_\mathrm{M} = \sum_{k=0}^\infty \bar{\phi}_{\mathrm{M}k}\leftR(\zeta^\alpha\rightR) \ChebT{k}{\theta\leftR(\zeta^3\rightR)}$, Eq.~\eqref{eq:DBC_Spar} becomes 
\begin{align}
    \phi_0\leftR(\zeta^\alpha\rightR) = \bar{\phi}_{\mathrm{M}0}\leftR(\zeta^\alpha\rightR)~,\quad \forall \zeta^\alpha \in \partial\Omega_{0\mathrm{D}}~, \label{eq:DBC_Spar_red}
\end{align}
where $\partial\Omega_{0\mathrm{D}}$ is the part of the parametric domain corresponding to $\partial\mathcal{S}_{0\mathrm{D}} = \partial \mathcal{S}_0 \cap \mathcal{S}_{||\mathrm{D}}$. Similarly, with the series expansion of the electric field component $\bar{e} = \sum_{k=0}^\infty \bar{e}_{k}\leftR(\zeta^\alpha\rightR) \ChebT{k}{\theta\leftR(\zeta^3\rightR)}$, Eq.~\eqref{eq:NBC_Spar} becomes
\begin{align}
    -{\nu}^\alpha \left( \phi_{0,\alpha}\leftR(\zeta^\alpha\rightR) + \frac{\delta}{4} \phi_{1,\beta}\leftR(\zeta^\alpha\rightR) b^\beta_\alpha + \frac{\delta^2}{16} \phi_{2,\beta}\leftR(\zeta^\alpha\rightR) b^\beta_\gamma b^\gamma_\alpha \right) = \bar{e}_0~, \quad \forall \zeta^\alpha \in \partial\Omega_{0\mathrm{N}}~,\label{eq:NBC_Spar_red}
\end{align}
where $\partial\Omega_{0\mathrm{N}}$ is the part of the parametric domain corresponding to $\partial\mathcal{S}_{0\mathrm{N}} = \partial \mathcal{S}_0 \cap \mathcal{S}_{||\mathrm{N}}$, and $\nu^\alpha$ and $b_\alpha^\beta$ are the components of $\bm{\nu}$ and the curvature tensor, respectively, and Einstein's summation convention applies. A detailed derivation of Eqs.~\eqref{eq:DBC_Spar_red} and~\eqref{eq:NBC_Spar_red} is provided in Sec.~2.3 of the SM. %\textspace
Equations~\eqref{eq:gauss_bulkm},~\eqref{eq:cont_Sm},~\eqref{eq:cont_Sp},~\eqref{eq:gauss_bulkp},~\eqref{eq:surface_gauss},~\eqref{eq:phi1}, and~\eqref{eq:phi2} together with the boundary conditions Eq.~\eqref{eq:DBC_Spar_red} and~\eqref{eq:NBC_Spar_red} form a closed set of equations that is independent of the parametric direction $\zeta^3$, while explicitly preserving effects due to the finite thickness of $\mathcal{M}$. This set of equations constitutes the dimensionally-reduced, $(2+\delta)$-dimensional theory of the electrostatics of thin films and is summarized in Tab.~\ref{tab:2pd_summary}. \textspace

\begin{table}[]
    \centering
\begin{tabular}{AA|AA}
\toprule
\text{3-dimen} &\text{sional theory} & &  &  (2+\delta)&\text{-dimensional theory} && \\
\toprule
%%%%%%%%%%%%%%%%%%%%%
\Delta \check{\phi}_{\mathrm{B}^-} &= -q_{\mathrm{B}^-}/\varepsilon_{\mathrm{B}^-} &
\bm{x} &\in \mathcal{B}^- &
\Delta \check{\phi}_{\mathrm{B}^-} &= -q_{\mathrm{B}^-}/\varepsilon_{\mathrm{B}^-} & 
\bm{x} &\in \mathcal{B}^- \\[3pt] %\midrule
%%%%%%%%%%%%%%%%%%%%%
\llbracket \check{\phi} \rrbracket &= 0 & 
\bm{x} &\in \mathcal{S}^- &
\llbracket \check{\phi} \rrbracket &= 0 & 
\bm{x} &\in \mathcal{S}^-  \\[3pt] %\midrule
%%%%%%%%%%%%%%%%%%%%%
\bar{\bm{n}}^+ \cdot \llbracket \varepsilon \check{\bm{e}} \rrbracket &= \sigma^- 
& \bm{x} &\in \mathcal{S}^- &
\phi_1 &= -\frac{1}{2 C_\mathrm{M}} \left(\bm{n} \cdot \langle \varepsilon_\mathrm{B} \bm{e}_{\mathrm{B}} \rangle^\mathrm{M} - \frac{1}{2}\left( \sigma^+ - \sigma^- \right)\right) & 
\\[3pt] 
 %\midrule
%%%%%%%%%%%%%%%%%%%%%
\varepsilon_\mathrm{M}\Delta \check{\phi}_\mathrm{M} &= 0 &
\bm{x} &\in \mathcal{M} &
\varepsilon_\mathrm{M} & \laplaces{\phi_{0}} - 4 C_\mathrm{M} \phi_{1} H +  \frac{16}{\delta} C_\mathrm{M} \phi_{2} = 0 &
\bm{x} &\in \mathcal{S}_0\\[3pt] %\midrule
%%%%%%%%%%%%%%%%%%%%%
\bar{\bm{n}}^- \cdot \llbracket \varepsilon \check{\bm{e}} \rrbracket &= \sigma^+ & 
\bm{x} &\in \mathcal{S}^+ &
\phi_2 &=  -\frac{1}{16 C_\mathrm{M}} \left(\bm{n} \cdot \llbracket \varepsilon_\mathrm{B}\bm{e}_{\mathrm{B}} \rrbracket^\mathrm{M} - \left(\sigma^+ + \sigma^-\right)\right) & \\[3pt] %\midrule
%%%%%%%%%%%%%%%%%%%%%
\llbracket \check{\phi} \rrbracket &= 0 &
\bm{x} &\in \mathcal{S}^+ &
\llbracket \check{\phi} \rrbracket &= 0 & 
\bm{x} &\in \mathcal{S}^- \\[3pt]% \midrule
%%%%%%%%%%%%%%%%%%%%%
\Delta \check{\phi}_{\mathrm{B}^+} &= -q_{\mathrm{B}^+}/\varepsilon_{\mathrm{B}^+} & 
\bm{x} &\in \mathcal{B}^+ & 
\Delta \check{\phi}_{\mathrm{B}^+} &= -q_{\mathrm{B}^+}/\varepsilon_{\mathrm{B}^+} & 
\bm{x} &\in \mathcal{B}^+\\
\bottomrule
\end{tabular}
\caption{Corresponding equations between the three-dimensional and $(2+\delta)$-dimensional theories. Note that the equations for $\phi_1$ and $\phi_2$ have not been assigned a location in the physical domain. This is due to the electric field being evaluated on both $\mathcal{S}^+$ and $\mathcal{S}^-$ in the expressions for $\phi_1$ and $\phi_2$. }
\label{tab:2pd_summary}
\end{table}
The expansion of the potential in Eq.~\eqref{eq:chebphi} allows finding the potential drop across the thin film:
% From the expansion of the potential in Eq.~\eqref{eq:chebphi}, we find the potential drop across the thin film to be 
\begin{align}
    \llbracket \phi\leftR(\zeta^\alpha\rightR) \rrbracket^\mathrm{M} &= 2 \phi_1\leftR(\zeta^\alpha\rightR) = -\frac{1}{C_\mathrm{M}} \left(\bm{n} \cdot \langle \varepsilon_\mathrm{B} \bm{e}_\mathrm{B} \rangle^\mathrm{M} - \frac{1}{2}\left( \sigma^+ - \sigma^- \right)\right)~, \label{eq:jump_phiM}
\end{align}
which is a generalization of an expression derived in Refs.~\cite{leonetti2004pattern, lacoste2009electrostatic,Ziebert2010}. Equation~\eqref{eq:jump_phiM} can also be written as
\begin{align}
    \llbracket \phi\leftR(\zeta^\alpha\rightR) \rrbracket^\mathrm{M} = \frac{ \Sigma_\mathrm{eff}\leftR(\zeta^\alpha\rightR)}{ C_\mathrm{M}}~,
\end{align} 
with the effective surface charge density $\Sigma_\mathrm{eff} = -\bm{n} \cdot \langle \varepsilon_\mathrm{B} \bm{e} \rangle^\mathrm{M} + \frac{1}{2}\left( \sigma^+ - \sigma^- \right)$, indicating an analogy to a parallel plate capacitor. 
Similarly, for two parallel, charged surfaces a distance $\delta$ apart and with constant charge density $q$ in the space between the plates, the second order Chebyshev coefficient of the potential is $\hat{\phi}_2 = -\frac{Q}{16C}$, where $Q = q\delta$ and $C$ is the capacitance per unit area. This motivates the definition of an effective charge density $Q_\mathrm{eff}$:
\begin{align}
    Q_\mathrm{eff}\leftR(\zeta^\alpha\rightR) = -16C_\mathrm{M} \phi_2\leftR(\zeta^\alpha\rightR) = \bm{n} \cdot \llbracket \varepsilon_\mathrm{B}\bm{e}_\mathrm{B} \rrbracket^\mathrm{M} - \left(\sigma^+ + \sigma^-\right)~. \label{eq:Qeff_def}
\end{align}
which, upon substitution in Eq.~\eqref{eq:surface_gauss}, yields
\begin{align}
    \varepsilon_\mathrm{M} \laplaces{\phik{0}} - 4 C_\mathrm{M} \phik{1} H  = \frac{Q_\mathrm{eff}\leftR(\zeta^\alpha\rightR)}{\delta}~, \quad \forall \zeta^\alpha \in \Omega~, \label{eq:surface_gauss_Qeff} 
\end{align}
where $Q_\mathrm{eff}/\delta$ appears like a charge density in Gauss' law. From Eqs.~\eqref{eq:phi1} and~\eqref{eq:phi2} and the definition of $C_\mathrm{M}$, we find that both $\phik{1}$ and $\phik{2}$ are $\mathcal{O}\leftR(\delta\rightR)$ terms, suggesting that $Q_\mathrm{eff}$ is $\mathcal{O}\leftR(1\rightR)$. Thus, for Eq.~\eqref{eq:surface_gauss_Qeff} to remain well-posed in the limit of $\delta\rightarrow 0$, we require $Q_\mathrm{eff} \propto \phi_2\leftR(\zeta^\alpha\rightR) \rightarrow 0$, implying 
\begin{equation}
   \bm{n} \cdot \llbracket  \varepsilon_\mathrm{B}  \bm{e} \rrbracket^\mathrm{M} = \sigma^+ + \sigma^-, \label{eq:shell_jump}
\end{equation}
which is the jump condition in Eq.~\eqref{eq:ME_jump_dot} for a surface with charge density $\sigma^+ + \sigma^-$. The same condition has been used for lipid membranes in Refs.~\cite{bensimon1990stability,fogden1991bending}. However, we generally do not consider this limit and instead work with the full expression in Eq.~\eqref{eq:surface_gauss}. \textspace

\begin{figure}[t]
     \centering
     \begin{subfigure}[c]{0.4\textwidth}
        \centering
        \imagebox{0.8\textwidth}{\begin{tikzpicture}[font=\normalfont, x=\textwidth, y=0.65\textwidth]

% adjust bounding box since some control points are outside the [0,1]
\useasboundingbox (0,0) rectangle (1,1);

% offsets
\newcommand\CPix{0.4}
\newcommand\CPiy{-0.15}
\newcommand\CPiix{-0.4}
\newcommand\CPiiy{0.15}
\newcommand\SNcenter{0.5}

% reference points
\coordinate (S0left) at (0,0.45);
\coordinate (S0right) at (1,0.55);
\coordinate (Spleft) at (0,0.54);
\coordinate (Spright) at (1,0.64);
\coordinate (Smleft) at (0,0.36);
\coordinate (Smright) at (1,0.46);

\coordinate (Sm1left) at (0,0.24);
\coordinate (Sm1right) at (1,0.34);
\coordinate (Sm2left) at (0,0.07);
\coordinate (Sm2right) at (1,0.17);
\coordinate (Sp1left) at (0,0.68);
\coordinate (Sp1right) at (1,0.78);
\coordinate (Sp2left) at (0,0.83);
\coordinate (Sp2right) at (1,0.93);

% mesh dash pattern
\newcommand\OnP{4pt}
\newcommand\OffP{3pt}

% top, bottom and mid-surface
\draw[smooth, thick] (Spleft) .. controls +(\CPix, \CPiy) and +(\CPiix, \CPiiy) .. (Spright);
\draw[smooth, thick] (Smleft) .. controls +(\CPix, \CPiy) and +(\CPiix, \CPiiy) .. (Smright);
\draw[smooth, very thick, dashed, dash pattern = on 2pt off 2pt] (S0left) .. controls +(\CPix, \CPiy) and +(\CPiix, \CPiiy) .. (S0right);

% membrane
\begin{scope}
\draw[smooth, fill=gray,fill opacity=.3,draw=none]  (Spleft) .. controls +(\CPix, \CPiy) and +(\CPiix, \CPiiy) .. (Spright) .. controls (S0right) .. (Smright) .. controls +(\CPiix, \CPiiy) and +(\CPix,\CPiy) .. (Smleft) ;
\end{scope}
% add white membrane below to hide mesh lines below
\begin{pgfonlayer}{bg1}
\begin{scope}
    \draw[smooth, fill=white,draw=none]  (Spleft) .. controls +(\CPix, \CPiy) and +(\CPiix, \CPiiy) .. (Spright) .. controls (S0right) .. (Smright) .. controls +(\CPiix, \CPiiy) and +(\CPix,\CPiy) .. (Smleft) ;
\end{scope}
\end{pgfonlayer}

% mesh below
\draw[smooth, thick, dashed, dash pattern= on \OnP off \OffP] (Sm1left) .. controls +(\CPix, \CPiy) and +(\CPiix, \CPiiy) .. (Sm1right);
\draw[smooth, thick, dashed, dash pattern= on \OnP off \OffP] (Sm2left) .. controls +(\CPix, \CPiy) and +(\CPiix, \CPiiy) .. (Sm2right);
\begin{pgfonlayer}{bg0}
    \draw[thick, dashed,dash phase=0pt, dash pattern= on \OnP off \OffP] (0.1, 0.5) -- +(-0.05,-0.5);
    \draw[thick, dashed,dash phase=0pt, dash pattern= on \OnP off \OffP] (0.25, 0.5) -- +(0.07,-0.5);
    \draw[thick, dashed, dash phase=3pt, dash pattern= on \OnP off \OffP] (0.48, 0.5) -- +(0.07,-0.5);
    \draw[thick, dashed,dash phase=4pt, dash pattern= on \OnP off \OffP] (0.7, 0.5) -- +(0.04,-0.5);
    \draw[thick, dashed,dash phase=4pt, dash pattern= on \OnP off \OffP] (0.95, 0.5) -- +(-0.05,-0.5);
\end{pgfonlayer}

% mesh above
\draw[smooth, thick, dashed, dash pattern= on \OnP off \OffP] (Sp1left) .. controls +(\CPix, \CPiy) and +(\CPiix, \CPiiy) .. (Sp1right);
\draw[smooth, thick, dashed, dash pattern= on \OnP off \OffP] (Sp2left) .. controls +(\CPix, \CPiy) and +(\CPiix, \CPiiy) .. (Sp2right);
\begin{pgfonlayer}{bg0}
    \draw[thick, dashed,dash phase=4pt, dash pattern= on \OnP off \OffP] (0.05, 0.5) -- +(0.05,0.5);
    \draw[thick, dashed,dash phase=4pt, dash pattern= on \OnP off \OffP] (0.3, 0.5) -- +(-0.03,0.5);
    \draw[thick, dashed,dash phase=4pt, dash pattern= on \OnP off \OffP] (0.49, 0.5) -- +(-0.04,0.5);
    \draw[thick, dashed,dash phase=1pt, dash pattern= on \OnP off \OffP] (0.7, 0.5) -- +(-0.05,0.5);
    \draw[thick, dashed,dash phase=0pt, dash pattern= on \OnP off \OffP] (0.85, 0.5) -- +(0.04,0.5);
\end{pgfonlayer}

% labels
\node[] at (0.18,0.58) {$\mathcal{S}^+$};
\node[] at (0.2,0.27) {$\mathcal{S}^-$};
\draw (0.31,0.43) .. controls +(0.0,0.05) and +(-0.05,0) .. ++(0.06,0.19) node[anchor=west, outer sep  = 0, inner sep =0.2em] (nodeSp) {$\mathcal{S}_0$}; 

\end{tikzpicture}}
        \caption{Meshing with explicit thickness.}%
        \label{subfig:meshing_w_thickness}
     \end{subfigure}\hspace{0.1\textwidth}%
     \begin{subfigure}[c]{0.4\textwidth}
         \centering
         \imagebox{0.8\textwidth}{\begin{tikzpicture}[font=\normalfont, x=\textwidth, y=0.8\textwidth]

% adjust bounding box since some control points might be outside the [0,1]
\useasboundingbox (0,0) rectangle (1,1);

% reference points
\coordinate (S0left) at (0,0.45);
\coordinate (S0right) at (1,0.55);

\coordinate (Sm1left) at (0,0.33);
\coordinate (Sm1right) at (1,0.43);
\coordinate (Sm2left) at (0,0.16);
\coordinate (Sm2right) at (1,0.26);
\coordinate (Sp1left) at (0,0.59);
\coordinate (Sp1right) at (1,0.69);
\coordinate (Sp2left) at (0,0.74);
\coordinate (Sp2right) at (1,0.84);

% offsets
\newcommand\CPix{0.4}
\newcommand\CPiy{-0.15}
\newcommand\CPiix{-0.4}
\newcommand\CPiiy{0.15}
\newcommand\SNcenter{0.5}

% mesh points
\coordinate (mesh1) at (0.1, 0.42);
\coordinate (mesh2) at (0.25, 0.42);
\coordinate (mesh3) at (0.48, 0.49);
\coordinate (mesh4) at (0.7, 0.57);
\coordinate (mesh5) at (0.95, 0.565);
\newcommand\lowb{0.09}
\newcommand\upb{0.91}

% mesh dash pattern
\newcommand\OnP{4pt}
\newcommand\OffP{3pt}

% top, bottom and mid-surface
\draw[smooth, very thick, dashed, dash pattern = on 2pt off 2pt] (S0left) .. controls +(\CPix, \CPiy) and +(\CPiix, \CPiiy) .. (S0right);

% mesh below
\draw[smooth, thick, dashed, dash pattern= on \OnP off \OffP] (Sm1left) .. controls +(\CPix, \CPiy) and +(\CPiix, \CPiiy) .. (Sm1right);
\draw[smooth, thick, dashed, dash pattern= on \OnP off \OffP] (Sm2left) .. controls +(\CPix, \CPiy) and +(\CPiix, \CPiiy) .. (Sm2right);

\draw[thick, dashed, dash pattern= on \OnP off \OffP] (mesh1) -- (0.05,\lowb);
\draw[thick, dashed, dash pattern= on \OnP off \OffP] (mesh2) -- (0.32,\lowb);
\draw[thick, dashed, dash pattern= on \OnP off \OffP] (mesh3) -- (0.55,\lowb);
\draw[thick, dashed, dash pattern= on \OnP off \OffP] (mesh4) -- (0.74,\lowb);
\draw[thick, dashed, dash pattern= on \OnP off \OffP] (mesh5) -- (1,\lowb);

% mesh above
\draw[smooth, thick, dashed, dash pattern= on \OnP off \OffP] (Sp1left) .. controls +(\CPix, \CPiy) and +(\CPiix, \CPiiy) .. (Sp1right);
\draw[smooth, thick, dashed, dash pattern= on \OnP off \OffP] (Sp2left) .. controls +(\CPix, \CPiy) and +(\CPiix, \CPiiy) .. (Sp2right);
\begin{pgfonlayer}{bg1}
    \draw[thick, dashed, dash pattern= on \OnP off \OffP] (mesh1) -- (0.05,\upb);
    \draw[thick, dashed, dash pattern= on \OnP off \OffP] (mesh2) -- (0.26,\upb);
    \draw[thick, dashed, dash pattern= on \OnP off \OffP] (mesh3) -- (0.44,\upb);
    \draw[thick, dashed, dash pattern= on \OnP off \OffP] (mesh4) -- (0.65,\upb);
    \draw[thick, dashed, dash pattern= on \OnP off \OffP] (mesh5) -- (0.99,\upb);
\end{pgfonlayer}

% labels
\node[] at (0.32,0.49) {$\mathcal{S}_0$};

\end{tikzpicture}}
         \caption{Meshing without explicit thickness.}
        \label{subfig:meshing_wo_thickness}
     \end{subfigure}
     \ifcap{\caption{When solving the $(2+\delta)$-dimensional theory numerically, a discretization that explicitly accounts for the finite thickness of the membrane, as shown in (a), may be cumbersome to implement, in particular for moving meshes. Alternatively, the mesh on the bounding surfaces, $\mathcal{S}^-$ and $\mathcal{S}^+$, can be collapsed onto the membrane mid-surface $\mathcal{S}_0$, as shown in (b). }}{\vspace{1cm}}
     \label{fig:meshing_thickness}
\end{figure}
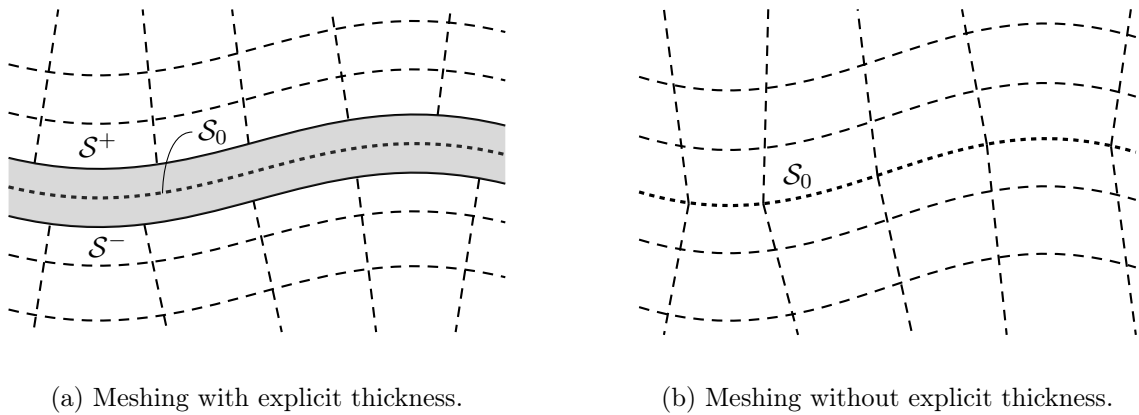

Lastly, note that we have defined Eqs.~\eqref{eq:surface_gauss}--\eqref{eq:phi2} on the mid-surface $\mathcal{S}_0$. However, the average $\langle \varepsilon_\mathrm{B} \bm{e}_\mathrm{B} \rangle^\mathrm{M}$ and jump $\llbracket \varepsilon_\mathrm{B}\bm{e}_\mathrm{B} \rrbracket^\mathrm{M}$ in Eqs.~\eqref{eq:phi1} and~\eqref{eq:phi2}, respectively, require evaluation of the electric field on $\mathcal{S}^+$ and $\mathcal{S}^-$ as shown in Fig.~\ref{subfig:meshing_w_thickness}. In practice, when solving the governing equations numerically, the membrane could be treated as a surface such that the interface conditions would be enforced on either side of $\mathcal{S}_0$ instead. This viewpoint is illustrated in Fig.~\ref{subfig:meshing_wo_thickness}. Treating the lipid membrane as a surface when creating the discretization introduces an error of order $\mathcal{O}\leftR(\delta \kappa\rightR)$. However, the $(2+\delta)$-dimensional theory is truncated at order $\mathcal{O}\leftR( \left(\delta \kappa\right)^2\rightR)$, suggesting that the error resulting from treating the lipid membrane as a surface can become dominant at large curvatures.

\section{Comparison to Three-Dimensional Gauss Law}
\label{sect: comp_to_3DGauss}
In this section, the accuracy of the $\left(2+\delta\right)$-dimensional theory is tested on flat geometries, cylinders, and spheres, which are common lipid membrane geometries encountered in both theory and experiments \cite{sens2002undulation, dimova2009vesicles, sinha2013electric, sahu2020geometry, salipante2014vesicle}. Section~\ref{sect:comp_analytical} considers examples with analytical solutions while Sec.~\ref{sect:comp_numerical} presents a numerical comparison for examples without analytical solutions but relevant for lipid membranes.

\subsection{Analytical Comparison} \label{sect:comp_analytical}

We begin by applying the $\left(2+\delta\right)$-dimensional theory to examples of thin films embedded in dielectric bulk media with univariate potentials. In the interest of clarity, many of the details of the analytical solutions are described in Sec.~3 of the SM. For the examples considered, we find that the pointwise, relative error between the exact and $(2+\delta)$-dimensional theories does not exceed $2\%$. For cylinders and spheres, the error decreases rapidly with increasing radius.  \textspace

\subsubsection{Flat Geometry}
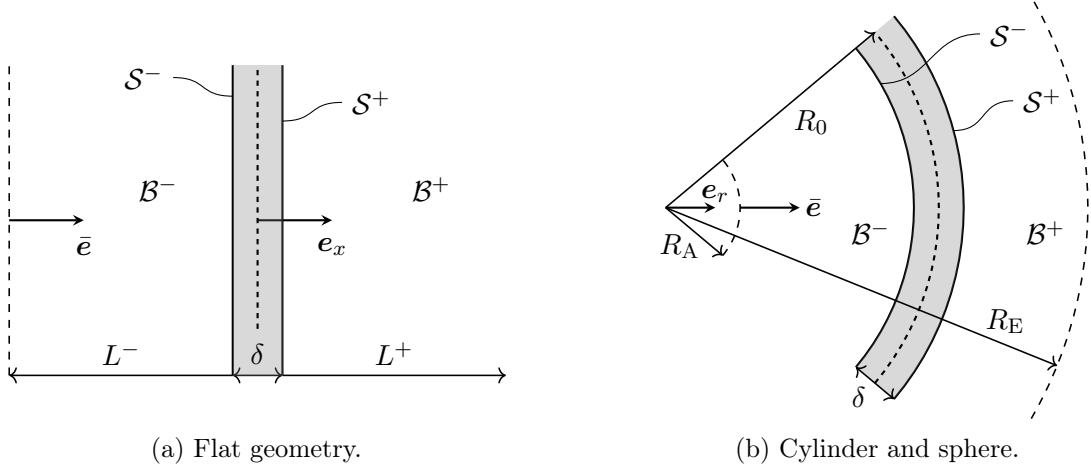
\begin{figure}[t]
     \centering
     \begin{subfigure}[b]{0.4\textwidth}
        \centering
        \begin{tikzpicture}[font=\normalfont, x=\textwidth, y=0.625\textwidth]

% reference points
\coordinate (Smbottom) at (0.45,0);
\coordinate (Spbottom) at (0.55,0);
\coordinate (S0bottom) at (0.5,0.15);
\coordinate (leftBCbottom) at (0,0);
\coordinate (rightBCbottom) at (1,0);

% axis arrows
\draw[<->,semithick] (Smbottom) -- (Spbottom);
\draw[<-,semithick] (leftBCbottom) -- (Smbottom);
\draw[->,semithick] (Spbottom) -- (rightBCbottom);

% left boundary
\draw[dashed,semithick] (leftBCbottom) -- +(0,1);

% e_x vector
\draw[thick, ->, >=stealth] (0.5,0.5) -- +(0.15,0.) node[anchor=north,outer sep = 3pt] (nodeex) {$\bm{e}_x$};

% e-field vector
\draw[thick, ->, >=stealth] (0.,0.5) -- +(0.15,0.) node[anchor=north,outer sep = 3pt] (nodee) {$\bar{\bm{e}}$};

% membrane 
\draw[thick] (Smbottom) -- +(0,1);
\draw[thick] (Spbottom) -- +(0,1);
\draw[fill=gray,fill opacity=.3,draw=none] (Smbottom) -- +(0,1) -- +(0.1,1) -- (Spbottom) -- (Smbottom);

% mid-surface
\draw[smooth, thick, dashed, dash pattern = on 2pt off 2pt] (S0bottom) -- (0.5,1);

% labels
\node[] at (0.3,0.6) {$\mathcal{B}^-$};
\node[] at (0.225,0.07) {$L^-$};
\node[] at (0.85,0.6) {$\mathcal{B}^+$};
\node[] at (0.775,0.07) {$L^+$};
\node[] at (0.5,0.07) {$\delta$};
\draw (0.45,0.9) .. controls +(-0.07,0) and +(0.07,-0.) .. ++(-0.12,0.06) node[anchor=east] (nodeSm) {$\mathcal{S}^-$}; 
\draw (0.55,0.82) .. controls +(0.07,0) and +(-0.07,-0.) .. ++(0.12,0.06) node[anchor=west] (nodeSp) {$\mathcal{S}^+$}; 

\end{tikzpicture}
        \caption{Flat geometry.}
        \label{fig:flat_analytical_setup}
     \end{subfigure}\hspace{0.1\textwidth}%
     \begin{subfigure}[b]{0.4\textwidth}
         \centering
         \begin{tikzpicture}[font=\normalfont, x=\textwidth, y=\textwidth]

% definitions
\newcommand\RA{0.15}
\newcommand\RE{0.85}
\newcommand\Rzero{0.55}
\newcommand\Rinner{0.5}
\newcommand\Router{0.6}
\newcommand\anglep{40}
\newcommand\anglem{-40}
\newcommand\ycenter{0.5}

% reference points
\coordinate (Center) at (0,\ycenter); 

% membrane outline
\draw [thick,domain=\anglem:\anglep] plot ({\Rinner*cos(\x)}, {\ycenter + \Rinner*sin(\x)});
\draw [thick,domain=\anglem:\anglep] plot ({\Router*cos(\x)}, {\ycenter + \Router*sin(\x)}); 
% membrane fill
\draw[draw=none,fill=gray,fill opacity=.3,domain=\anglem:\anglep] 
    plot ({\Rinner*cos(\x)}, {\ycenter + \Rinner*sin(\x)}) --
    ({\Router*cos(\anglep)}, {\ycenter + \Router*sin(\anglep)}) --
    plot ({\Router*cos(-\x)}, {\ycenter + \Router*sin(-\x)}) --
    ({\Router*cos(\anglem)}, {\ycenter + \Router*sin(\anglem)});
% mid-surface
\draw [thick, dashed, dash pattern = on 2pt off 2pt,domain=\anglem:\anglep] plot ({\Rzero*cos(\x)}, {\ycenter + \Rzero*sin(\x)}); 

% inner boundary condition
\draw [semithick, domain=\anglem:\anglep,dashed] plot ({\RA*cos(\x)}, {\ycenter + \RA*sin(\x)}); 

% inner boundary condition
\draw [semithick, domain=(\anglem+10):(\anglep-10),dashed] plot ({\RE*cos(\x)}, {\ycenter + \RE*sin(\x)}); 

% geometry arrows
\draw[->,semithick] (Center) -- +({\RA*cos(\anglem)}, {\RA*sin(\anglem)}) node[anchor=south east, outer xsep = 0.7em, outer ysep = -0.2em, inner sep = 0.1em] (nodeRA) {$R_\mathrm{A}$};
\draw[->,semithick] (Center) -- +({\Rzero*cos(\anglep)}, {\Rzero*sin(\anglep)});
\draw[->,semithick] (Center) -- +({\RE*cos(\anglem+18)}, {\RE*sin(\anglem+18)}) node[anchor=south east, outer sep = 0.8em] (nodeRE) {$R_\mathrm{E}$};
\draw[<->,semithick] ({\Rinner*cos(\anglem)}, {\ycenter + \Rinner*sin(\anglem)}) -- ({\Router*cos(\anglem)}, {\ycenter + \Router*sin(\anglem)});

% e_r vector
\draw[thick, ->, >=stealth] (Center) -- +(0.1,0.) node[anchor=south ,outer sep = 0pt, inner sep = 2 pt] (nodeex) {$\bm{e}_r$};

% e field vector
\draw[thick, ->, >=stealth] (\RA,\ycenter) -- +(0.12,0.) node[anchor=west, outer sep = 0pt, inner sep = 0.2em] (nodee) {$\bar{\bm{e}}$};

% labels
\node[] at ({0.75*\Rzero},0.45) {$\mathcal{B}^-$};
\node[] at ({0.9*\RE},0.45) {$\mathcal{B}^+$};
\node[] at ({0.6*\Rzero*cos(\anglep)+0.04}, {\ycenter + 0.6*\Rzero*sin(\anglep)-0.04}) {$R_0$};
\node[] at ({\Rzero*cos(\anglem-5)}, {\ycenter + \Rzero*sin(\anglem-4)}) {$\delta$};
\draw ({(\Rzero + 0.05)*cos(\anglep-25)},{\ycenter + (\Rzero + 0.05)*sin(\anglep-25)}) .. controls +(0.07,0) and +(-0.07,-0.) .. ++(0.12,0.06) node[anchor=west] (nodeSp) {$\mathcal{S}^+$};
\draw ({(\Rzero - 0.05)*cos(\anglep-10)},{\ycenter + (\Rzero - 0.05)*sin(\anglep-10)}) .. controls +(0.1,0) and +(-0.1,-0.) .. ++(0.2,0.1) node[anchor=west] (nodeSp) {$\mathcal{S}^-$};

\end{tikzpicture}
         \caption{Cylinder and sphere.}
         \label{fig:cyl_sph_analytical_setup}
     \end{subfigure}
     \caption{Setup for the flat geometry (a), cylinder and sphere (b). The bulk domains are dielectric materials without any free charge. On one boundary, an electric field is prescribed, while the potential is fixed on the other. }
     \label{fig:FC_setup}
\end{figure}

Consider the flat, thin film shown in Fig.~\ref{fig:flat_analytical_setup} where the potential only depends on the $x$-direction and $q_{\mathrm{B}^\pm} =0$. The electric field is prescribed on the left-hand side boundary of the domain and the potential is fixed on the right-hand side boundary of the domain: 
\begin{alignat}{2}
    -\frac{\ddiff \check{\phi}_{\mathrm{B}^-}}{\ddiff x} &= \bar{e}~,\quad && x=-\delta/2 - L^-~, \label{eq:flat_e_BC} \\
    \check{\phi}_{\mathrm{B}^+} &= 0~,\quad && x = \delta/2 + L^+~. \text{\footnotemark}
    \label{eq:flat_phi_BC} 
\end{alignat}%
%%%%
%%%%%%%
\footnotetext{\text{Equations~\eqref{eq:flat_e_BC} and~\eqref{eq:flat_phi_BC} hold analogously for the $(2+\delta)$-dimensional theory.}}%
By simplifying Eq.~\eqref{eq:gauss_M}, it becomes apparent that the solution to the exact theory is at most linear in $x$ within the thin film. Since linear solutions can be represented exactly in the $(2+\delta)$-dimensional theory, the $\left(2+\delta\right)$-dimensional theory recovers the exact solution. The governing equations and solutions for both the exact and $\left(2+\delta\right)$-dimensional theory for this case can be found in Sec.~3.1 of the SM.

\subsubsection{Cylinders} \label{subsubsect:ana_cylinders}

The next example is similar to the one discussed in the previous section but with the flat geometry replaced by a cylinder with mid-surface radius $R_0$. The setup, depicted in Fig.~\ref{fig:cyl_sph_analytical_setup}, is axisymmetric and homogeneous along the cylinder's axis such that the potential only depends on the radial direction. Similar to before, we fix the potential to be zero at $r=R_\mathrm{E}$ and impose the electric field at $r=R_\mathrm{A} > 0$, with  $R_\mathrm{E}$ and $R_\mathrm{A}$ shown in Fig.~\ref{fig:cyl_sph_analytical_setup}:
\begin{align}
    -\frac{\ddiff \check{\phi}_{\mathrm{B}^-}}{\ddiff r} &= \bar{e}~,\quad r=R_\mathrm{A}~, \label{eq:cyl_ana_full_BCRA} \\
    \check{\phi}_{\mathrm{B}^+} &= 0~,\quad r = R_\mathrm{E}~. \label{eq:cyl_ana_full_BCRE}
\end{align}
The simplified equations and corresponding solutions for both the three-dimensional and $(2+\delta)$-dimensional theories are presented in Sec.~3.2 of the SM. In contrast to the flat geometry, the potential is no longer linear within the membrane and the $(2+\delta)$-dimensional theory does not reproduce the exact solution. %
To assess the differences between the exact and $(2+\delta)$-dimensional solutions, we introduce the following non-dimensional quantities: 
\begin{alignat*}{4}
    &r^\ast = \frac{r}{\delta}~, \hspace{2cm} 
    &&\delta^\ast = 1~, \hspace{2cm} %
    &&\varepsilon_\mathrm{M}^\ast = \frac{\varepsilon_\mathrm{M}}{\varepsilon_0}~, \hspace{2cm} %
    &&\varepsilon_\mathrm{B}^\ast = \frac{\varepsilon_\mathrm{B}}{\varepsilon_0}~, \\
    &\phi^\ast = \frac{\phi \varepsilon_0}{\delta \sigma^+}~, \hspace{2cm} %
    &&\sigma^{\pm \ast} = \frac{\sigma^\pm}{\sigma^+}~, \hspace{2cm} %
    &&\bar{e}^\ast = \frac{\bar{e} \varepsilon_0}{\sigma^+}~. &&
\end{alignat*}
This non-dimensionalization does not carry physical meaning but is merely chosen for convenience. We consider two different parameter choices, cases A and B, defined in Tab.~\ref{tab:comparison_cases}. %
For case A, the dielectric constants are the same throughout the entire domain but the surface charge densities on the inner and outer surface of the thin film differ in magnitude and sign. For case B, the dielectric constants in the thin film and bulk domains differ, and the surface charge densities have different magnitudes. The non-dimensional mid-surface radius $R_0^\ast$ is varied and thus not listed in Tab.~\ref{tab:comparison_cases}. \textspace

\begin{table}[t!]
    \centering
    \begin{tabular}{c|ccccccc} \toprule
         case & $\varepsilon_\mathrm{M}^\ast$ & $\varepsilon_\mathrm{B}^\ast$ & $\sigma^{+\ast}$ & $\sigma^{-\ast}$ & $\bar{e}^\ast$ & $R_\mathrm{A}^\ast$ & $R_\mathrm{E}^\ast$ \\ \midrule
         A & $1$ & $1$ & $1$ & $-1$ & $1$ & $1$ & $R_0^\ast + 10$ \\ \hline
         B & $2$ & $80$ & $1$ & $100$ & $-10$ & $1$ & $R_0^\ast + 10$ \\ \bottomrule %\hline
    \end{tabular}
    \caption{Non-dimensional quantities for the two analytical test cases for cylinders and spheres.}
    \label{tab:comparison_cases}
\end{table}%

\begin{figure}[h]
    \centering
    \begin{subfigure}[b]{0.48\linewidth}
         \centering
         \includegraphics[]{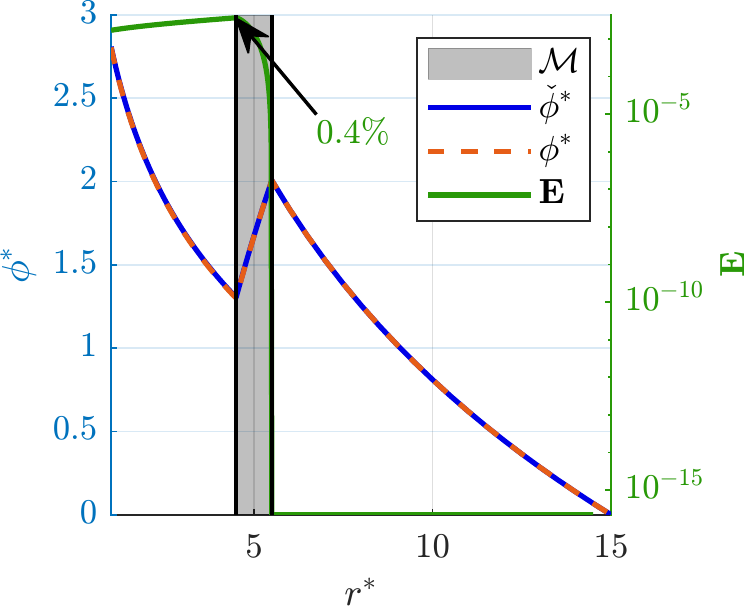}
         \caption{Cylinder, case A.}
         \label{subfig:cylinder_phi_caseA}
     \end{subfigure}\hspace{0.04\linewidth}%
     \begin{subfigure}[b]{0.48\linewidth}
         \centering
         \includegraphics{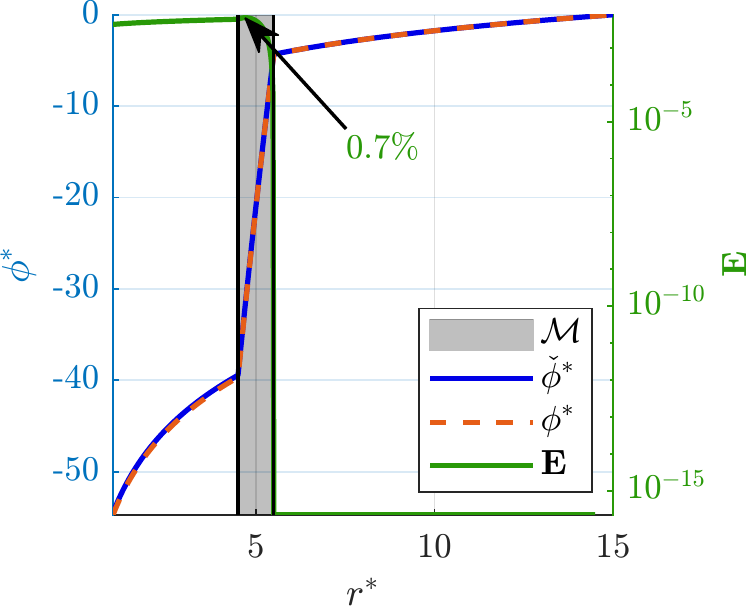}
         \caption{Cylinder, case B.}
         \label{subfig:cylinder_phi_caseB}
     \end{subfigure}%
     \caption{Comparison between the exact and $(2+\delta)$-dimensional theories on the cylinder for the two test cases described in Tab.~\ref{tab:comparison_cases}. }
     \label{fig:c_phi_caseAB}
\end{figure}%

\begin{figure}[H]
     \centering
     \begin{subfigure}[b]{0.5\textwidth}
         \centering         \includegraphics{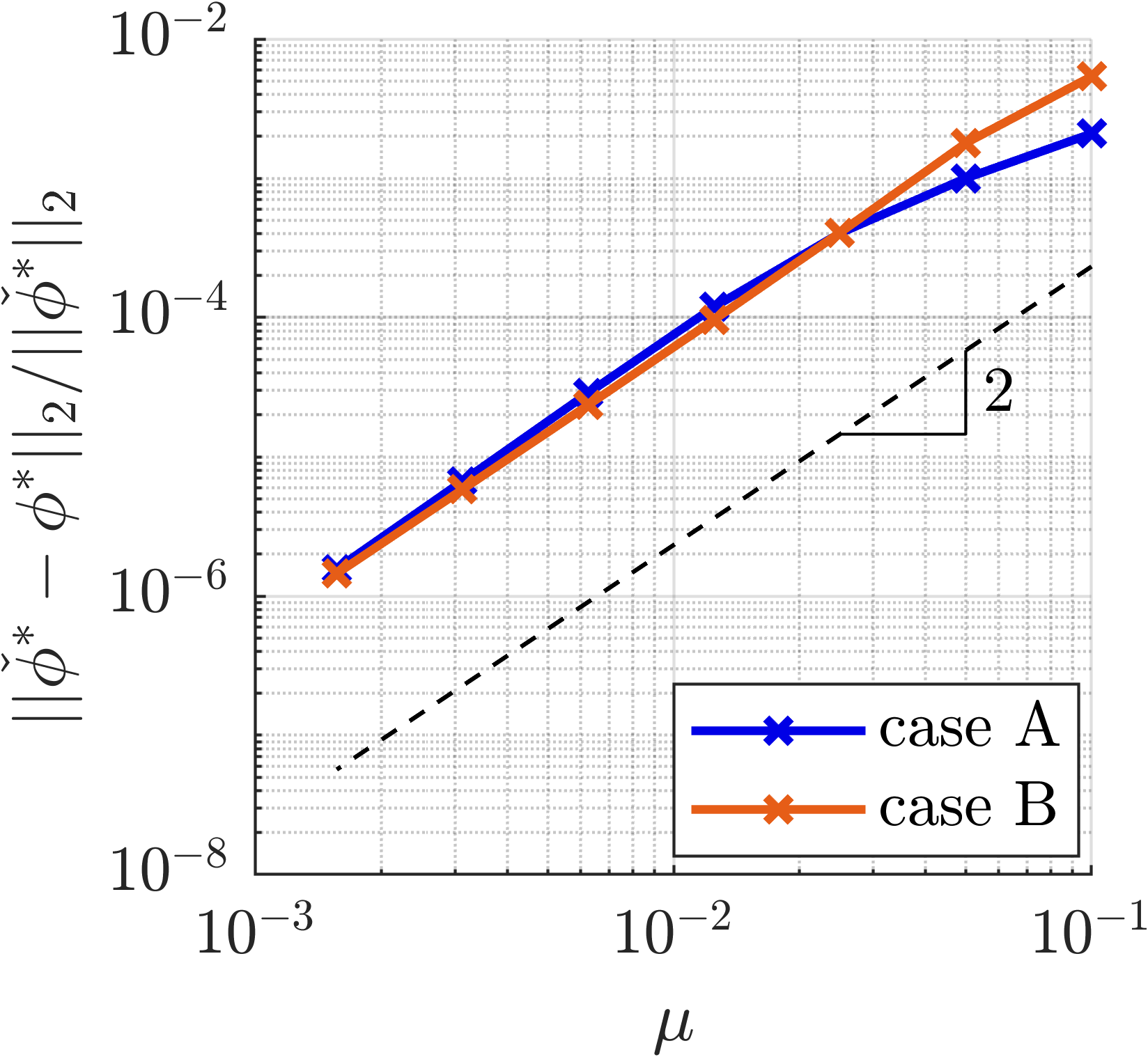}
         \caption{Cylinder.}
         \label{subfig:cylinder_mu_caseAB}
     \end{subfigure}%
     \hfill
     \begin{subfigure}[b]{0.5\textwidth}
         \centering        \includegraphics{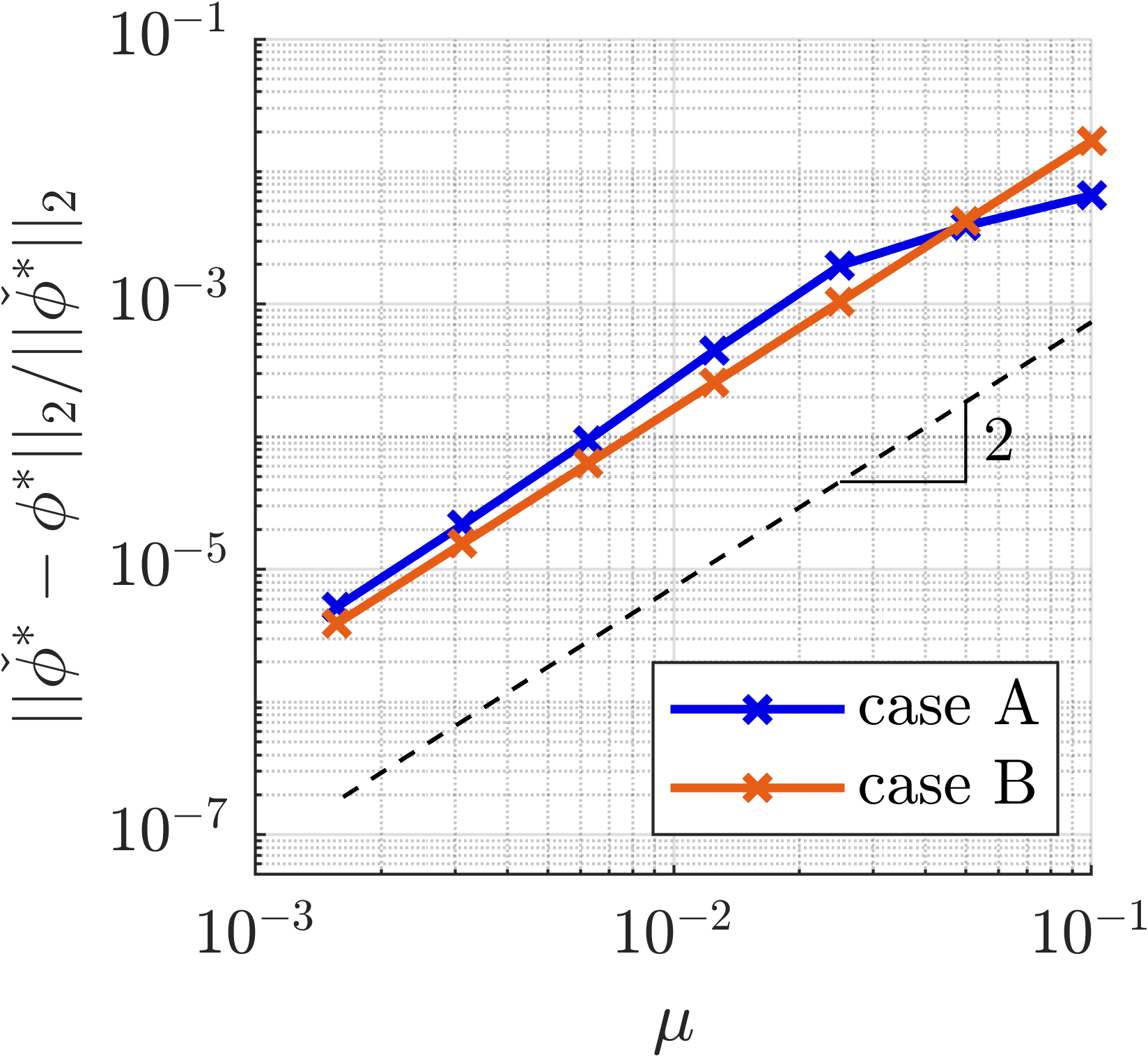}
         \caption{Sphere.}
         \label{subfig:sphere_mu_caseAB}
     \end{subfigure}
     \caption{Dependence of the $L_2$-error on the non-dimensional curvature $\mu = \delta/(2R_0)$ for the cylinder (a) and sphere (b) for cases A and B. The error is found to reduce quadratically with $\mu$.}
     \label{fig:cs_caseAB}
\end{figure}

Figures~\ref{subfig:cylinder_phi_caseA} and~\ref{subfig:cylinder_phi_caseB} show the potential and error profiles for cases A and B, respectively, with $R_0^\ast = 5$ and the pointwise relative error defined as
\begin{align}
    \bf{E} = \frac{|\check{\phi}^\ast-\phi^\ast|}{|\check{\phi}^\ast|}~.
\end{align}
The potential profiles from the exact and $\left(2+\delta\right)$-dimensional theory agree closely for both cases, with a maximum error of less than $1\%$. We note that in Figs.~\ref{subfig:cylinder_phi_caseA} and~\ref{subfig:cylinder_phi_caseB}, the radius of the mid-surface is only five times the thickness, even though, in the derivation of the $(2+\delta)$-dimensional theory, we used the assumption that the thickness is small compared to the radius of curvature (Eq.~\eqref{eq:dR_ll_1}). Figure~\ref{subfig:cylinder_mu_caseAB} shows that the $L_2$-error in the potential decreases quadratically with the non-dimensional curvature $\mu$,
\begin{align}
    \mu = \frac{\delta/2}{R_0}~,
\end{align}
consistent with Eq.~\eqref{eq:dR_ll_1}. In Sec.~3.2 of the SM, this result is confirmed by comparing the exact and $(2+\delta)$-dimensional solutions analytically.

\begin{figure}[h]
    \begin{subfigure}[b]{0.48\linewidth}
         \centering
         \includegraphics[width=\textwidth]{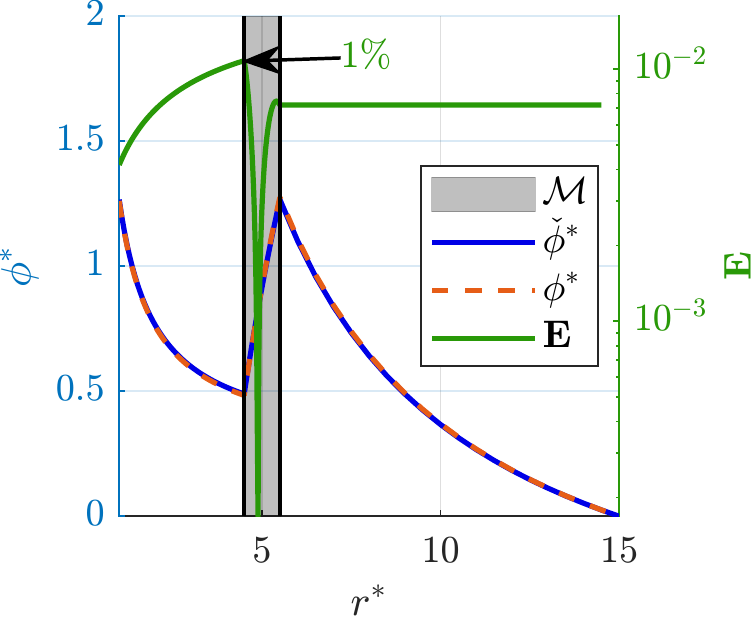}
         \caption{Sphere, case A.}
         \label{subfig:sphere_phi_caseA}
     \end{subfigure}\hspace{0.04\linewidth}%
     \begin{subfigure}[b]{0.48\linewidth}
         \centering
         \includegraphics[width=\textwidth]{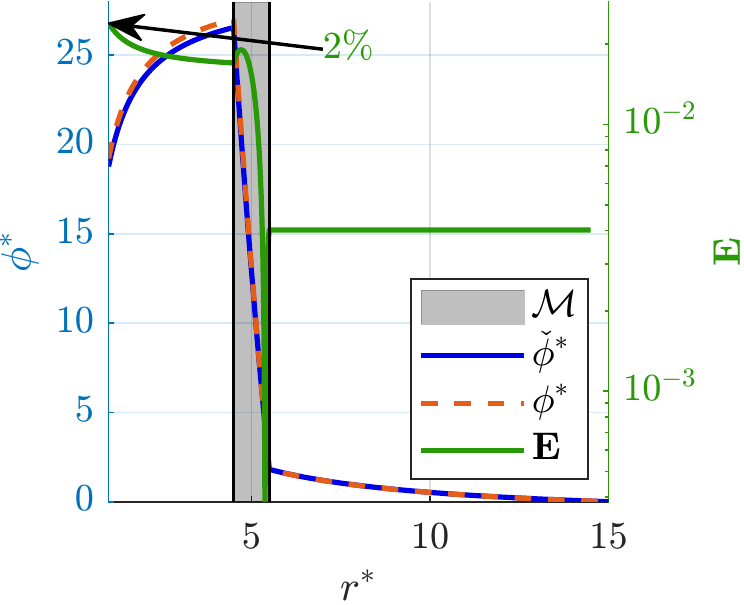}
         \caption{Sphere, case B.}
         \label{subfig:sphere_phi_caseB}
     \end{subfigure}%
    \caption{Comparison between the exact and $(2+\delta)$-dimensional theories on the sphere for the two test cases described in Tab.~\ref{tab:comparison_cases}. }
     \label{fig:s_phi_caseAB}
\end{figure}%

\subsubsection{Spheres}
We consider a sphere with axisymmetry along both the azimuthal and polar angle, such that the potential only depends on the radial direction, similar to the setup described for cylinders in Sec.~\ref{subsubsect:ana_cylinders}, Fig.~\ref{fig:cyl_sph_analytical_setup}. The boundary conditions are the same as in Eqs.~\eqref{eq:cyl_ana_full_BCRA} and~\eqref{eq:cyl_ana_full_BCRE} and the governing equations and 
corresponding solutions for both the exact and $\left(2+\delta\right)$-dimensional theory are shown in Sec.~3.3 of the SM. To compare the exact and $\left(2+\delta\right)$-dimensional solutions, we again consider the two test cases in Tab.~\ref{tab:comparison_cases}. For case A and B, the potential profiles and relative errors are plotted in Figs.~\ref{subfig:sphere_phi_caseA} and~\ref{subfig:sphere_phi_caseB}, respectively. As for the cylindrical case, the pointwise, relative error does not exceed $1\%$ in either case and the $L_2$-error decreases quadratically with the non-dimensional curvature $\mu$, as shown in Fig.~\ref{subfig:sphere_mu_caseAB}. 

\newpage
\subsection{Numerical Solutions} \label{sect:comp_numerical}

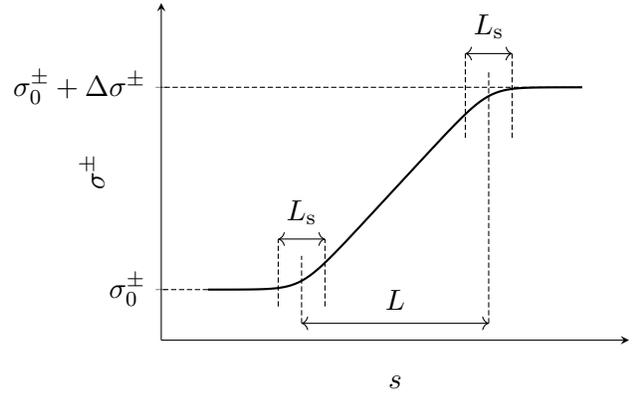
\begin{wrapfigure}{r}{0.525\textwidth}
    \centering
    \begin{tikzpicture}[font=\normalfont, x=0.95\linewidth, y=0.80\linewidth]
% if you change y here, also change in height of axis

% definitions
\newcommand{\xstart}{0.1}
\newcommand{\ystart}{0.1}
\newcommand{\Cstartx}{0.1}
\newcommand{\Cendx}{0.9}
\newcommand{\Cstarty}{0.15}
\newcommand{\Cendy}{0.75}
\newcommand{\LC}{0.4}
\newcommand{\Ls}{0.1}

% reference points
\coordinate(CScenter) at (\xstart,\ystart);
\coordinate(CSxend) at (1,\ystart);
\coordinate(CSyend) at (\xstart,1);

% charge curve
\begin{axis}[
    axis lines = left,
    xlabel = \(s\),
    ylabel = {\(\sigma^\pm\)},
    y label style={at={(-0.1,0.5)}},
    xticklabels={,,},
    yticklabels={$\sigma_0^\pm$, $\sigma_0^\pm + \Delta \sigma^\pm$},
    ytick={\Cstarty,\Cendy},
    xtick style={draw=none},
    ymin=0, 
    ymax = 1,
    xmin = 0,
    xmax = 1,
    at={(\xstart,\ystart)},
    width = {(1-\xstart)*\linewidth},
    height = {(0.80-\ystart)*\linewidth},
    line width = 0.4pt
]
\addplot[line width = 0.8pt, domain=\Cstartx:\Cendx,samples = 100]{ \Cstarty + 0.5*(\Cendy-\Cstarty)*(1 +  (ln(cosh( (\LC + 2*(\x - (\Cstartx + \Cendx)/2))/\Ls )) - ln(cosh( (\LC - 2*(\x - (\Cstartx + \Cendx)/2))/\Ls )) )/( ln(cosh( 5*\LC/\Ls )) - ln(cosh( 3*\LC/\Ls )) ) )};

% aid lines horizontal, charge densities
\draw[dashed,thin,dash pattern = on 2pt off 1pt] (\Cstartx,\Cstarty) -- (0,\Cstarty);
\draw[dashed,thin,dash pattern = on 2pt off 1pt] (\Cendx,\Cendy) -- (0,\Cendy);

% aid lines vertical, Ls left
\draw[dashed,thin,dash pattern = on 2pt off 1pt] ( {-\LC/2 - \Ls/2 + (\Cstartx + \Cendx)/2}, {\Cstarty-0.05}) -- ( {-\LC/2 - \Ls/2 + (\Cstartx + \Cendx)/2}, {\Cstarty+0.15});
\draw[dashed,thin,dash pattern = on 2pt off 1pt] ( {-\LC/2 + \Ls/2 + (\Cstartx + \Cendx)/2}, {\Cstarty-0.05}) -- ( {-\LC/2 + \Ls/2 + (\Cstartx + \Cendx)/2}, {\Cstarty+0.15});

% aid lines vertical, Ls right
\draw[dashed,thin,dash pattern = on 2pt off 1pt] ( {\LC/2 - \Ls/2 + (\Cstartx + \Cendx)/2}, {\Cendy-0.15}) -- ( {\LC/2 - \Ls/2 + (\Cstartx + \Cendx)/2}, {\Cendy+0.1});
\draw[dashed,thin,dash pattern = on 2pt off 1pt] ( {\LC/2 + \Ls/2 + (\Cstartx + \Cendx)/2}, {\Cendy-0.15}) -- ( {\LC/2 + \Ls/2 + (\Cstartx + \Cendx)/2}, {\Cendy+0.1});

% aid lines vertical, L
\draw[dashed,thin,dash pattern = on 2pt off 1pt] ( {-\LC/2 + (\Cstartx + \Cendx)/2}, {\Cstarty-0.1}) -- ( {-\LC/2 + (\Cstartx + \Cendx)/2}, {\Cstarty+0.1});
\draw[dashed,thin,dash pattern = on 2pt off 1pt] ( {\LC/2 + (\Cstartx + \Cendx)/2}, {\Cstarty-0.1}) -- ( {\LC/2 + (\Cstartx + \Cendx)/2}, {\Cendy+0.05});

% Ls arrow left
\draw[<->] ( {-\LC/2 - \Ls/2 + (\Cstartx + \Cendx)/2}, {\Cstarty+0.15}) -- ( {-\LC/2 + \Ls/2 + (\Cstartx + \Cendx)/2}, {\Cstarty+0.15});
\node[] (Lslnode) at ( {-\LC/2 + (\Cstartx + \Cendx)/2}, {\Cstarty+0.23}) {$L_\mathrm{s}$};

% Ls arrow right
\draw[<->] ( {\LC/2 - \Ls/2 + (\Cstartx + \Cendx)/2}, {\Cendy+0.1}) -- ( {\LC/2 + \Ls/2 + (\Cstartx + \Cendx)/2}, {\Cendy+0.1});
\node[] (Lsrnode) at ( {\LC/2 + (\Cstartx + \Cendx)/2}, {\Cendy+0.18}) {$L_\mathrm{s}$};

% L arrow
\draw[<->] ( {-\LC/2 + (\Cstartx + \Cendx)/2}, {\Cstarty-0.1}) -- ( {\LC/2 + (\Cstartx + \Cendx)/2}, {\Cstarty-0.1});
\node[] (Lnode) at ( {(\Cstartx + \Cendx)/2}, {\Cstarty-0.03}) {$L$};

\end{axis}

\end{tikzpicture}
    \caption{The surface charge density is changing from $\sigma_0^\pm$ to $\sigma_0^\pm + \Delta \sigma^\pm$ over a length $L$. $L_\mathrm{s}$ denotes the length of the smooth transition region between the constant and linearly varying surface charge density.}
    \label{fig:surface_charge_varying}
\end{wrapfigure}
We now test the $\left(2+\delta\right)$-dimensional theory numerically on examples without analytical solutions but motivated by lipid membranes. Namely, we consider flat, cylindrical, and spherical lipid membranes---typical shapes in biological systems---embedded in a symmetric, monovalent electrolyte. The lipid membranes are equipped with spatially varying surface charge densities, modeling charged lipids or charges accumulated on the interfaces between the electrolyte and lipid membrane \cite{szekely2011structure}. 
The surface charge densities are screened by charges in the electrical double layers in the electrolyte, as described by the Poisson-Boltzmann equation \cite{andelman2006introduction}. Accordingly, the charge densities in the bulk domains, required in Eqs.~\eqref{eq:gauss_bulkm} and~\eqref{eq:gauss_bulkp}, are given by
\begin{align}
    q = -\frac{\varepsilon_\mathrm{B} k_\mathrm{B}T}{e \lambda_\mathrm{D}^2} \, \sinh{ \frac{e \phi}{k_\mathrm{B}T} }~, \label{eq:PB_charge}
\end{align}
with $e$ being the elementary charge and $\lambda_\mathrm{D}$ the Debye length, determined by the bulk electrolyte concentration \cite{andelman2006introduction}. In physical systems, the surface charge densities on lipid membranes can vary spatially, consequently leading to in-plane variations of the electric potential. To test the accuracy of the $(2+\delta)$-dimensional theory under different characteristic in-plane length scales, we prescribe the surface charge densities by univariate functions along the direction $s$:
\begin{align}
    \sigma^\pm = \sigma_0^\pm + \frac{1}{2}\Delta \sigma^\pm \left( 1 + \frac{\ln{ \leftR( \cosh \frac{L + 2s}{L_\mathrm{s}}  \rightR)} - \ln{ \leftR( \cosh \frac{L - 2s}{L_\mathrm{s}}  \rightR)}}{\ln{ \leftR( \cosh \frac{5 L }{L_\mathrm{s}} \rightR)} - \ln{ \leftR( \cosh \frac{3L}{L_\mathrm{s}}  \rightR)}} \right)~, \label{eq:sigma_varying}
\end{align}
schematically shown in Fig.~\ref{fig:surface_charge_varying}. The surface charge densities change from their constant value $\sigma_0^\pm$ to varying linearly over a length $L$ until reaching the constant value $\sigma_0^\pm + \Delta \sigma^\pm$. The transition between constant and linearly varying surface charge densities is smoothed over the length $L_\mathrm{s}$. By varying $L$ and $L_\mathrm{s}$, we can study the effects of different in-plane length scales on the accuracy of the $(2+\delta)$-dimensional theory. A more detailed description of the setup is presented in the respective geometry sections, Secs.~\ref{sec:num_flat}--\ref{sec:num_sphere}. The differential equations governing the exact and $(2+\delta)$-dimensional theories are solved using a second order finite difference scheme, with the interface conditions evaluated on $\mathcal{S}^\pm$, as described in Fig.~\ref{subfig:meshing_w_thickness}.

\subsubsection{Flat Geometry} \label{sec:num_flat}

\begin{figure}[h]
    \centering
    \begin{subfigure}[t]{0.5\textwidth}
    \centering
    \imagebox{2.1in}{
    \begin{annotationimage}[]{width=2.75in}{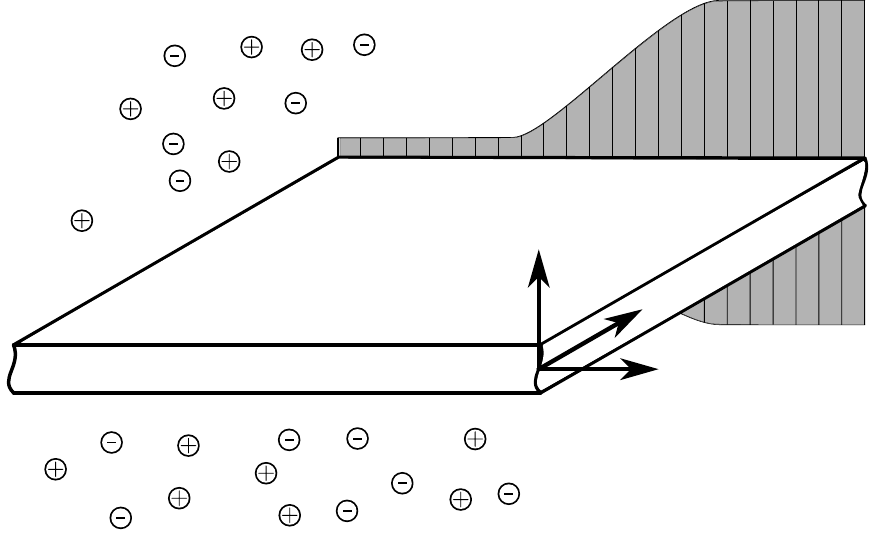}
        \imagelabelset{
                coordinate label style/.style = {
                rectangle,
                fill = none,
                text = black,
                font = \normalfont
        }}
        \draw[coordinate label = {$\sigma^+\leftR(x\rightR)$ at (0.58,0.87)}];
        \draw[coordinate label = {$\sigma^-\leftR(x\rightR)$ at (0.93,0.31)}];
        \draw[coordinate label = {$\bm{e}_x$ at (0.79,0.33)}];
        \draw[coordinate label = {$\bm{e}_y$ at (0.7,0.5)}];
        \draw[coordinate label = {$\bm{e}_z$ at (0.57,0.55)}];
    \end{annotationimage}}
    \ifcap{\caption{Schematic of a flat lipid membrane with spatially varying surface charges on the top and bottom surfaces.}}{\vspace{1cm}}
    \label{fig:flat_num_setup}
    \end{subfigure}%
    \hfill
    \begin{subfigure}[t]{0.45\textwidth}
        \centering
        \imagebox{2.1in}{
        \includegraphics[scale=1]{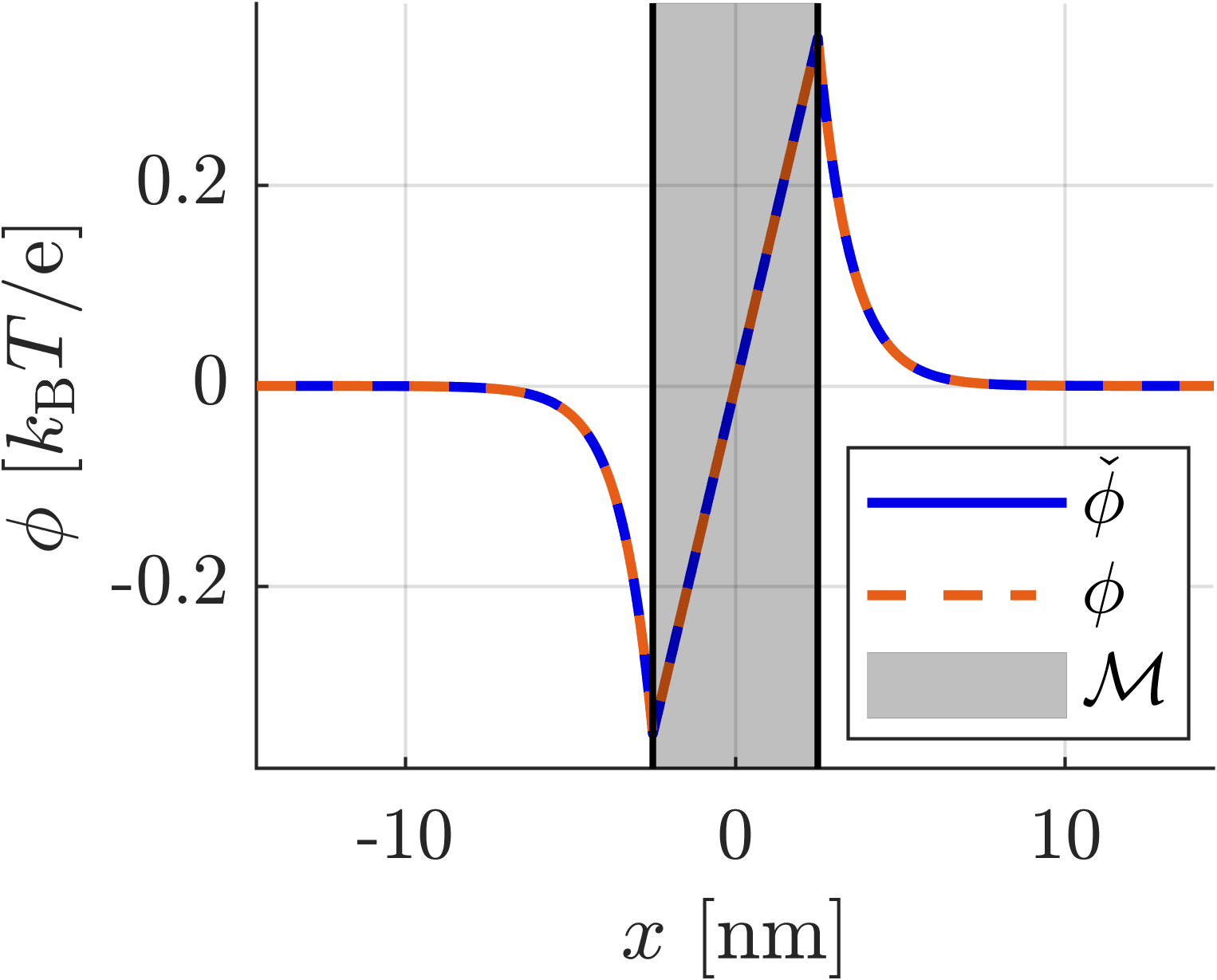}}
        \caption{Potential profiles for constant surface charge densities for the exact and $(2+\delta)$-dimensional theories.}
        \label{fig:flat_caseA_phi}
    \end{subfigure}
    \caption{Schematic setup (a) and potential profiles (b) for the flat lipid membrane embedded in a symmetric, monovalent electrolyte. The exact and $(2+\delta)$-dimensional theories agree to machine precision.}
    \label{fig:flat_schematic_profile}
\end{figure}
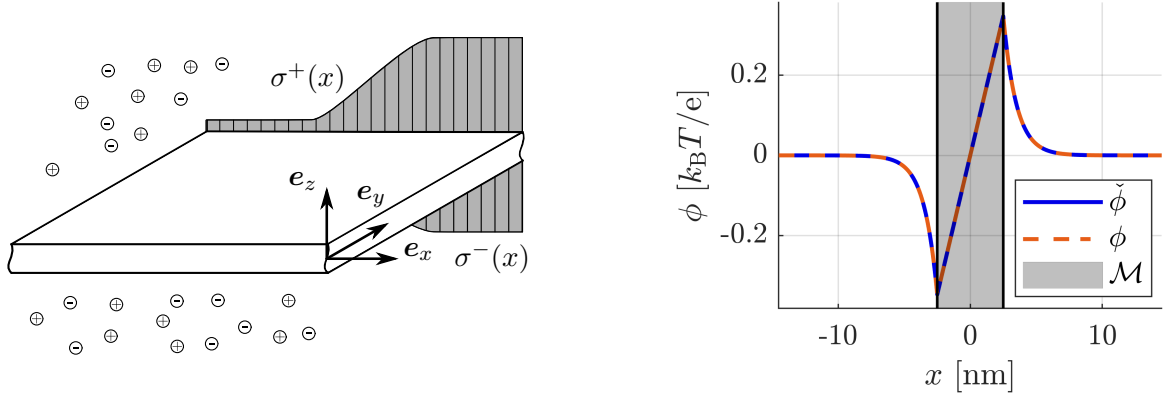

Consider a flat lipid membrane whose mid-surface lies in the $x$-$y$-plane, schematically shown in Fig.~\ref{fig:flat_num_setup}. The surface charge densities vary only along the $x$-direction, i.e. $x \equiv s$ in Eq.~\eqref{eq:sigma_varying}, rendering the potential independent of the $y$-direction. The problem is subjected to the boundary conditions 
\begin{alignat}{2}
    \frac{\partial \phi}{\partial z} \biggr\rvert_{z = -\delta/2 - L_{\mathcal{B}2}} &= 0~, \hspace{3cm}
    \phi \bigr\rvert_{z = \delta/2 + L_{\mathcal{B}2}} &&= 0~, \\[8pt]
    \frac{\partial \phi}{\partial x} \biggr\rvert_{x = 0} &= 0~, \hspace{3cm}
    \frac{\partial \phi}{\partial x} \biggr\rvert_{x = L_{\mathcal{B}1}} &&= 0~,
\end{alignat}
where $L_{\mathcal{B}1}$ is the domain size along the $x$-direction and $L_{\mathcal{B}2}$ is the domain size above and below the membrane, with the mid-surface located at $z = 0$. We consider two different cases: In case A, the charge densities are constant while in case B, the charge densities change from $\pm 1~\mathrm{mC/m^2}$ to $\pm 40~\mathrm{mC/m^2}$ along the $x$-direction, centered at $x = L_{\mathcal{B}2}/2$. The two cases A and B are summarized in Tab.~\ref{tab:charge_props} and the remaining geometric and material parameters are listed in Tab.~\ref{tab:geom_mat_props}. \textspace

\begin{table}[t]
    \centering
    \begin{tabular}{c|cccc}\toprule
         case & $\sigma_0^+~[\mathrm{mC/m^2}]$ & $\sigma_0^-~[\mathrm{mC/m^2}]$ & $\Delta \sigma^+~[\mathrm{mC/m^2}]$ & $\Delta \sigma^-~[\mathrm{mC/m^2}]$ \\ \midrule
         A & 40 & -40 & 0 & 0 \\ \hline
         B & 1  & -1 & 39 & -39 
         \\ \bottomrule
    \end{tabular}
    \caption{Surface charge densities on the top ($\sigma^+$) and bottom ($\sigma^-$) surface for cases A and B. }
    \label{tab:charge_props}
\end{table}

\begin{table}[t]
    \centering
    \begin{tabular}{ccccccccc}\toprule
         $L_{\mathcal{B}1}~[\mathrm{nm}]$ & $L_{\mathcal{B}2}~[\mathrm{nm}]$ & $R_0~[\mathrm{nm}$]  & $\delta~[\mathrm{nm}]$ & $L~[\mathrm{nm}]$ & $L_\mathrm{s}~[\mathrm{nm}]$ & $\varepsilon_\mathrm{B}~[\varepsilon_0]$ & $\varepsilon_\mathrm{M}~[\varepsilon_0]$ &  $\lambda_\mathrm{D}~[\mathrm{nm}]$ \\ \midrule
         $75$ & $12$ & 25 & $5$ & $5$ & $2.5$ & $80$ & $2$ & 1 \\ \bottomrule
    \end{tabular}
    \caption{Geometric and material parameters for the flat, cylindrical, and spherical test cases. The length scales and parameters are typical for lipid membranes.}
    \label{tab:geom_mat_props}
\end{table}

\begin{figure}[t]
     \centering
     \begin{subfigure}[b]{0.475\textwidth}
         \centering
         \includegraphics{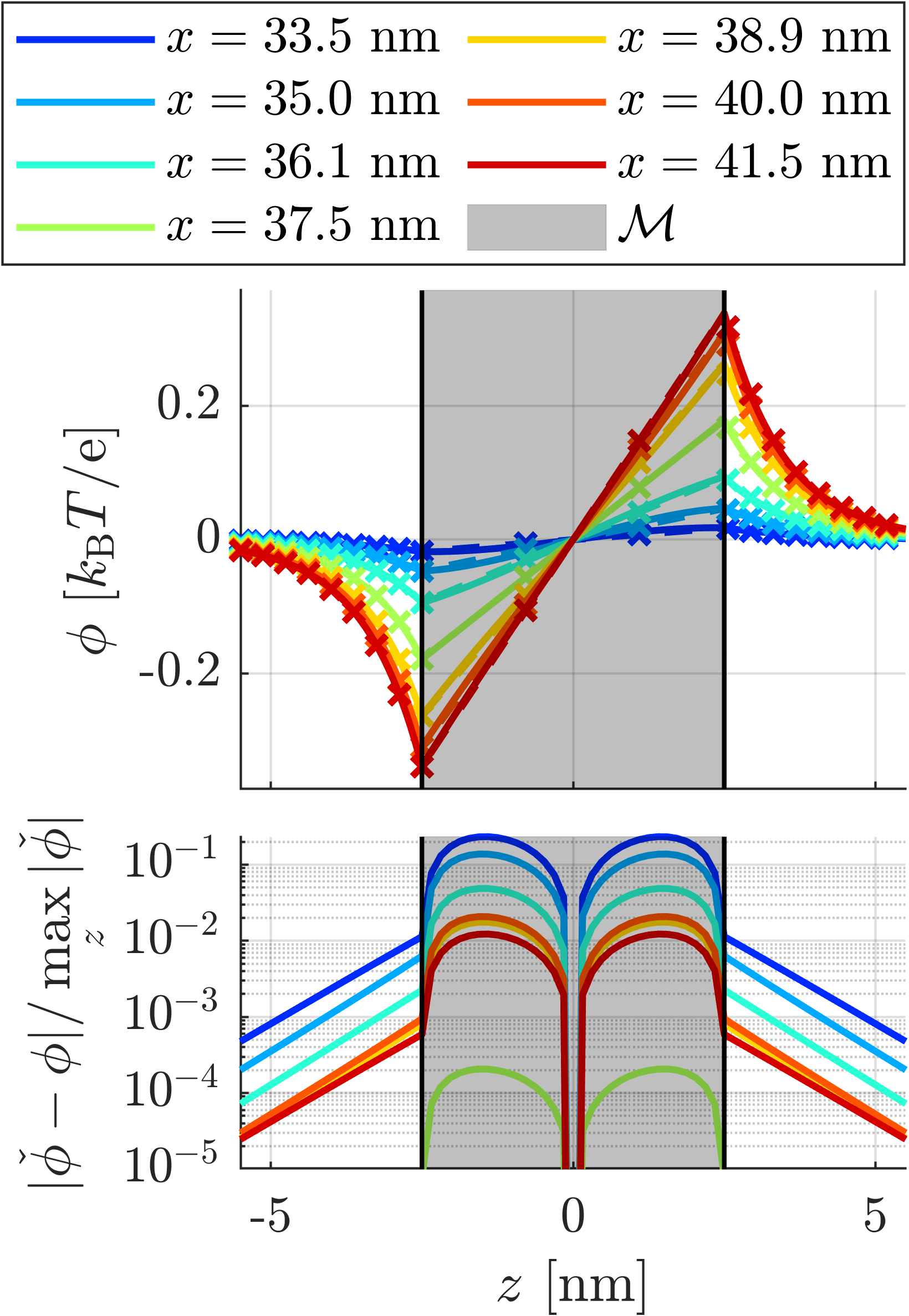}
         \caption{Potential profiles in the region of varying surface charge densities.}
         \label{fig:flat_inc40_PP}
     \end{subfigure}
     \hfill
     \begin{subfigure}[b]{0.475\textwidth}
         \centering
         \includegraphics{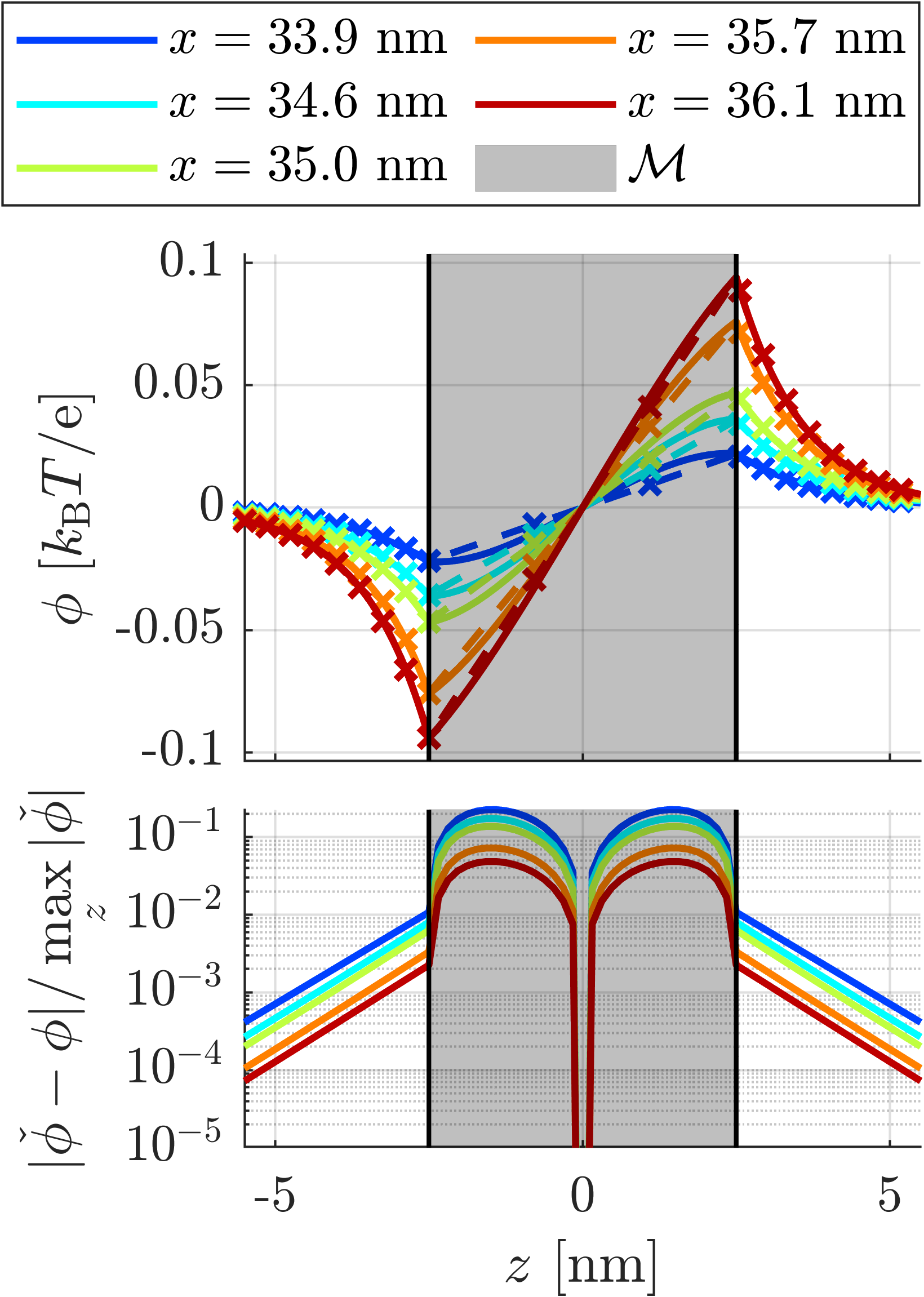}
         \caption{Potential profiles in the left transition region of the varying surface charge densities.}
         \label{fig:flat_inc40_PPTR_left}
     \end{subfigure}
     \caption{Potential profiles (top) and relative errors (bottom) plotted at discrete values of $x$ for case B of the flat membrane. The full lines represent the exact theory and the dashed lines (nearly indistinguishable in (a)) represent the $(2+\delta)$-dimensional theory. The error is only plotted down to $10^{-5}$. In (a), the values of $x$ are taken from the entire region of varying surface charge densities while (b) shows potential profiles from the left transition region between constant and linearly varying surface charge densities.}
     \label{fig:flat_profile_PP_PPTR}
\end{figure}

The potential profile corresponding to case A, shown in Fig.~\ref{fig:flat_caseA_phi}, is linear within the membrane and exponentially decays to zero in the bulk domains. Due to the non-zero surface charge density and different permittivities in the bulk and membrane, the slope of the potential is discontinuous on the top and bottom boundaries of the membrane. Since the solution is linear in the membrane, the exact and $(2+\delta)$-dimensional theories agree to machine precision. \textspace

In Fig.~\ref{fig:flat_profile_PP_PPTR}, the results for case B are presented. Figure~\ref{fig:flat_inc40_PP} (top) shows the potential in the region of varying surface charge densities at discrete values of $x$. The exact theory is plotted with full lines while the $(2+\delta)$-dimensional theory is plotted with dashed lines and $\bm{\times}$-markers, revealing excellent qualitative agreement between the exact and $(2+\delta)$-dimensional theories across all values of $x$. Figure~\ref{fig:flat_inc40_PP} (bottom) shows the corresponding relative, pointwise error, which remains below $\approx20\%$ throughout the entire domain. To find where the error is largest, Fig.~\ref{fig:flat_inc40_PPTR_left} shows the potential and error where the surface charge densities change from constant values of $\pm 1~\mathrm{mC}/\mathrm{m^2}$ to varying linearly along $x$. In this narrow transition region of length $L_\mathrm{s}$, the potential is small and the deviations from the exact solutions are large compared to other regions of the domain. However, the qualitative behavior of the membrane is still well-captured by the $(2+\delta)$-dimensional theory. \textspace
\begin{figure}
     \centering
     \begin{subfigure}[b]{0.475\textwidth}
         \centering
         \includegraphics{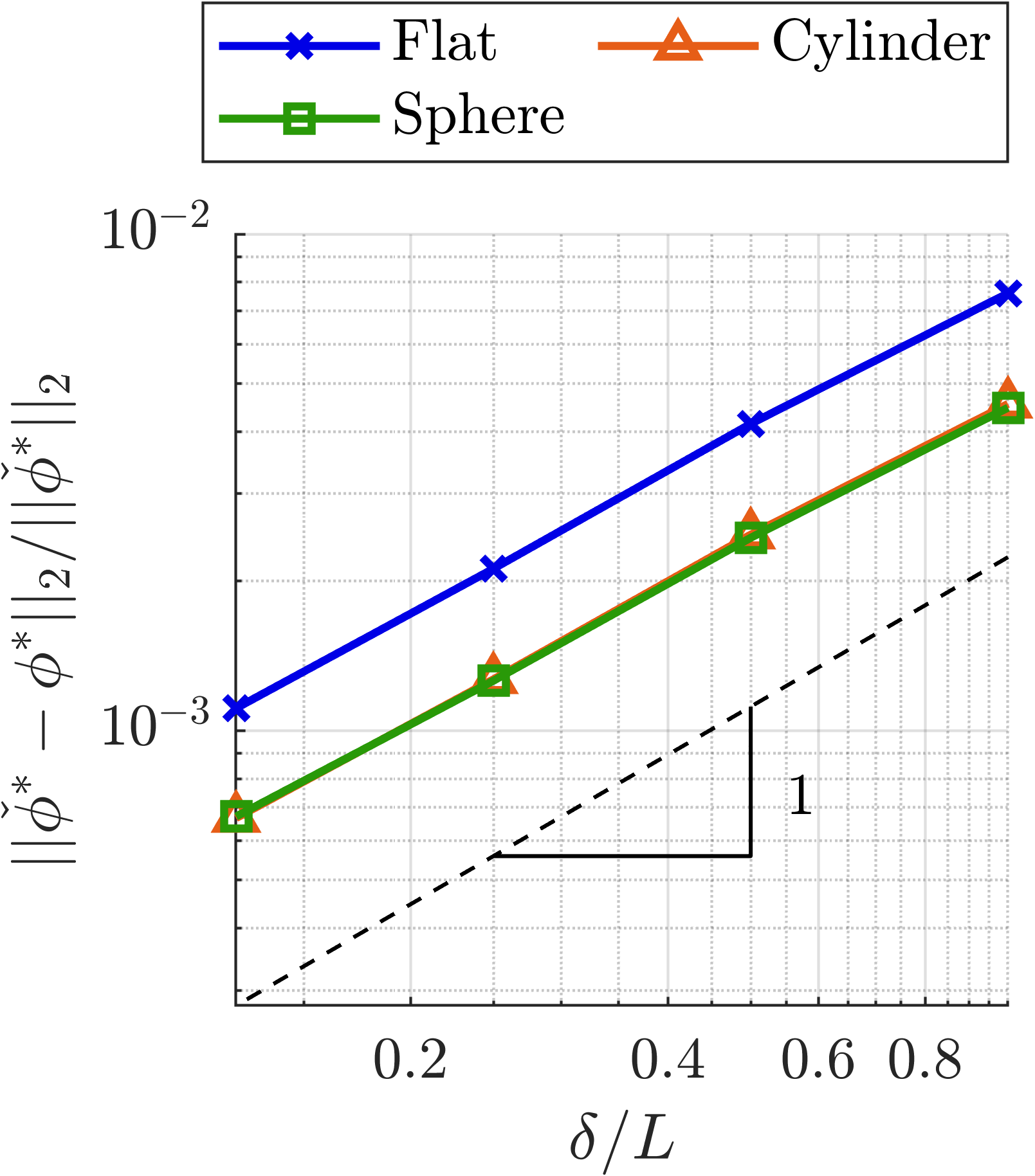}
         \caption{Error in $L_2$-norm for varying $L$.}
         \label{fig:fcs_inc40_err_L}
     \end{subfigure}
     \hfill
     \begin{subfigure}[b]{0.475\textwidth}
         \centering
         \includegraphics{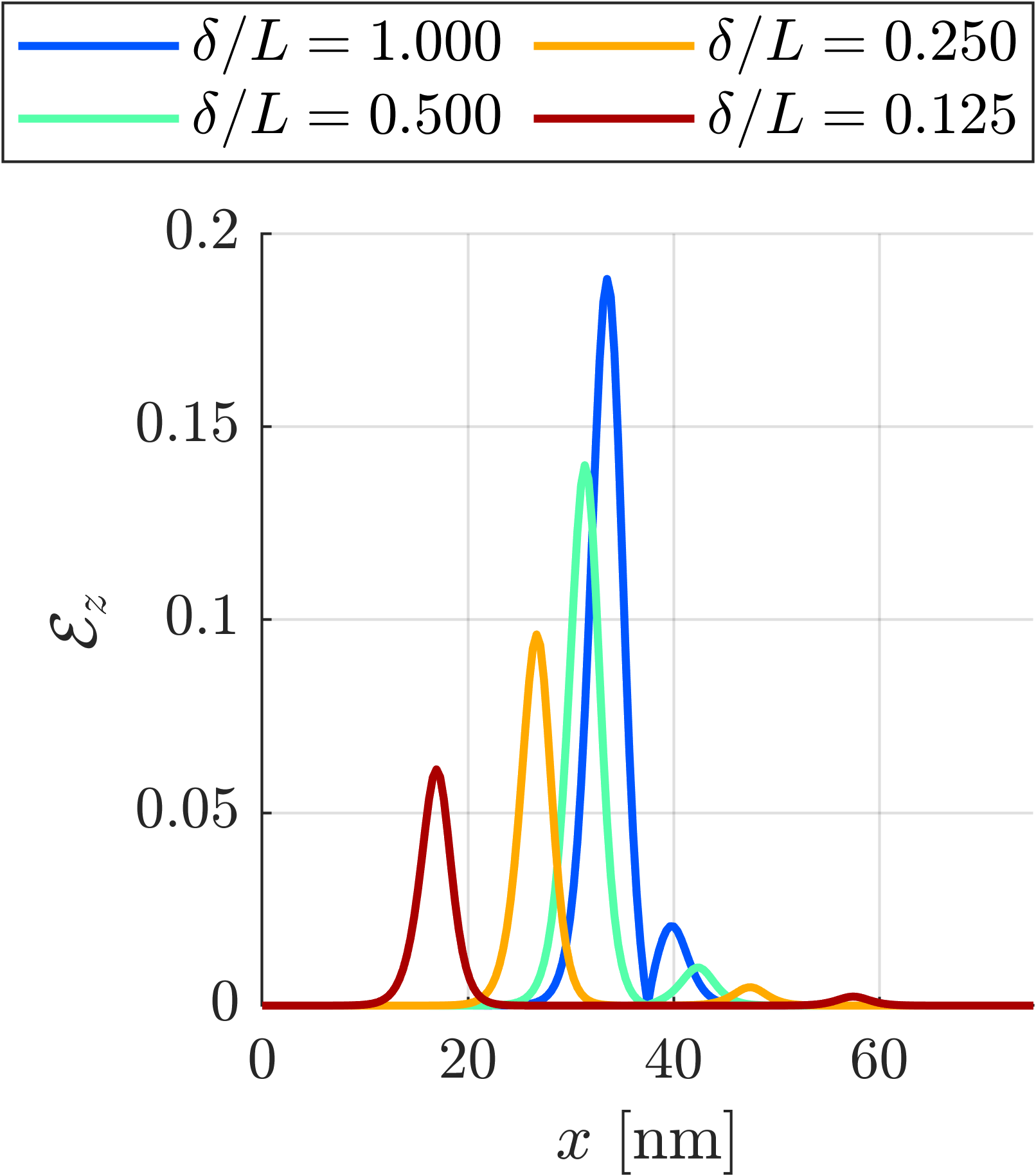}
         \caption{Error in $\mathcal{E}_z$-norm for varying $L$.}
         \label{fig:flat_L2x_LwidthU}
     \end{subfigure}
     
     \vspace{0.5cm}
     \begin{subfigure}[b]{0.475\textwidth}
         \centering
         \includegraphics{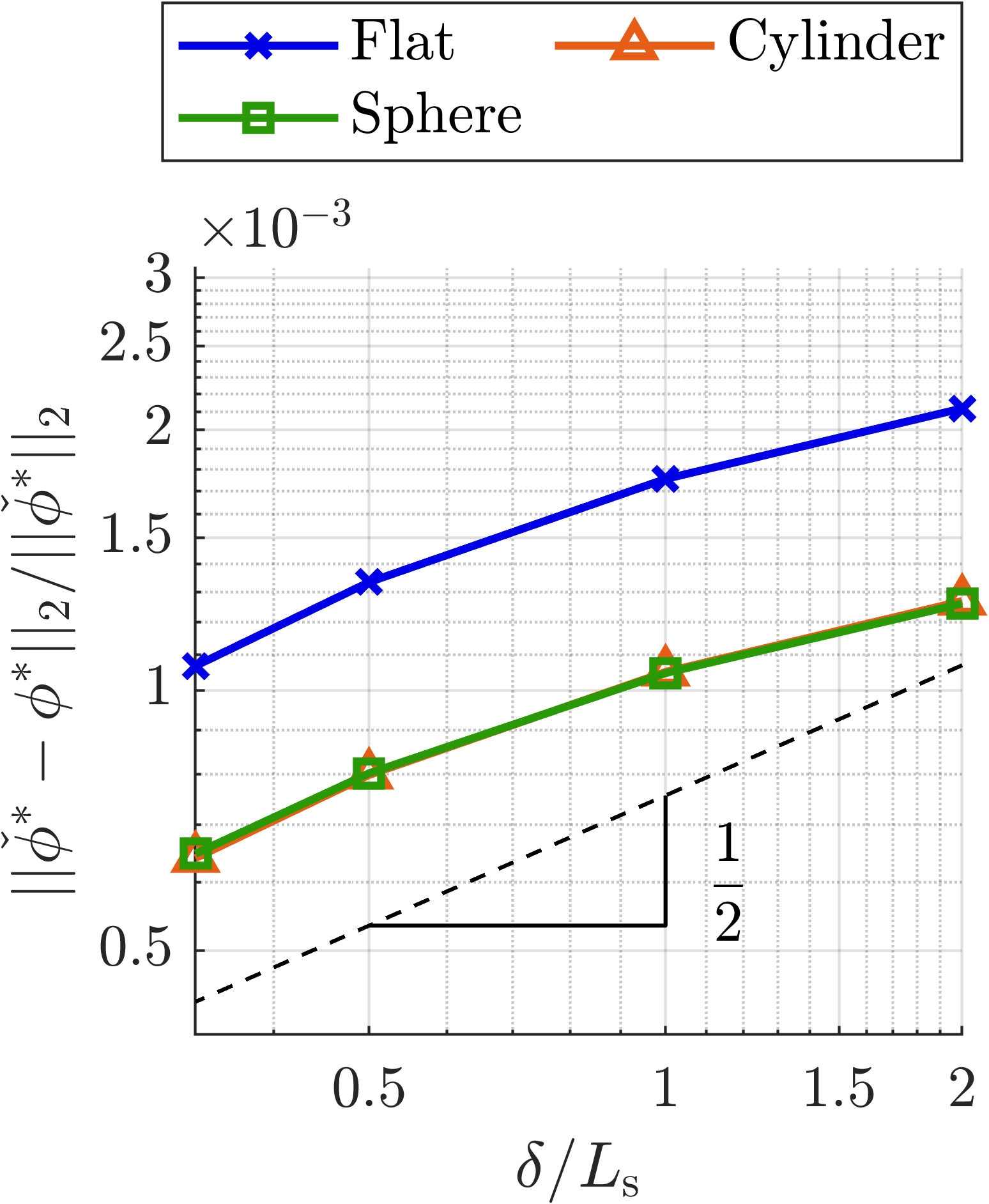}
         \caption{Error in $L_2$-norm for varying $L_\mathrm{s}$.}
         \label{fig:fcs_inc40_err_Ls}
     \end{subfigure}
     \hfill
     \begin{subfigure}[b]{0.475\textwidth}
         \centering
         \includegraphics{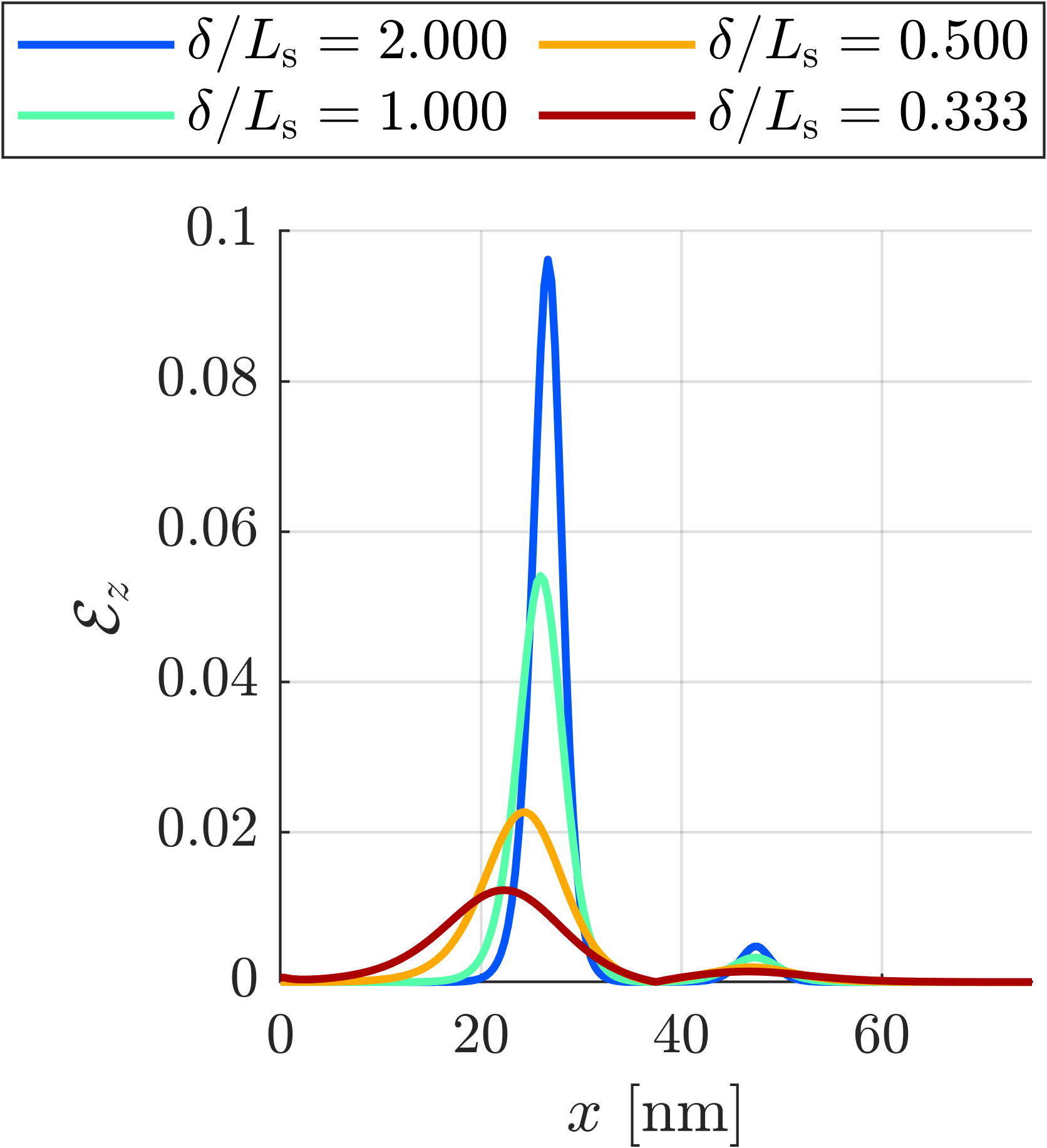}
         \caption{Error in $\mathcal{E}_z$-norm for varying $L_\mathrm{s}$.}
         \label{fig:flat_L2x_LsmoothU}
     \end{subfigure}
     \caption{Error in the $L_2$-norm for case B against varying $L$ (a) and $L_\mathrm{s}$ (c) for all geometries and error in the $\mathcal{E}_z$-norm, as defined in Eq.~\eqref{eq:definition_eps_z}, for the flat geometry ((b) and (d)). For the cylinder and sphere, the mid-surface radius is fixed at $R_0 = 200~\mathrm{nm}$ and for the case of varying $L_\mathrm{s}$, the length over which the surface charge densities vary linearly is fixed at $L = 20~\mathrm{nm}$. The remaining parameters are given in Tabs.~\ref{tab:charge_props} and~\ref{tab:geom_mat_props}.}
     \label{fig:error_L}
\end{figure}

According to Tab.~\ref{tab:geom_mat_props}, the length over which the surface charge densities vary linearly, $L$, as well as the smoothing length, $L_\mathrm{s}$, are on the order of the thickness $\delta$. This violates the assumption of the $(2+\delta)$-dimensional theory that the characteristic in-plane length scale is much larger than the thickness, Eq.~\eqref{eq:dL_ll_1}. This motivates examining the error in the $(2+\delta)$-dimensional theory under varying $L_\mathrm{s}$ and $L$. Figure~\ref{fig:fcs_inc40_err_L} shows that the $L_2$-error decreases linearly with $L$ while $L_\mathrm{s}=2.5~\mathrm{nm}$ is fixed. To show that this is due to the decrease in the error in the transition region between constant and linearly varying surface charge densities, Fig.~\ref{fig:flat_L2x_LwidthU} shows the $L_2$-error along $z$ as a function of $x$, defined as 
\begin{align}
    \mathcal{E}_{z}\leftR(x\rightR) = \sqrt{\frac{\int_{-\delta/2-L_{\mathcal{B}2}}^{\delta/2+L_{\mathcal{B}2}} \left(\check{\phi} - \phi\right)^2 \mathrm{d}z}{\int_{-\delta/2-L_{\mathcal{B}2}}^{\delta/2+L_{\mathcal{B}2}} \check{\phi}^2 \, \mathrm{d}z}}~, \label{eq:definition_eps_z}
\end{align}
where we find that the peak in the error in the transition region decays quickly as $L$ is increased. Similarly, varying the smoothing length $L_\mathrm{s}$ while fixing $L=20~\mathrm{nm}$ yields an error that decreases with order $1/2$ (Fig.~\ref{fig:fcs_inc40_err_Ls}). As seen in Fig.~\ref{fig:flat_L2x_LsmoothU}, this is again a result of the decrease in error in the transition region between constant and linearly varying surface charge densities. Thus, we conclude that the error of the $(2+\delta)$-dimensional theory becomes small when the characteristic in-plane length scales become large compared to the thickness of the membrane.

\subsubsection{Cylinder} \label{sec:num_cyl}
\begin{figure}
    \centering
    \begin{subfigure}[t]{0.48\textwidth}
    \centering
    \imagebox{3.1in}{
    \begin{annotationimage}[]{width=2.75in}{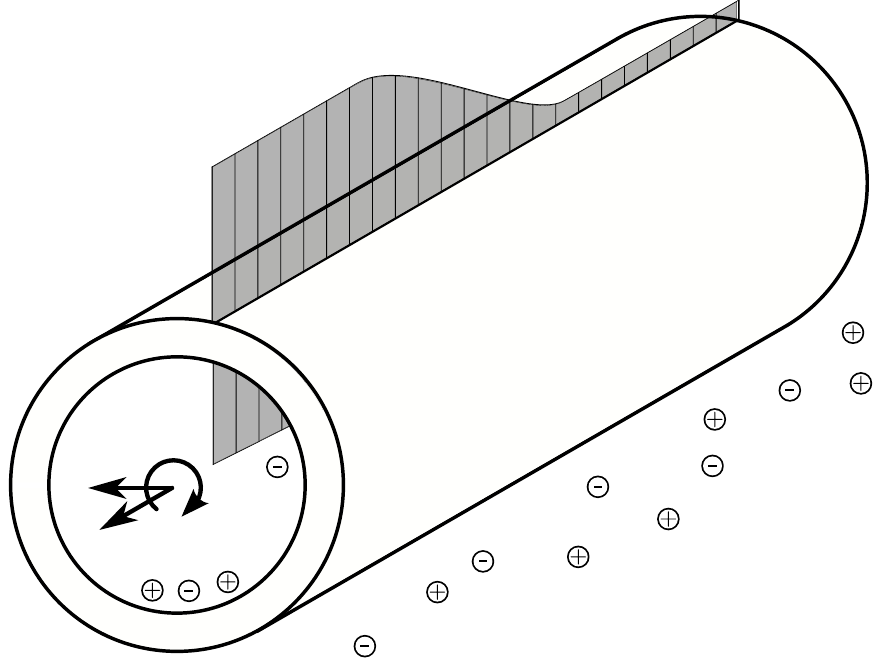}
        \imagelabelset{
                coordinate label style/.style = {
                rectangle,
                fill = none,
                text = black,
                font = \normalfont
        }}
        \draw[coordinate label = {$\sigma^+\leftR(z\rightR)$ at (0.58,0.95)}];
        \draw[coordinate label = {$\sigma^-\leftR(z\rightR)$ at (0.17,0.39)}];
        \draw[coordinate label = {$\bm{e}_r$ at (0.09,0.3)}];
        \draw[coordinate label = {$\bm{e}_z$ at (0.12,0.16)}];
        \draw[coordinate label = {$\theta$ at (0.24,0.2)}];
    \end{annotationimage}}
    \ifcap{\caption{Schematic of a cylinder with spatially varying surface charge densities.}}{\vspace{1cm}}
    \label{fig:cyl_num_setup}
    \end{subfigure}%
    \hfill
    \begin{subfigure}[t]{0.47\textwidth}
        \centering
        \imagebox{3.1in}{
        \includegraphics[scale=1]{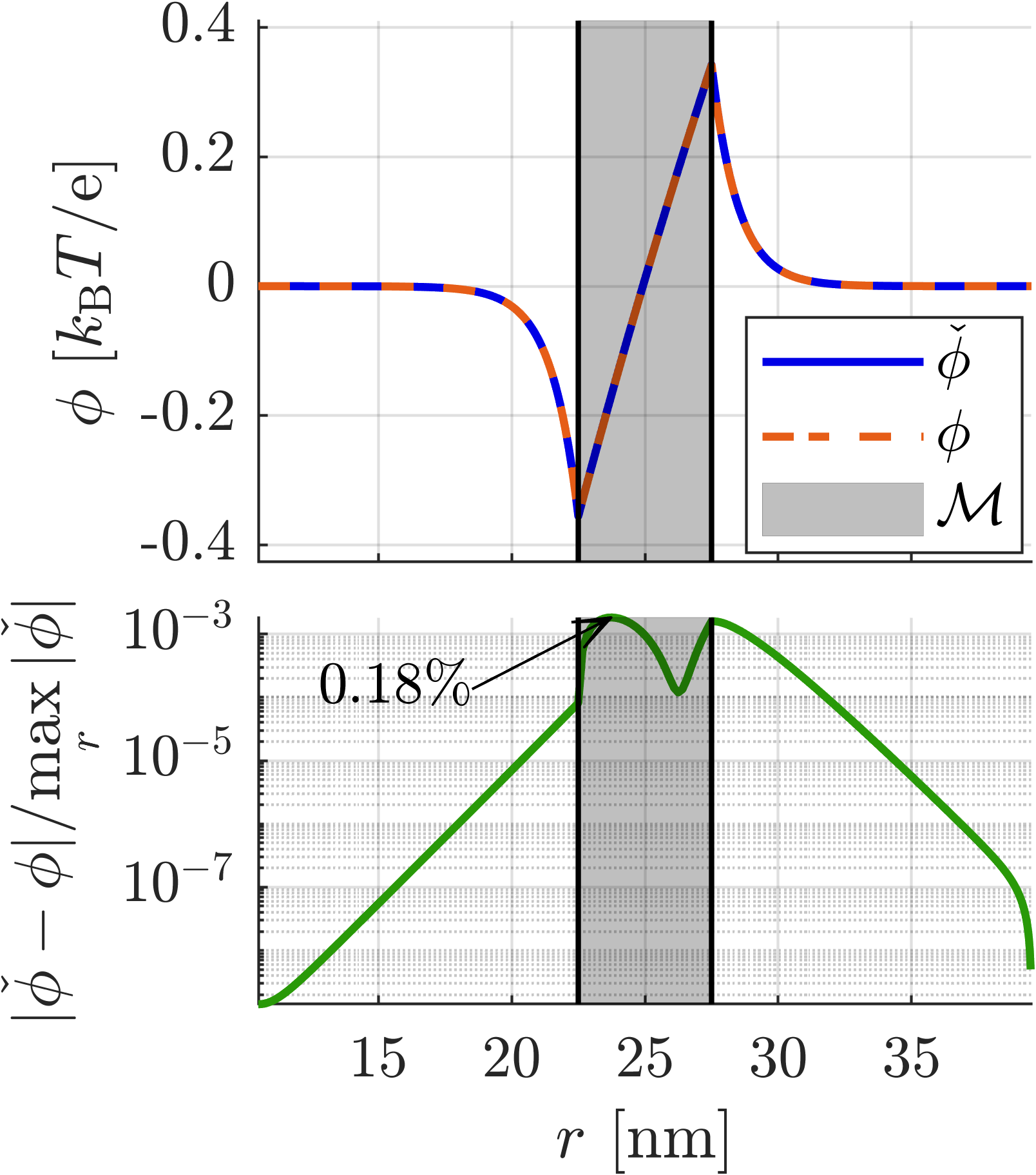}}
        \caption{Potential profiles (top) and relative error (bottom) for constant surface charge densities for the exact and $(2+\delta)$-dimensional theories.}
        \label{fig:cyl_caseA_phi}
    \end{subfigure}
    \caption{Schematic setup (a) and potential profiles (b) for the cylindrical lipid membrane embedded in a symmetric, monovalent electrolyte.}
    \label{fig:cyl_schematic_profile}
\end{figure}

\begin{figure}[h]
     \centering
     \begin{subfigure}[b]{0.475\textwidth}
         \centering
         \includegraphics{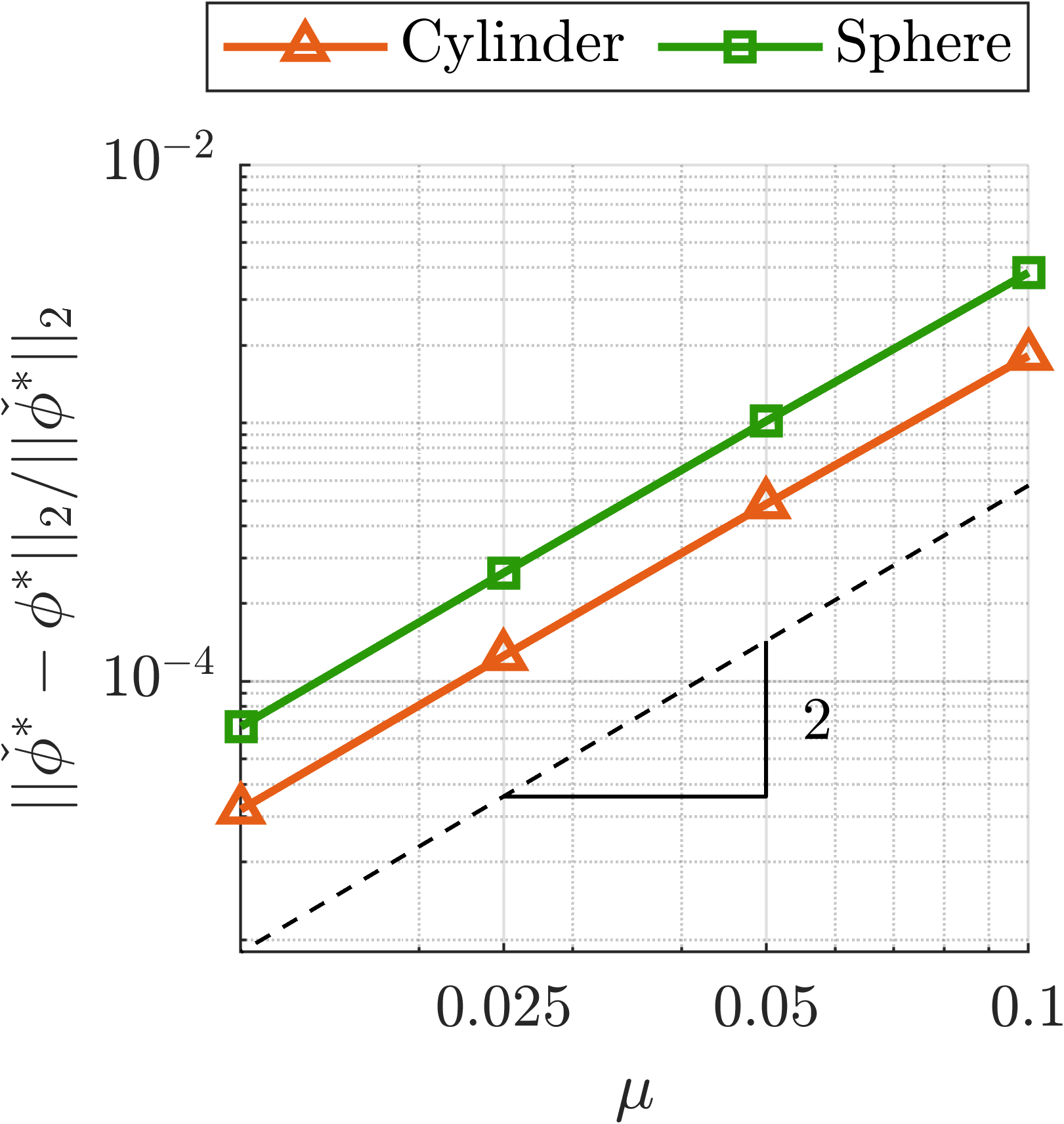}
         \caption{Case A.}
         \label{fig:cs_error_conv_R0_caseA}
     \end{subfigure}
     \hfill
     \begin{subfigure}[b]{0.475\textwidth}
         \centering
         \includegraphics{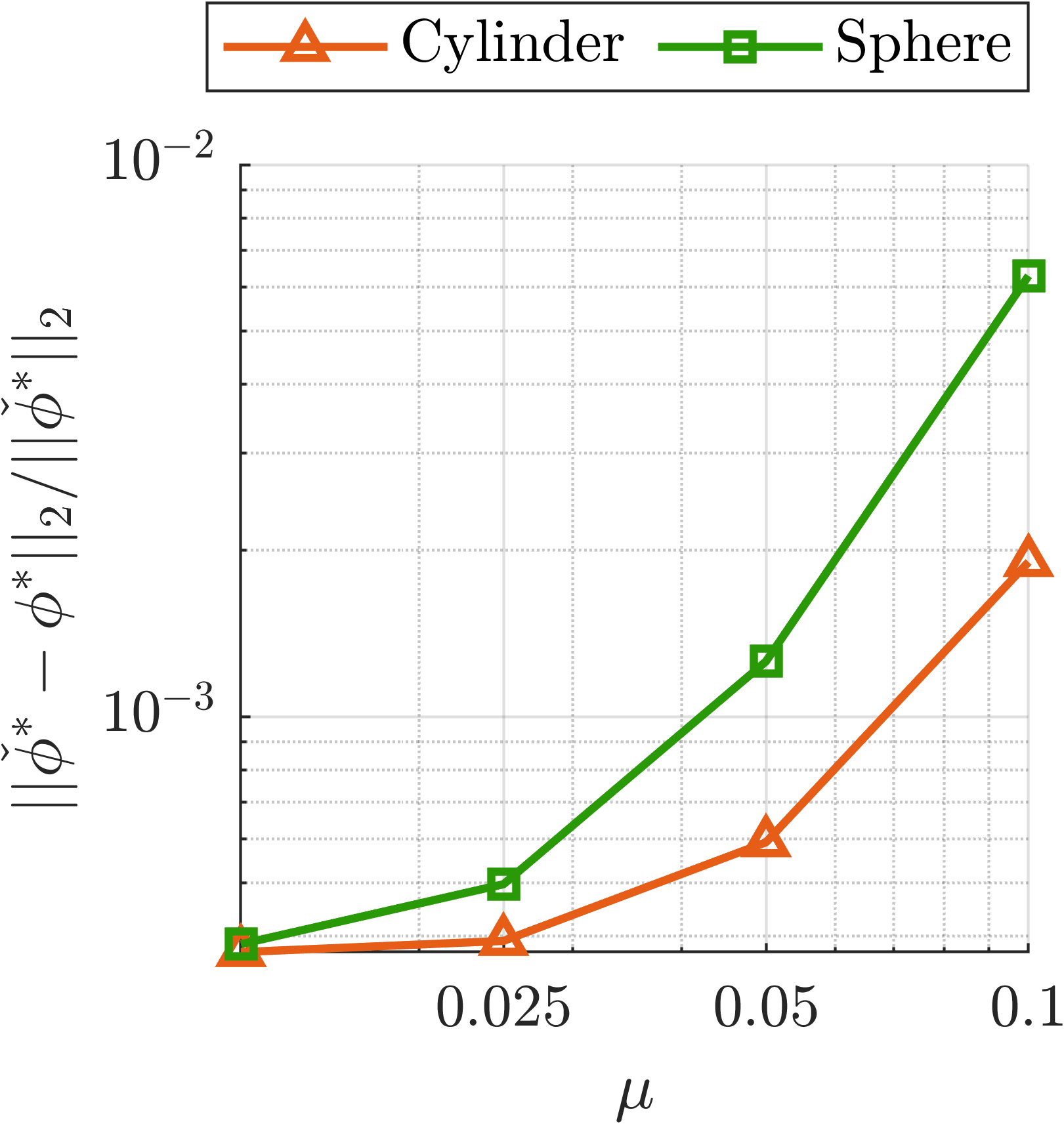}
         \caption{Case B.}
         \label{fig:cs_error_conv_R0_caseB}
     \end{subfigure}
     \caption{Error convergence in the $L_2$-norm with respect to the non-dimensional curvature $\mu = \frac{\delta}{2 R_0}$ for cylinders and spheres for cases A (a) and B (b), with $L = 20~\mathrm{nm}$ and $L_\mathrm{s} = 10~\mathrm{nm}$. For case A, the expected quadratic convergence is observed while for case B, the error saturates as a result of the dominating error from the spatially varying surface charge densities.}
     \label{fig:L2_R0}
\end{figure}

Consider a cylindrical lipid membrane with mid-surface radius $R_0$, schematically shown in Fig.~\ref{fig:cyl_num_setup}. We choose a surface charge density that varies only along the $z$-direction of the cylinder such that the setup is axisymmetric, i.e. $s\equiv z$ in Eq.~\eqref{eq:sigma_varying}. The boundary conditions remain similar to the flat case:
\begin{alignat}{2}
    \frac{\partial \phi}{\partial r} \biggr\rvert_{r = 0} &= 0~, \hspace{2cm}
    \phi \bigr\rvert_{r = R_0 + \delta/2 + L_{\mathcal{B}2}} &&= 0~, \\[8pt]
    \frac{\partial \phi}{\partial z} \biggr\rvert_{z = 0} &= 0~, \hspace{2cm}
    \frac{\partial \phi}{\partial z} \biggr\rvert_{z = L_{\mathcal{B}1}} &&= 0~.
\end{alignat}
As compared to the  flat case, however, the first boundary condition is replaced by a symmetry condition in the center of the cylinder. All geometric and material properties remain as before and are listed in Tabs.~\ref{tab:charge_props} and~\ref{tab:geom_mat_props}. Additionally, the mid-surface radius is fixed at $R_0 = 25~\mathrm{nm}$, unless stated otherwise.\textspace

\begin{figure}[h]
     \centering
     \begin{subfigure}[b]{0.475\textwidth}
         \centering
         \includegraphics{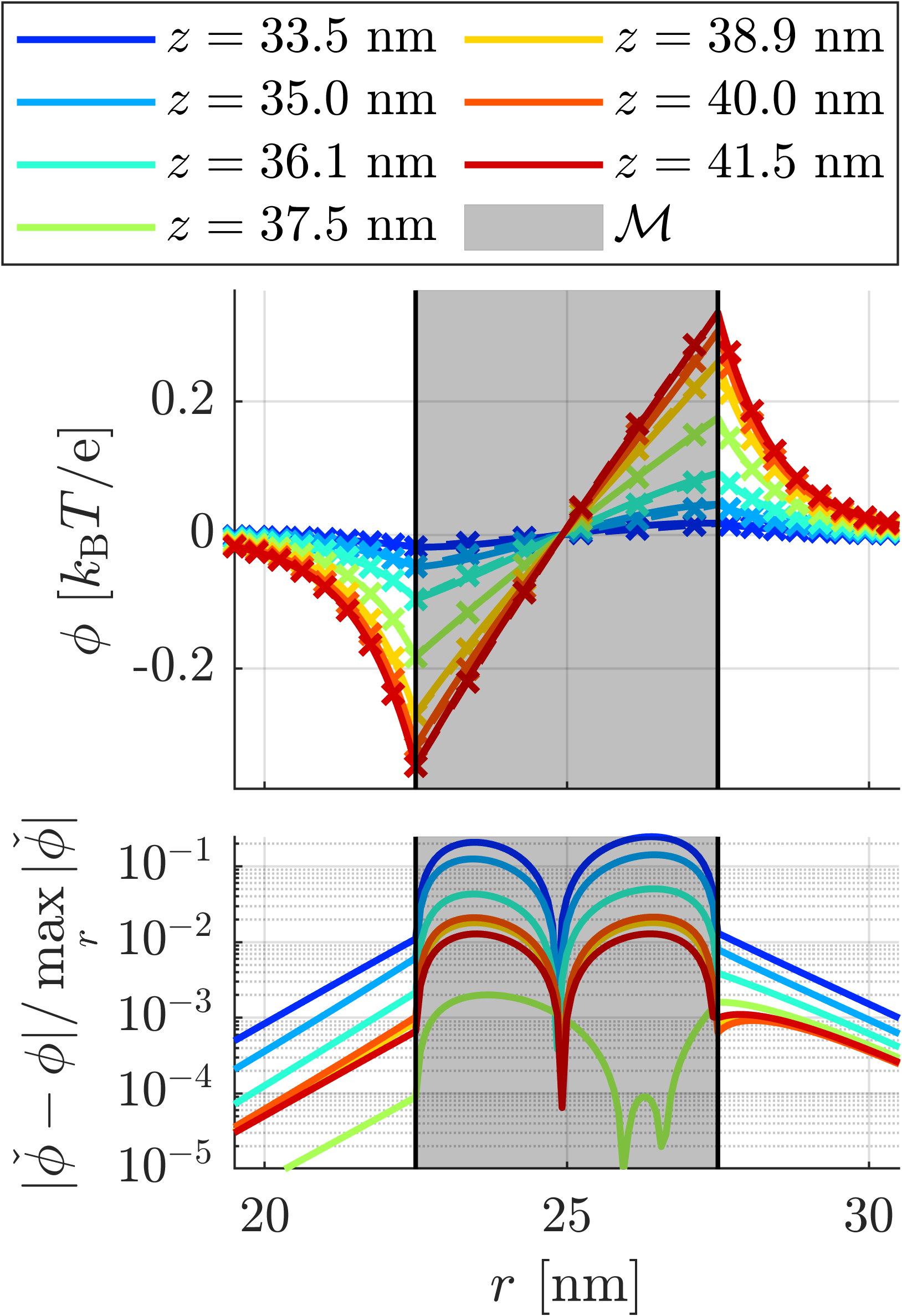}
         \caption{Potential profiles in the region of varying surface charge densities.}
         \label{fig:cyl_inc40_PP}
     \end{subfigure}
     \hfill
     \begin{subfigure}[b]{0.475\textwidth}
         \centering
         \includegraphics{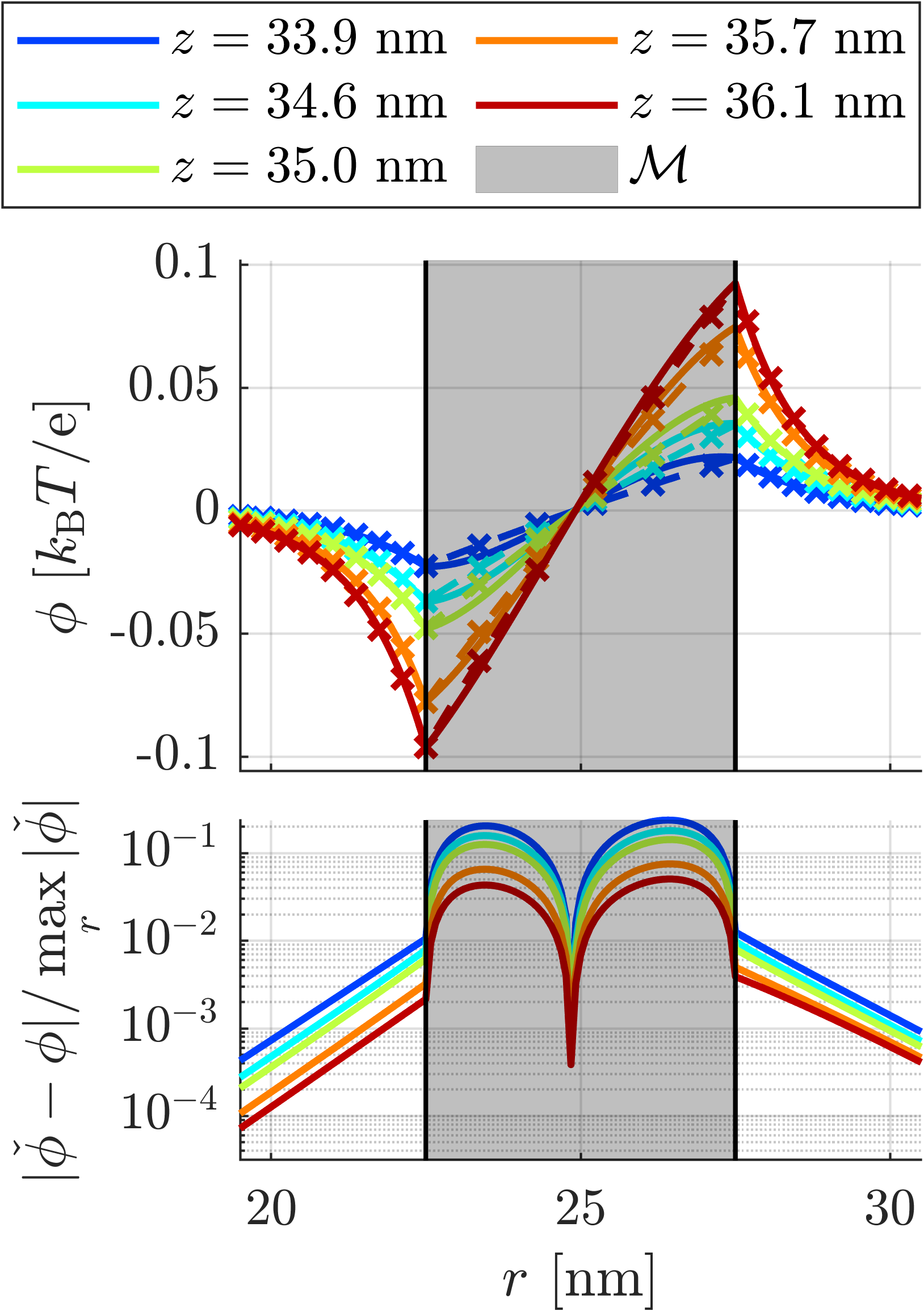}
         \caption{Potential profiles in the left transition region of the varying surface charge densities.}
         \label{fig:cyl_inc40_PPTR_left}
     \end{subfigure}
     \caption{Potential profiles (top) and relative errors (bottom) plotted at discrete values of $x$ for case B of the cylindrical membrane. The full lines represent the exact theory and the dashed lines (nearly indistinguishable in (a)) represent the $(2+\delta)$-dimensional theory. The error is only plotted down to $10^{-5}$. In (a), the values of $x$ are taken from the entire region of varying surface charge densities while (b) shows profiles from the left transition region between constant and linearly varying surface charge densities.}
     \label{fig:cyl_profile_PP_PPTR}
\end{figure}

The potential profile for case A is shown in Fig.~\ref{fig:cyl_caseA_phi} (top). Due to the cylinder's curvature, the solution is no longer linear in the membrane and the $(2+\delta)$-dimensional theory does not capture the solution exactly. However, the relative error, plotted in Fig.~\ref{fig:cyl_caseA_phi} (bottom), does not exceed $0.2\%$. Figure~\ref{fig:cs_error_conv_R0_caseA} shows that the $L_2$-error decreases quadratically with increasing mid-surface radius $R_0$. This is consistent with the assumption of the $(2+\delta)$-dimensional theory that the radius of curvature is large compared to the thickness, Eq.~\eqref{eq:dR_ll_1}. \textspace

In Fig.~\ref{fig:cyl_inc40_PP} (top), the potential profiles for case B are plotted at discrete values of $z$ along the cylinder. Again, the qualitative behavior of the potential is well approximated by the $(2+\delta)$-dimensional theory. The relative error is plotted in Fig.~\ref{fig:cyl_inc40_PP} (bottom), and, as before, does not exceed $20\%$ anywhere in the domain despite the additional error introduced by the curvature of the geometry. The largest error again appears in the transition region where the potential is small, as is shown in Fig.~\ref{fig:cyl_inc40_PPTR_left}. Similar to the flat case, the $L_2$-error decreases with order $1$ and about $1/2$ with increasing $L$ and $L_\mathrm{s}$, respectively, as shown in Figs.~\ref{fig:fcs_inc40_err_L} and~\ref{fig:fcs_inc40_err_Ls}. The error $\mathcal{E}_{z}\leftR(r\rightR)$ decreases similarly to Figs.~\ref{fig:flat_L2x_LwidthU} and~\ref{fig:flat_L2x_LsmoothU} and is thus omitted here. Therefore, we conclude, as before, that the error is small when the characteristic in-plane length scale is large compared the thickness of the membrane. Figure~\ref{fig:cs_error_conv_R0_caseB} shows how the error for case B changes with increasing radius, for $L = 20~\mathrm{nm}$ and $L_\mathrm{s} = 10~\mathrm{nm}$. As compared to case A, the error does not converge quadratically but instead saturates. This is a result of the error due to in-plane surface charge density changes dominating over the error due to the curvature of the cylinder, and we expect the same scaling as in Fig.~\ref{fig:cs_error_conv_R0_caseA} for smaller radii of curvature.\textspace
\subsubsection{Spheres} \label{sec:num_sphere}

\begin{figure}[ht]
    \centering
    \begin{subfigure}[t]{0.48\textwidth}
    \centering
    \imagebox{3.1in}{
    \begin{annotationimage}[]{width=2.75in}{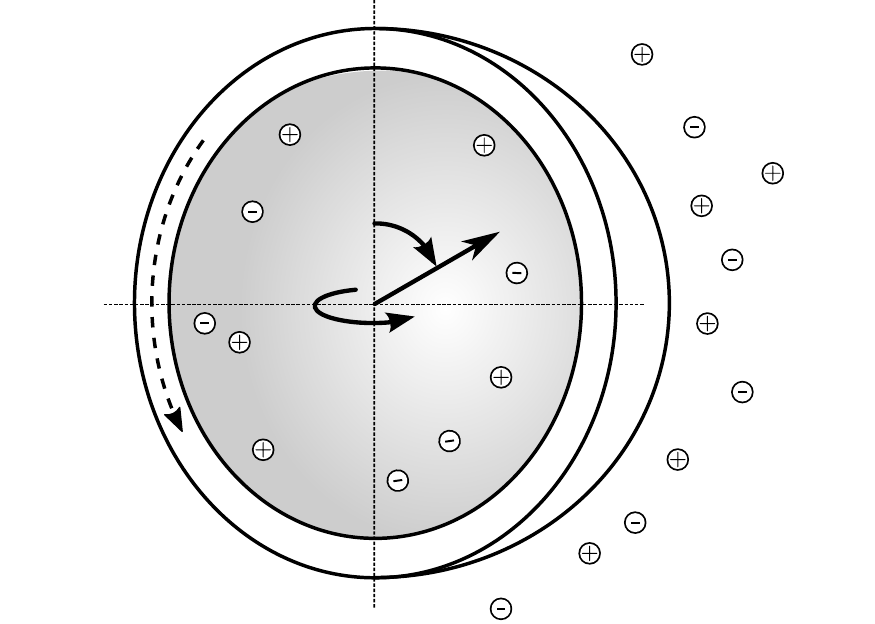}
        \imagelabelset{
                coordinate label style/.style = {
                rectangle,
                fill = none,
                text = black,
                font = \normalfont
        }}
        \draw[coordinate label = {$\sigma^\pm\leftR(\Theta\rightR)$ at (0.07,0.6)}];
        \draw[coordinate label = {$\Theta$ at (0.49,0.67)}];
        \draw[coordinate label = {$\bm{e}_r$ at (0.59,0.65)}];
        \draw[coordinate label = {$\Phi$ at (0.37,0.44)}];
    \end{annotationimage}}
    \ifcap{\caption{Schematic of a sphere with spatially varying surface charge densities.}}{\vspace{1cm}}
    \label{fig:sphere_num_setup}
    \end{subfigure}%
    \hfill
    \begin{subfigure}[t]{0.47\textwidth}
        \centering
        \imagebox{3.1in}{
        \includegraphics[scale=1]{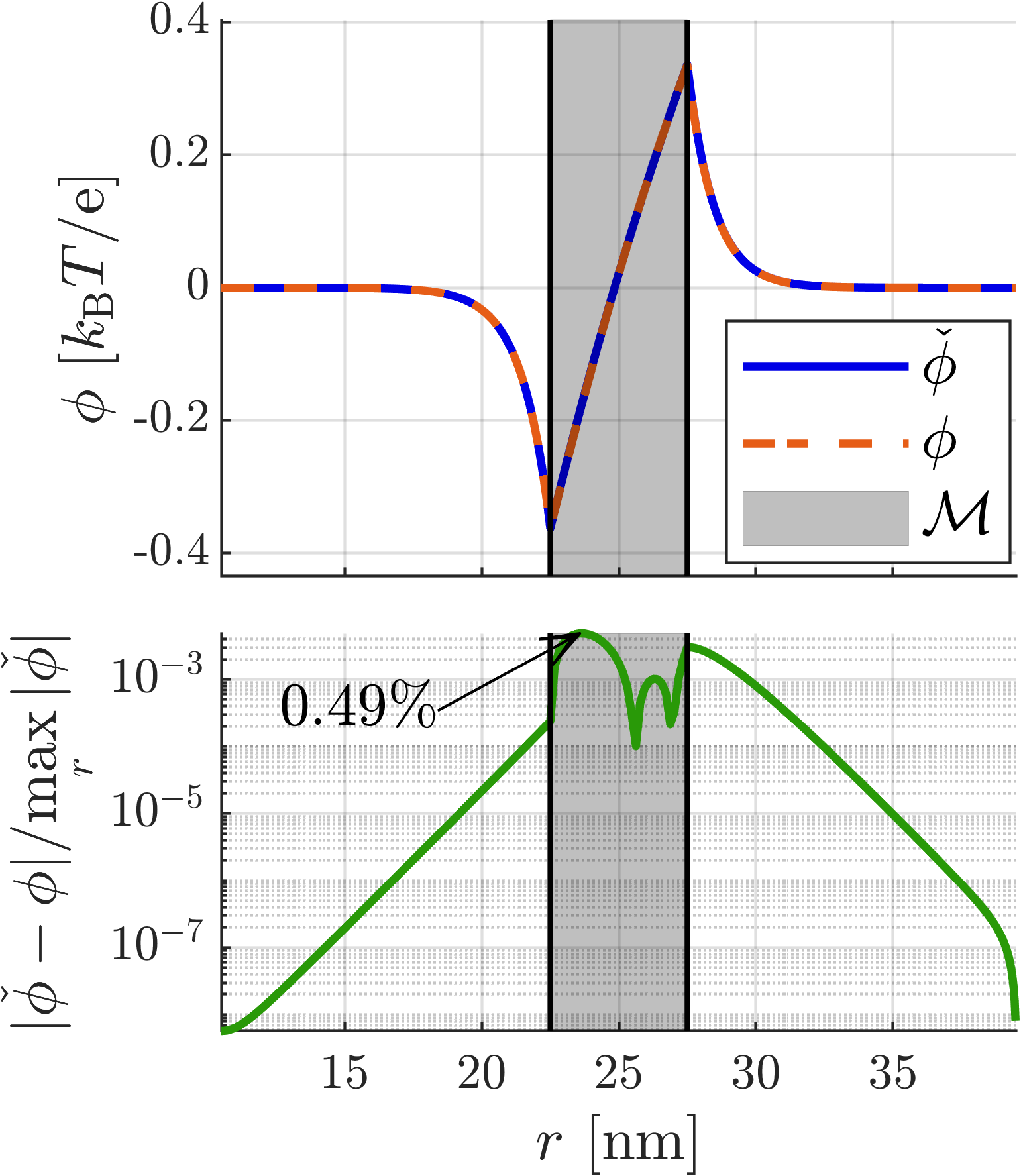}}
        \caption{Potential profiles (top) and relative error (bottom) for constant surface charge densities for the exact and $(2+\delta)$-dimensional theories.}
        \label{fig:sphere_caseA_phi}
    \end{subfigure}
    \caption{Schematic setup (a) and potential profiles (b) for the spherical lipid membrane embedded in a symmetric, monovalent electrolyte.}
    \label{fig:sphere_schematic_profile}
\end{figure}

Consider a sphere with mid-surface radius $R_0$, shown schematically in Fig.~\ref{fig:sphere_num_setup}. The surface charge density is chosen to only depend on the $\Theta$-direction, i.e. $s \equiv \Theta R_0 $ in Eq.~\eqref{eq:sigma_varying}, and is thus axisymmetric along the $\Phi$-direction. Similar to the cylindrical membrane, the problem is closed with the boundary conditions
\begin{alignat}{2}
    \frac{\partial \phi}{\partial r} \biggr\rvert_{r = 0} &= 0~, \hspace{2cm}
    \phi \bigr\rvert_{r = R_0 + \delta/2 + L_{\mathcal{B}2}} &&= 0~, \\[8pt]
    \frac{\partial \phi}{\partial \Theta} \biggr\rvert_{\Theta R_0 = (\pi R_0 - L_{\mathcal{B}1})/2} &= 0~, \hspace{2cm}
    \frac{\partial \phi}{\partial \Theta} \biggr\rvert_{\Theta R_0 = (\pi R_0 + L_{\mathcal{B}1})/2} &&= 0~.
\end{alignat}
The radius of the sphere's mid-surface is chosen as $R_0 = 25~\mathrm{nm}$ and the remaining geometric and material parameters are listed in Tabs.~\ref{tab:charge_props} and~\ref{tab:geom_mat_props}.  \textspace

\begin{figure}[t]
     \centering
     \begin{subfigure}[b]{0.475\textwidth}
         \centering
         \includegraphics{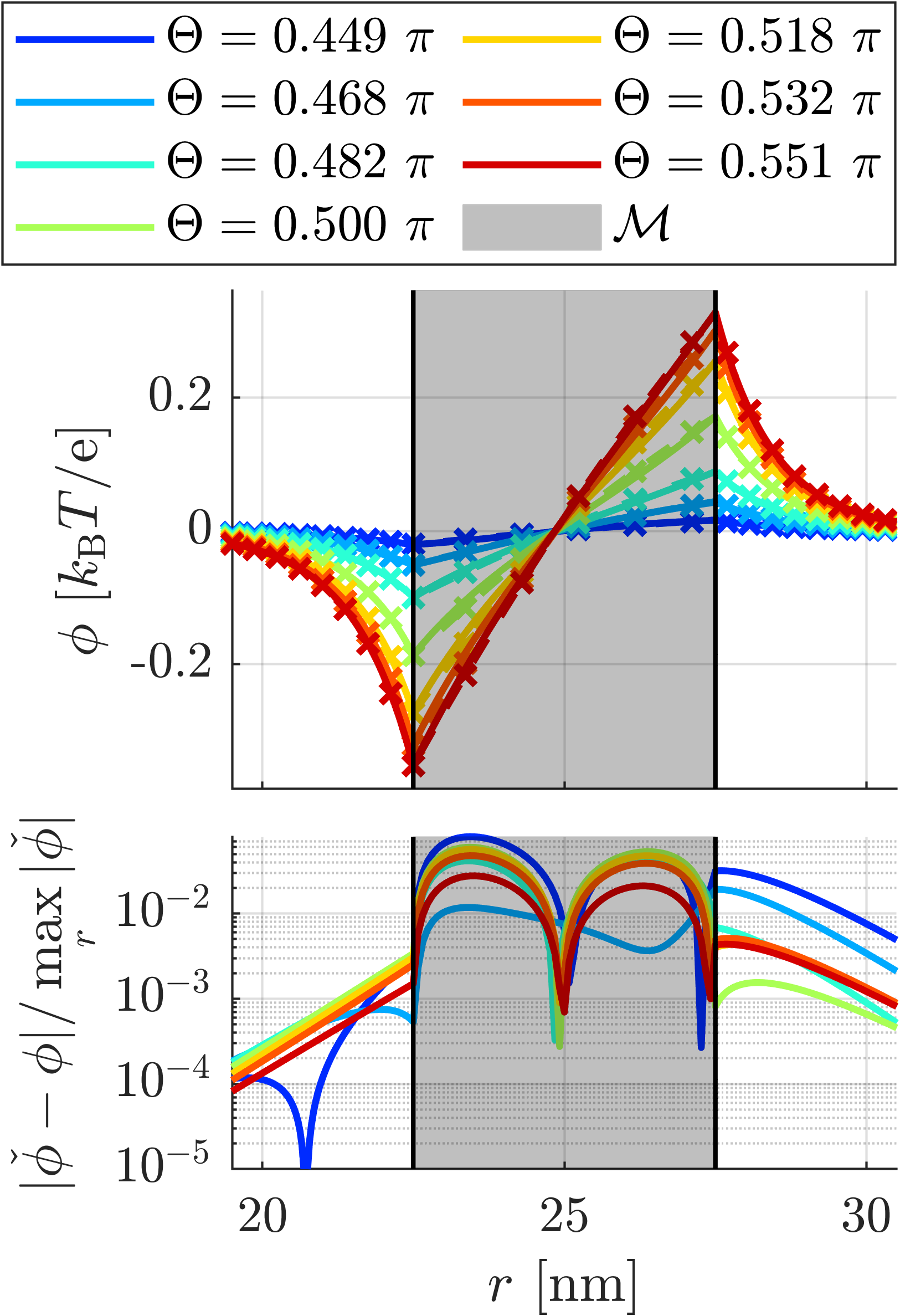}
         \caption{Potential profiles in the region of varying surface charge densities.}
         \label{fig:sphere_inc40_PP}
     \end{subfigure}
     \hfill
     \begin{subfigure}[b]{0.475\textwidth}
         \centering
         \includegraphics{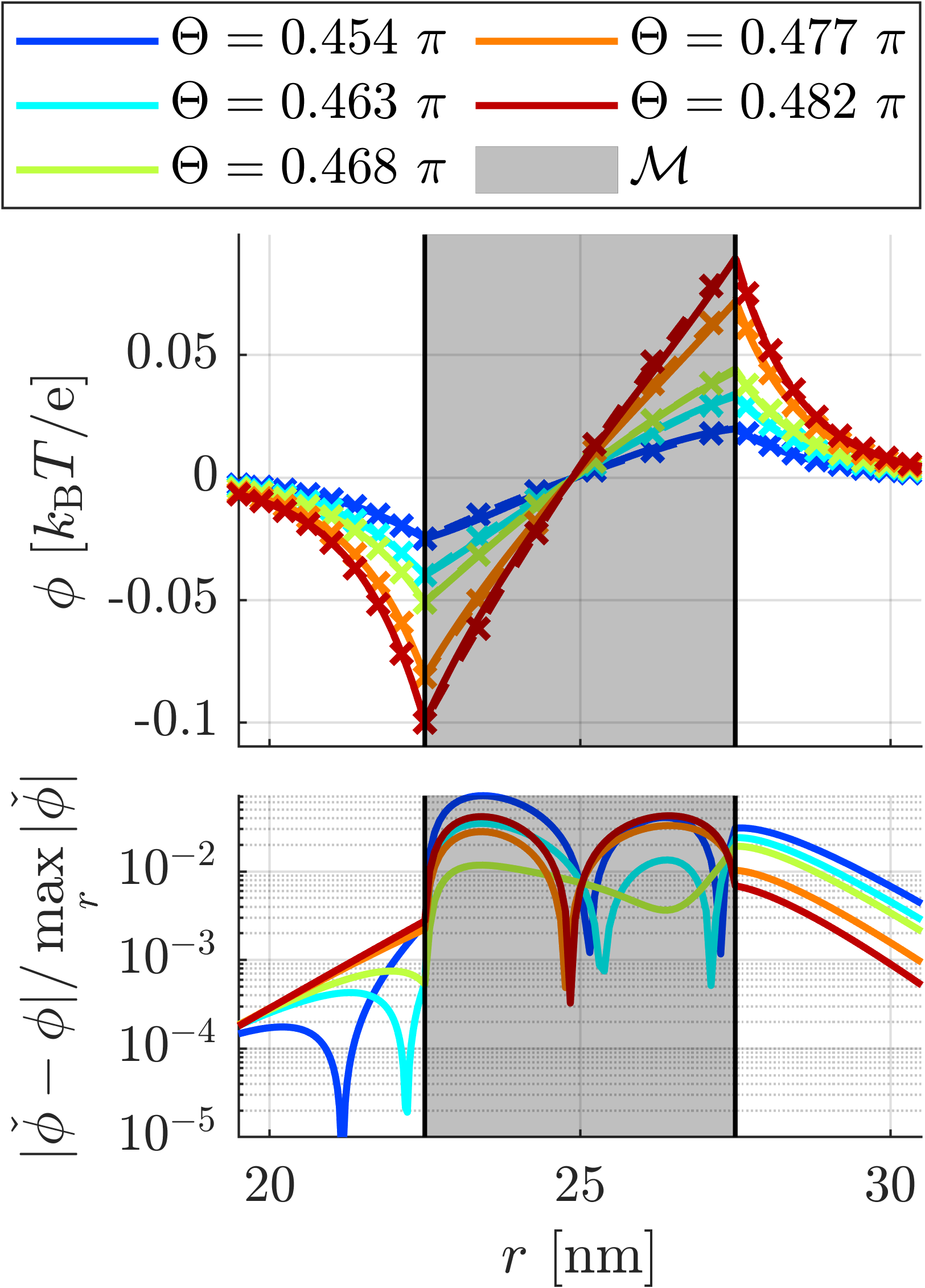}
         \caption{Potential profiles in the left transition region of the varying surface charge densities.}
         \label{fig:sphere_inc40_PPTR_left}
     \end{subfigure}
     \caption{Potential profiles (top) and relative errors (bottom) plotted at discrete values of $x$ for case B of the spherical membrane. The full lines represent the exact theory and the dashed lines (nearly indistinguishable) represent the $(2+\delta)$-dimensional theory. The error is only plotted down to $10^{-5}$. In (a), the values of $x$ are taken from the entire region of varying surface charge densities while (b) shows profiles from the left transition region between constant and linearly varying surface charge densities.}
     \label{fig:sphere_profile_PP_PPTR}
\end{figure}

For case A, Fig.~\ref{fig:sphere_caseA_phi} (top) shows excellent qualitative agreement between the exact and $(2+\delta)$-dimensional theories while Fig.~\ref{fig:sphere_caseA_phi} (bottom) shows that the relative error does not exceed $0.5\%$. As with the cylinder, the error reduces quadratically with increasing radius, as shown in Fig.~\ref{fig:cs_error_conv_R0_caseA}. For case B, the potential profiles for the exact and $(2+\delta)$-dimensional theories at discrete values of $\Theta$ are plotted in Fig.~\ref{fig:sphere_inc40_PP} (top), showing good qualitative agreement. Figure~\ref{fig:sphere_inc40_PP} (bottom) shows that the corresponding relative error does not exceed $10\%$, with the error again being largest in the transition region, as seen in Fig.~\ref{fig:sphere_inc40_PPTR_left}. The decrease in error with increasing $L$, $L_\mathrm{s}$, and $R_0$ is consistent with the results for the flat and cylindrical geometries, as shown in Figs.~\ref{fig:fcs_inc40_err_L},~\ref{fig:fcs_inc40_err_Ls}, and~\ref{fig:cs_error_conv_R0_caseB}, respectively.

\section{Conclusion and Outlook}
A theory describing the electromechanics of lipid membranes requires resolving the electric potential across their thickness. This requirement is incompatible with treating lipid membranes as strictly two-dimensional surfaces, a common approach to modeling lipid membrane mechanics. Nonetheless, surface theories have both analytical and numerical advantages, motivating the derivation of a novel, effective surface theory for the electromechanics of lipid membranes in this sequence of articles. \textspace

We start from a three-dimensional model and propose a new dimension reduction procedure that assumes a low-order solution expansion along the lipid membrane thickness. Expanding using orthogonal polynomials allows us to derive new differential equations for the expansion coefficients. These equations are not dependent on the thickness direction but account for the finite thickness of lipid membranes. Therefore, we refer to such dimensionally-reduced, effective surface theory as $(2+\delta)$-dimensional. Applying the proposed dimension reduction procedure to the electrostatics of lipid membranes yields an effective surface form of Gauss' law. Using both analytical and numerical comparisons, we show excellent qualitative agreement between the three-dimensional and $(2+\delta)$-dimensional theories. The two theories also show excellent quantitative agreement when the electric potential changes over length scales larger than the lipid membrane thickness, consistent with the assumptions of the theory. \textspace

Similar approaches to derive dimensionally-reduced theories for the electromechanics of thin films were proposed by Green and Naghdi \cite{green1983electromagnetic} and Khoma \cite{khoma1983construction} based on Legendre polynomials. However, the authors do not make their order of expansion precise, only giving general equations for the expansion coefficients. This generality makes their theories largely intractable, and we are unaware of any practical applications beyond the examples discussed in Ref. \cite{green1983electromagnetic}. \textspace

%Green and Naghdi \cite{green1983electromagnetic} and Khoma \cite{khoma1983construction} independently formulated similar approaches based on Legendre polynomials to describe the electromechanics of thin films. However, the authors do not make their order of expansion precise, only giving general equations for the expansion coefficients. This generality makes their theories largely intractable, and we are unaware of any practical applications beyond the examples discussed in Ref. \cite{green1983electromagnetic}. \textspace

Edmiston and Steigmann \cite{edmiston2011analysis} also derive a dimensionally-reduced theory for the electrostatics of thin films but consider the limit of vanishing thickness, $\delta \rightarrow 0$. The authors assume equal and opposite surface charge densities on $\mathcal{S}^+$ and $\mathcal{S}^-$ and neglect fields external to the thin film. However, their theory can be easily generalized to account for arbitrary surface charge densities and external electric fields. This allows comparing the Edmiston-Steigmann theory to the $(2+\delta)$-dimensional theory in the limit of vanishing thickness. In this limit, the $(2+\delta)$-dimensional theory produces the same normal component of the electric field as the Edmiston-Steigmann theory. The $\phik{2}$ contribution in Eq.~\eqref{eq:surface_gauss} does not appear in the Edmiston-Steigmann theory, which is expected considering $\phik{2} \rightarrow 0$ as $\delta \rightarrow 0$. However, the $\phik{1}$ contribution in Eq.~\eqref{eq:surface_gauss} is also absent in the Edmiston-Steigmann theory even though it remains non-zero in the limit of vanishing thickness. Thus, we find that generalizing the theory of Edmiston and Steigmann \cite{edmiston2011analysis} does not correspond to the $(2+\delta)$-dimensional theory in the limit of vanishing thickness.\textspace

The leaky dielectric model (LDM), originally devised by Melcher and Taylor for droplets in weak electrolytes \cite{taylor1966studies,melcher1969electrohydrodynamics}, is often invoked to describe lipid vesicles in an external electric field \cite{vlahovska2019electrohydrodynamics,wubshet2022differential}. The LDM describes a droplet or vesicle with radius $R$ much larger than the Debye length $\lambda_\mathrm{D}$, exposed to an electric field that is large compared to the thermal voltage (Baygents-Saville limit). The Baygents-Saville limit allows for a macroscopic description that coarsegrains the genuine interface and its diffuse layer into an effective interface \cite{saville1997electrohydrodynamics,schnitzer2015taylor}, thus not capturing electrokinetic effects on the length scale of the diffuse layer. In contrast, the $(2+\delta)$-dimensional theory takes a microscopic perspective and describes a material interface without making any assumption about the bulk fluid domains. Hence, the LDM and $(2+\delta)$-dimensional theory describe electric field effects on different length scales and are thus not comparable. Instead, the $(2+\delta)$-dimensional theory should serve as a starting point for deriving the LDM for lipid vesicles, a derivation currently missing from the literature. \textspace 
%Finally, we note that we are currently unaware of any derivation of the LDM specific to lipid vesicles. 

Recently, Ma et al. \cite{ma2022model} proposed a model similar to the LDM, specific to lipid vesicles but valid in the strong electrolyte limit---as opposed to the LDM which is valid in the weak electrolyte limit. Their microscopic electrostatics model assumes equal and opposite surface charge densities and continuous electric displacements across the membrane. However, according to the $(2+\delta)$-dimensional theory, the latter would only be valid in the limit of vanishing thickness. Furthermore, their microscopic electrostatics model is not consistent with the potential drop derived in Eq.~\eqref{eq:jump_phiM}. The effect of adopting the $(2+\delta)$-dimensional theory as a starting point in the derivation of the model by Ma et al. \cite{ma2022model} as well as the LDM (see \cite{baygents1990circulation, saville1997electrohydrodynamics,zholkovskij2002electrokinetic,schnitzer2015taylor, mori2018electrodiffusion}) currently remains an open question and merits future investigation.\textspace

This article is the first in a series of three that systematically derives the governing equations describing the electromechanics of lipid membranes. In subsequent articles, the dimension reduction procedure proposed in this article is applied to the mechanical balance laws and appropriate constitutive equations, yielding a complete and self-consistent theory of the electromechanics of lipid membranes. 

\section*{Acknowledgements}
This work was
supported by the Director, Office of Science, Office of Basic Energy Sciences, of the U.S. Department of Energy under Contract No. DEAC02-05CH1123. KKM also acknowledges the support from the Hellmann Fellowship.

	%
	% *** REFERENCES
	%

	\newpage
	\addcontentsline{toc}{section}{References}
	\bibliographystyle{electrostatics-article/bibliography/bibStyle}

	\bibliography{electrostatics-article/bibliography/bibliography_main.bib}

\end{document}